\documentclass[iop,apj]{emulateapj}
\usepackage{amsmath}
\usepackage{txfonts}
\usepackage{graphicx}
\usepackage{threeparttable}
\usepackage{latexsym}
\usepackage{amsbsy}
\usepackage{float}

\begin{document}

\title{Mask effects on cosmological studies with weak lensing peak statistics}
\author{Xiangkun Liu$^1$, Qiao Wang$^2$, Chuzhong Pan$^1$, Zuhui Fan$^1$}
\affil{$^1$Department of Astronomy, Peking University, Beijing 100871, China; fanzuhui@pku.edu.cn\\
$^2$National Astronomical Observatories, Chinese Academy of Science, Beijing 100012, China}

\begin{abstract}
With numerical simulations, we analyze in detail how the bad data removal, i.e., the mask effect, can influence
the peak statistics of the weak lensing convergence field reconstructed from the shear measurement of background galaxies.
It is found that high peak fractions are systematically enhanced due to the
presence of masks, the larger the masked area, the higher the enhancement. In the case that the total masked area is
about $13\%$ of the survey area, the fraction of peaks with signal-to-noise ratio $\nu\ge 3$ is $\sim 11\%$ of the total
number of peaks, in comparison with $\sim 7\%$ of the mask-free case in our considered cosmological model.
This can have significant effects on cosmological studies with
weak lensing convergence peak statistics, inducing a large bias in the parameter constraints if the
effects are not taken into account properly. Even for a survey area of $9\hbox{ deg}^2$, the bias in $(\Omega_m, \sigma_8)$
is already intolerably large and close to $3\sigma$. It is noted that most of the affected peaks are close to
the masked regions. Therefore excluding peaks in those regions in the peak statistics can reduce the bias effect but at the
expense of losing usable survey areas. Further investigations find that the enhancement of the
number of high peaks around the masked regions can be largely attributed to the fewer number of galaxies usable in
the weak lensing convergence reconstruction leading to higher noise than that of the areas
away from the masks. We thus develop a model in which we exclude only those very large masks with radius larger than $3\arcmin$
but keep all the other masked regions in peak counting statistics. For the remained part,
we treat the areas close to and away from the masked regions separately with different noise levels.
It is shown that this two-noise-level model can account for the mask effect on peak statistics very well,
and the bias in cosmological parameters is significantly reduced if this model is applied in the parameter fitting.
%\textbf{(The final sentence is deleted.)}
\end{abstract}
%cosmology - dark matter - clusters: general - gravitational lensing: weak - large-scale structure of universe}
\maketitle

\section{Introduction}

Gravitationally induced weak lensing effects have emerged as one of the most important probes in cosmological studies
(e.g., Bartelmann \& Schneider 2001; Albrecht et al. 2006; Amendola et al. 2012; Abate et al. 2012;
Heymans et al. 2012; Erben et al. 2013; Simpson et al. 2013; Kilbinger et al. 2013). Besides the shear two-point
correlation analyses, weak lensing peak statistics can provide important and complementary information
especially considering that the structure formation is a nonlinear process (e.g., White et al. 2002; Hamana et al. 2004;
Tang \& Fan 2005; Hennawi \& Spergel 2005; Marian et al. 2009; Dietrich \& Hartlap 2010; Kratochvil et al. 2010;
Marian et al. 2012; Hilbert et al. 2012). Current observations have proved the feasibility
of detecting massive clusters from weak lensing peak identifications (e.g., Wittman et al. 2006; Gavazzi \& Soucail 2007;
Shan et al. 2012). Future weak lensing surveys will be able to provide a large number of peaks with high signal-to-noise ratio
and therefore their statistics should expectedly be able to contribute significantly to precision cosmological studies.
On the other hand, it is known that many effects can affect profoundly the
weak lensing peak statistics. The complex mass distribution of clusters of galaxies and the
projection effects of large-scale structures along line of sights prevent
us from linking weak lensing peaks to single clusters in a simple way (e.g., Tang \& Fan 2005; Yang et al. 2011;
Yang et al. 2012; Hamana et al. 2012). The intrinsic ellipticities of source galaxies generate large noise that not only
produces false peaks through their chance alignments (e.g., van Waerbeke 2000; Fan 2007) but also affects the
true peak signals from massive clusters significantly \citep[][hereafter F10]{F2010}. Furthermore, various observational
effects can also have large impacts on weak lensing peak statistics if they are
not taken into account properly. The full realization of the power of weak lensing analyses in
future cosmological studies relies on our thorough understandings about different systematics.

Weak lensing observations target at far away background galaxies, and bad data occurrences are unavoidable
and they should be masked out carefully (e.g., Heymans et al. 2012; Erben et al. 2013).
These masks can occupy $\sim 10\%$ to $\sim 20\%$ of the
total survey area and result irregular survey boundaries and artificial voids
in the background galaxy distribution, which in turn can affect the weak lensing analyses considerably.
The mask effects on the shear power spectrum estimation and on the weak lensing Minkowski Functionals
have been investigated recently (Hikage et al. 2011; Shirasaki et al. 2013).
%In Hikage et al. (2011), they investigate the mask effect on the shear power spectrum estimation and propose
%the pseudo-spectrum method to account for the effect.
Considering weak lensing peak statistics, it involves in one way or another the reconstruction of the mass distribution
from the shape measurements of background galaxies. The so called shear peak statistics is based on the aperture mass map
which is the smoothed convergence field with a compensated filter (e.g., Schneider et al. 1998; Marian et al 2012).
It is theoretically shown that the aperture mass at a spatial location $\boldsymbol x_0$ can be obtained by applying a suitable filter
to the tangential shear field with respect to $\boldsymbol x_0$ (e.g., Schneider 1996).
The filter to the tangential shear field can be derived from the compensated filter to the convergence field.
Alternatively, we can apply a filter to the full shear field (not the tangential component) to obtain the smoothed shear field
and then from it to reconstruct the smoothed convergence field (e.g., van Waerbeke et al. 2013).
It should be noted that in this later approach, the filtering process is also applied directly to the
shear field but not to the noisy convergence field reconstructed from the unsmoothed shear field.
%Considering convergence peaks, because the convergence field is reconstructed from the shape measurements of background galaxies,
Thus the devoidness of galaxies in masked regions affects inevitably the reconstructed mass map
and consequently the weak lensing peak statistics.
In this paper, with numerical simulations, we perform detailed studies of the mask effect on weak lensing peak statistics
and the derived cosmological parameter constraints.
%\textbf{We run the cosmological simulations and generate shear and convergence maps by Ray-tracing
%as the fiducial inputs for weak lensing signal calculations.}
Specifically, we run sets of dark-matter-only N-body simulations and generate shear and convergence maps by ray-tracing.
Background galaxies with intrinsic ellipticities are randomly populated
and `observed' ellipticities including the shear signals from simulations are then
constructed for each galaxies. The masks are generated according to the mask size distribution from Shan et al. (2012),
and are given spatial positions randomly in our statistical analyses. We then remove galaxies inside the masks.
To obtain the weak lensing mass distribution, we adopt the above mentioned second approach to
reconstruct the smoothed convergence field from the smoothed shear field obtained from the remaining galaxies.
The peak statistics is analyzed and compared with the case without masks.
%We further explore different methods to correct the mask effects.

The rest of the paper is organized as follows. In \S 2, we introduce the lensing basics related to
our studies, including the convergence reconstruction method. In \S 3, we describe the simulations and the ray-tracing method.
The generation of the `observed galaxy ellipticities' and the reconstruction of the convergence field from them without and with masks
are presented. In \S 4, the theoretical model of F10 used in our peak statistical analyses is given,
and its applicability is studied by comparing with the results from simulations.
\S 5 contains the main results of our analyses of the mask effects on weak lensing peak statistics. Summary and discussions are given in \S 6.

\section{Theoretical basics}
%\subsection{Convergence reconstruction}

Observationally, the weak lensing effect is mostly extracted from the shape distortion measurements of background
galaxies, which is directly related to the weak lensing shear components.
On the other hand, for weak lensing peak statistics, it targets at high peaks in the large-scale mass distribution,
and therefore is more directly linked to the lensing convergence which is the weighted projection of the density distribution along the line of sight.
The convergence and the shear are not independent quantities and all determined by the lensing potential.
Thus we can derive the mass distribution from the observed shape measurements as described in the following.
%On the other hand, the lensing convergence is the weighted projection of the density distribution along the line of sight, and thus more
%visually linked to the structures we are interested in. In this paper, we concentrate on the convergence peak analyses.

Considering small source galaxies, their linear-order image distortion from the gravitational lensing effect of a
single lens can be described by the Jacobian matrix given by (e.g., Schneider et al. 1992)
\begin{equation}
A=\bigg (\delta_{ij}-\frac {\partial^2 \psi(\boldsymbol{\theta})}{\partial \theta_i\partial \theta_j} \bigg )
=
\begin{pmatrix}
 1-\kappa-\gamma_1  & -\gamma_2 \\
 -\gamma_2 & 1-\kappa+\gamma_1
\end{pmatrix},
\label{eq1}
\end{equation}
where $\kappa$ is the lensing convergence and $\gamma_1$ and $\gamma_2$ are the two shear components with
\begin{equation}
\kappa=\frac{1}{2}\nabla^2\psi, \quad
\gamma_1=\frac{1}{2}\bigg (\frac{\partial^2\psi}{\partial^2\theta_1}-\frac{\partial^2\psi}{\partial^2\theta_2}\bigg ), \quad
\gamma_2=\frac{\partial^2\psi}{\partial \theta_1 \partial \theta_2}.
\label{eq2}
\end{equation}
The lensing potential $\psi$ is determined by the surface mass density of the lens through
\begin{equation}
\psi(\boldsymbol{\theta})={1\over \pi}\int d^2\boldsymbol {\theta^{'}}\frac{\Sigma(\boldsymbol {\theta^{'}})}{\Sigma_{cr}}
\ln |\boldsymbol {\theta}-\boldsymbol {\theta^{'}}|,
\label{eq3}
\end{equation}
where $\Sigma_{cr}$ is the critical surface mass density given by
\begin{equation}
\Sigma_{cr}={c^2\over 4\pi G}{D_s\over D_l D_{ls}}
\label{eq4}
\end{equation}
with $D_l, D_s$ and $D_{ls}$ being the angular diameter distances from the observer to the lens, to the source, and
between the lens and the source. It can be seen that $\kappa=\Sigma/\Sigma_{cr}$.
For weak lensing effects from large-scale structures beyond a single lens, under the Born approximation
the above formulations still hold except the lensing convergence is given by, in the case of a fixed source position
(e.g., Bartelmann \& Schneider 2001),
\begin{equation}
\kappa_{eff}=\frac{3H_0^2\Omega_m}{2c^2}\int_0^w dw'\frac{f_K(w')f_K(w-w')}{f_K(w)}\frac{\delta[f_K(w')\boldsymbol {\theta},w']}{a(w')},
\label{eq5}
\end{equation}
where $w$ is the comoving radial distance, $f_K$ is the comoving angular diameter distance, $a$ is the scale factor of the universe,
and $\delta$ is the density perturbation along the line of sight.

The image distortion is then described by $\kappa$ and $\gamma_{i}$ with the quantity
$(\det A)^{-1}=[(1-\kappa)^2-|\boldsymbol{\gamma}|^2]^{-1}$ giving rise to the flux magnification
($|\boldsymbol{\gamma}|=(\gamma_1^2+\gamma_2^2)^{1/2}$),
and the eigen values of $A$ related to the axial length.
Specifically, the lensing effect makes a circular source appear as an ellipse with the axial ratio of
\begin{equation}
\frac{a^2}{b^2}=\frac{1-\kappa-|\boldsymbol{\gamma}|}{1-\kappa+|\boldsymbol{\gamma}|}=\frac{1-|\boldsymbol{g}|}{1+|\boldsymbol{g}|},
\label{eq6}
\end{equation}
where $g_{i}=\gamma_i/(1-\kappa)$ is named as the reduced shear component.
Thus for ideally circular sources, we can obtain $\boldsymbol{g}$ by accurately measuring the shape of the sources, and further
reconstruct the convergence $\kappa$ from the relation between $\kappa$ and $\gamma$, which in the Fourier space can be
written as (e.g., Kaiser \& Squires 1993)
\begin{equation}
\hat{\gamma}(\boldsymbol{l})=\pi^{-1}\hat{D}(\boldsymbol{l})\hat{\kappa}(\boldsymbol{l}),
\label{rk}
\end{equation}
where $\hat{D}$ is given by
\begin{equation}
\hat{D}(\boldsymbol{l})=\pi\frac{l_{1}^{2}-l_{2}^{2}+2il_{1}l_{2}}{|\boldsymbol{l}|^2}.
\label{dl}
\end{equation}

However, galaxies have intrinsic ellipticities. The complex ellipticity of the lensing distorted image $\boldsymbol {\epsilon}$
with $|\boldsymbol {\epsilon}|=(1-b/a)/(1+b/a)$ is then related to the intrinsic ones $\boldsymbol{\epsilon}_{s}$ by the following relation
(e.g., Seitz \& Schneider 1997)
\begin{equation}
\boldsymbol {\epsilon}=\left\{ \begin{array}{ll} \frac{\boldsymbol{\epsilon}_s-\boldsymbol{g}}{1-\boldsymbol{g}^{*}\boldsymbol{\epsilon}_s}
& \textrm{for $\vert{\boldsymbol{g}}\vert\leq 1$}\\ \\\frac{1-\boldsymbol{g}\boldsymbol{\epsilon}_s^{*}}{\boldsymbol{\epsilon}_s^{*}-\boldsymbol{g}^{*}}
& \textrm{for $\vert{\boldsymbol{g}}\vert>1$}
 \end{array}\right.
\label{eobs}
\end{equation}
where `*' represents the complex conjugation. It has been shown that the average of $\boldsymbol{\epsilon}$ gives rise to the unbiased estimate of
$\boldsymbol{g}$ and $1/\boldsymbol{g}$ for $|\boldsymbol g|\le 1$ and $|\boldsymbol g|>1$, respectively (Seitz \& Schneider 1997).
In the case of $\kappa\ll1$ and $|\gamma|\ll1$, we have $\boldsymbol g\approx \boldsymbol \gamma$.

Eq.(\ref{rk}) and (\ref{dl}) show that theoretically we can obtain the mass distribution which is related to the convergence
field from the observed $\langle\boldsymbol \epsilon\rangle$. The aperture mass peak statistics is based on the quantity
$M_{ap}(\boldsymbol \theta)=\int d^2\boldsymbol \theta' \kappa(\boldsymbol \theta')U(|\boldsymbol \theta'-\boldsymbol \theta|)$,
where the function $U$ is a compensated filter satisfying $U(|\boldsymbol \theta|)=0$ for
$|\boldsymbol \theta| > |\boldsymbol \theta_0|$ and
$\int_0^{|\boldsymbol \theta_0|}\  |\boldsymbol \theta| d|\boldsymbol \theta|\  U(|\boldsymbol \theta|)=0$.
From the relation between $\kappa$ and $\boldsymbol \gamma$, it is shown that $M_{ap}$ can be obtained directly from the tangential component
of the shear with $M_{ap}(\boldsymbol \theta)=\int d^2\boldsymbol \theta' \gamma_t(\boldsymbol \theta'; \boldsymbol \theta)Q(|\boldsymbol \theta'|)$,
where $\gamma_t(\boldsymbol \theta'; \boldsymbol \theta)$ is the tangential shear component at $\boldsymbol \theta'$ with respect to
$\boldsymbol \theta$. The filter function $Q$ can be derived from $U$ with
$Q(|\boldsymbol \theta|)=2/|\boldsymbol \theta|^2\int_0^{|\boldsymbol \theta|} |\boldsymbol \theta'|d|\boldsymbol \theta'|
U(|\boldsymbol \theta'|)-U(|\boldsymbol \theta|)$ (e.g., Schneider 1996). Therefore if
the approximation $\langle\boldsymbol{\epsilon}\rangle\approx \boldsymbol \gamma$ is valid, one can directly obtain $M_{ap}$ from the
tangential component of the observed ellipticities $\langle\boldsymbol{\epsilon}_t\rangle$. For peak analyses, we are interested in high peaks
which are related to massive halos. In those regions, $\boldsymbol g\approx \boldsymbol \gamma$ is not a good approximation
and thus $M_{ap}$ obtained from $\langle\boldsymbol{\epsilon}_t\rangle$ with the filter function $Q$ is not exactly equivalent to
$M_{ap}$ defined through the convergence $\kappa$ with the filter function $U$. Thus there can have some complications if one
wants to theoretically link $M_{ap}$ from observations to the properties of $\kappa$ due to the nonlinear relation
between $\boldsymbol g$ and $\boldsymbol \gamma$.

Another approach to derive the mass distribution from the observed ellipticities $\boldsymbol \epsilon$ is to first obtain the
smoothed field of the full $\langle\boldsymbol \epsilon\rangle$, and then go through the nonlinear reconstruction process to get the smoothed
convergence field $\kappa$ (e.g., van Waebeke 2013). This is the approach we adopt in this paper. It is noted that the
smoothing/filtering here is still applied directly to the observed ellipticities.
From Eq.(\ref{eobs}), we can construct the distortion $\boldsymbol {\delta}$ by (e.g., Schneider \& Seitz 1995)
\begin{equation}
\boldsymbol{\delta}=\frac{2\langle\boldsymbol {\epsilon}\rangle}{1+|\langle\boldsymbol {\epsilon}\rangle|^2}=\frac{2\boldsymbol {g}}{1+|\boldsymbol {g}|^2}
\label{eq10}
\end{equation}
as the observed quantity, which is independent of $|\boldsymbol g|\le 1$ or $>1$. One can then solve for $\boldsymbol {\gamma}$
by
\begin{equation}
\boldsymbol {\gamma}=\frac{1-\kappa}{\boldsymbol{\delta}^*}\Big[1\pm\sqrt{1-|\boldsymbol{\delta}|^2}\Big],
\label{eq11}
\end{equation}
where the sign is determined by $- sign[\det (A)]$. We proceed the reconstruction of the lensing convergence
iteratively from the following relation
%\begin{eqnarray}
\begin{equation}
\kappa(\boldsymbol{\theta}) =-\frac{1}{\pi}\int_{R^{2}} d^{2}\boldsymbol{\theta}'
Re[D(\boldsymbol {\theta}-\boldsymbol{\theta}')\boldsymbol{\gamma}^{*}(\boldsymbol{\theta}')]
\label{eq12}
\end{equation}
%\end{eqnarray}
where $D(\boldsymbol{\theta})=(\theta_1^{2}-\theta_2^{2}+2i\theta_1\theta_2)/|\boldsymbol{\theta}|^{4}$.
Specifically, we start by assuming $\kappa^{(0)}=0$ and $|\boldsymbol {g}|\le 1$ everywhere, and thus (e.g., Bartelmann 1995)
\begin{equation}
\boldsymbol{\gamma}^{(0)}(\boldsymbol{\theta})=\frac{1-\sqrt{1-|\boldsymbol {\delta}(\boldsymbol{\theta})|^2}}
{\boldsymbol{\delta}^*(\boldsymbol{\theta})}.
\label{eq13}
\end{equation}
At ${n}$-th step, we obtain $\kappa^{(n)}$ from $\boldsymbol{\gamma}^{(n-1)}$ via Eq.(12) and
further calculate $\boldsymbol{\gamma}^{(n)}_{test}$ from $\kappa^{(n)}$ to determine the sign of $\det(A^{(n)})$ everywhere.
At $n+1$ step, we insert $\kappa^{(n)}$ into Eq. (11) to estimate $\boldsymbol{\gamma}^{(n)}$ by
considering the signs of $\det (A^{(n)})$ calculated in step $n$.

In the case of $\kappa\ll1$ and $|\boldsymbol{\gamma}|\ll1$, we have $\langle\boldsymbol{\epsilon}\rangle=-\boldsymbol{g}\approx -\boldsymbol{\gamma}$,
and thus the convergence reconstruction is a single-step linear process.

\section{Simulations}

To study the mask effects on weak lensing peak statistics and the corresponding cosmological parameter constraints derived from
the peak analyses, we carry out sets of dark-matter-only N-body simulations in the flat $\Lambda$CDM framework.
The fiducial model is taken to be $\Omega_\mathrm{m}=0.28$, $\Omega_\Lambda=0.72$, $\Omega_\mathrm{b}=0.046$,
$h=0.7$, $\sigma_8=0.82$, and $n_s=0.96$, where $\Omega_\mathrm{m}$, $\Omega_\Lambda$, $\Omega_\mathrm{b}$,
$h$ are the present dimensionless total matter density, energy density from the cosmological constant,
baryonic matter density and the Hubble constant in units of $100\hbox{ km/s/Mpc}$, respectively.
The parameter $n_s$ is the power index for the initial density perturbations and $\sigma_8$ is the root-mean-square
of the linear density perturbations extrapolated to present with the top-hat smoothing scale of $8h^{-1}\hbox{Mpc}$.
In order to test the applicability of our theoretical model for weak lensing peak statistics (F10),
we also run four sets of simulations with different $\Omega_m$ and $\sigma_8$ near the fiducial ones.
The detailed cosmological parameters for the simulations are listed in Table~\ref{tab:param}.
The conventional ray tracing algorithm is adopted to calculate the deflection of light rays and the corresponding shear and
convergence maps.

%In order to examine mask effect, we construct five groups of convergence and shear sky maps in five cosmologies. A conventional algorithm
%of weak lensing ray-tracing is adopted. The cosmic density fluctuation generated by N-body simulations provide a deflection and
%distortion of the bundle of light. In this section, we present that some details.

\begin{table}
  \caption{COSMOLOGY PARAMETERS}
  \label{tab:param}
  \begin{center}
    \leavevmode
	\begin{tabular}{l  c c c c c} \hline \hline
                        &  fiducial  & M1  & M2 & M3 & M4\\	\hline
	$\sigma_8$          & 0.82 & 0.77 & 0.87 & 0.82 & 0.82 \\
	$\Omega_\mathrm{m}$ & 0.28 & 0.28 & 0.28 & 0.25 & 0.31 \\
	$\Omega_\Lambda$    & 0.72 & 0.72 & 0.72 & 0.75 & 0.69\\
	$\Omega_\mathrm{b}$ & 0.046 & 0.046 & 0.046 & 0.046 & 0.046 \\
	$h$                 & 0.7  & 0.7 & 0.7 & 0.7 & 0.7 \\
	$n_s$               & 0.96  & 0.96 & 0.96 & 0.96 & 0.96 \\
	\hline
	\end{tabular}
    \begin{tabular}{lll} \hline \hline

%  \multicolumn{3}{l}{}  \\

    \end{tabular}
  \end{center}
\end{table}

%\subsection{Base simulations, algorithm and parameters}
\subsection{Base simulations}

\begin{figure}
\centering
\includegraphics[width=0.95\columnwidth]{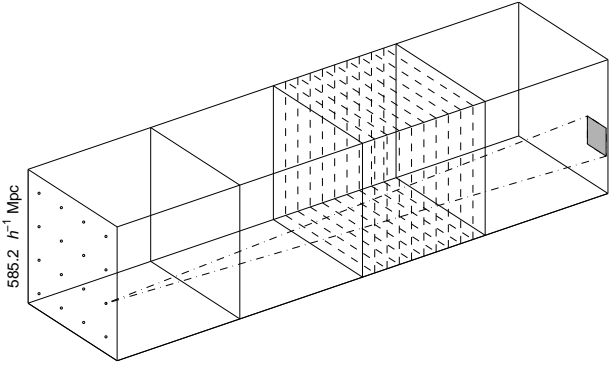}
\caption{Ray-tracing sketch.}.
%One simulation set includes 4 boxes. Every box is divided into 10 planes by equal comoving width
%along the line of sight as shown in the third box. The solid angle of
%$3\times3$ $\mathrm{deg^2}$ covers the shadow area at $z=1$. Sixteen light cone are arranged in one simulation set.}
\label{fig:raytracing}
\end{figure}

In our weak-lensing analyses, we take the source redshift $z_s=1$. For the fiducial cosmological model, the comoving distance to
$z_s=1$ is approximately $2.34h^{-1}\hbox{ Gpc}$. To balance the efficiency and the resolution, we bind four independent simulations
together to fill the range to $z_s=1$ as illustrated in Fig.\ref{fig:raytracing}. In other words, for each set of ray-tracing calculations,
we run four independent simulations with different realizations of the initial conditions. Each simulation is run in
a comoving cubic box of $585.2 h^{-1} \mathrm{Mpc}$ in size. Therefore in our setting, an individual simulation box occurs only once and
there are no repetitious structures along line of sights. Such a design allows us to pad the simulation boxes regularly without the need of
shifting and rotating to avoid the possibly multiple use of same structures in the ray tracing calculations.

%In our analysis, we construct lensing maps until the redshift of $z=1$. We use N-body code of Gadget-2\citep{GADGET2} to
%run 5 groups of flat $\Lambda$CDM cosmology simulations. The cosmological parameters for fiducial group are $\sigma_8 = 0.82$,
%$\Omega_\mathrm{m} = 0.28$, $\Omega_\mathrm{b}=0.046$,$h = 0.7$ and $n_s = 0.96$. We change either $\sigma_8$ or $\Omega_\mathrm{m}$
%two times to generate the cosmological parameters for other 4 groups around fiducial one.(see details in Table~\ref{tab:param}).

\begin{figure}
\centering
\includegraphics[width=1\columnwidth]{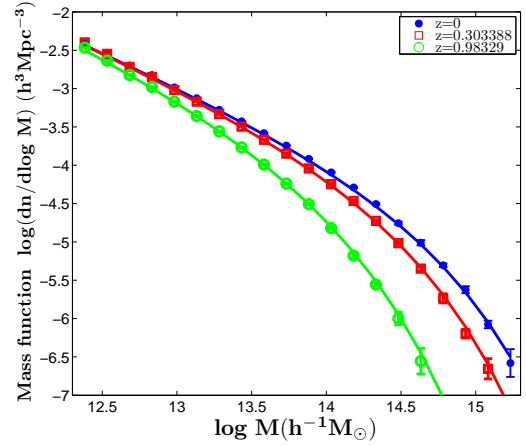}
\caption{Mass functions for the fiducial model at different redshifts. The blue dot, red square and green circle symbols with error bars are
for the simulation results at $z=0$, $z\approx 0.3$ and $z\approx 0.98$, respectively. The corresponding lines are the
theoretical results calculated from the Sheth \& Tormen mass function.}
\label{fig:MF}
\end{figure}

For each run, we use $640^3$ dark matter particles in the simulation box. The particle mass is $\sim$ $6\times 10^{10} h^{-1} \mathrm{M}_{\sun}$
for the fiducial model. The N-body code of Gadget-2 \citep{GADGET2} is used to run the simulations. The initial redshift is taken to be $z=50$.
The initial power spectrum is generated by CAMB \citep{CAMB}, and initial conditions are constructed by the code of 2LPTic \citep{2LPTic}.
The force softening length is $\sim$ 20 $h^{-1} \mathrm{kpc}$. The mass and the force resolutions should be good enough for our purpose of studies that
are mainly interested in high weak-lensing peaks corresponding to massive dark matter halos with mass $M>10^{13}h^{-1} \mathrm{M}_{\sun}$ along line of sights.
As a test, in Fig.~\ref{fig:MF}, we show the mass functions of halos identified with the Friends-of-Friends (FoF) algorithm
with the linking length of 0.18 of the average separation of dark matter particles, which is
suitable for the considered cosmological model \citep{Courtin2011}. The results at redshift $z=0$ (blue symbols),
$\sim 0.3$ (red symbols) and $\sim 0.98$ (green symbols) from our fiducial simulations are presented.
The corresponding solid lines are the results calculated from the Sheth-Tormen mass function (ST) \citep{ST1999}.
It is seen that our simulation results agree with those from ST very well.

For multiple-lens-plane ray tracing calculations to be described in the next subsection, we use $40$ planes corresponding to $40$ snapshots
equally distributed along the comoving distance to $z=1$. Therefore there are $10$ planes for each simulation box (see Fig.\ref{fig:raytracing}).
Each plane contains particles in a slice with a comoving volume of $ (58.52 \times 585.2 \times 585.2)\  h^{-3} \mathrm{Mpc}^3$.
These particles are projected
along the thickness of the slice into the plane with the comoving area of $(585.2 \times 585.2)\ h^{-2} \mathrm{Mpc}^2$.
The size of the simulations allows us to construct $16$ weak lensing maps of $3\times 3\deg^2$ each through one set of ray-tracing calculations
that is based on $4$ independent runs of N-body simulations with different realizations of the initial conditions.
For the fiducial model, we perform $8$ sets of ray-tracing simulations from total of $32$ runs of N-body simulations.
Therefore totally we have $8\times 16=128$ weak-lensing maps of $9\deg^2$.
For the other four cosmological models, $16$ N-body simulations are done to generate $4$ sets of ray tracing calculations, and thus
$4\times 16=64$ weak-lensing maps of $9\deg^2$ for each model.

\subsection{The multiple-lens-plane ray-tracing calculations}
For the ray-tracing calculations, we follow closely the method of \cite {Hilbert2009}. We use $40$ different snapshots to
construct $40$ lens planes evenly distributed in the comoving distance to $z=1$. Dark matter particles within the slice of thickness of
$58.52h^{-1}\mathrm{Mpc}$ around the lens plane $k$ are projected onto the plane. The two-dimensional density fluctuation field
$\Sigma^{(k)}$ on a regular mesh is then constructed from the projected particle positions by the Cloud-in-Cell scheme.
The potential $\hat \psi^{(k)}$ on the lens plane is calculated from the two-dimensional Poisson equation
\begin{equation}
%\nabla^2_{\boldsymbol \theta^{(k)}} \hat \psi^{(k)} (\boldsymbol \theta^{(k)})= 3H_0^2\Omega_m\frac{f_K^{(k)}}{a^{(k)}} \Sigma^{(k)}.
\nabla^2 \hat \psi^{(k)} = 3H_0^2\Omega_m\frac{f_K^{(k)}}{a^{(k)}} \Sigma^{(k)},
\label{eq:psik}
\end{equation}
where $f_K^{(k)}$ and $a^{(k)}$ are the comoving angular diameter distance to the $k$th plane and the scale factor of the
universe at the epoch corresponding to the $k$th plane, and the operation $\nabla^2$ is taken with respect to the angular scale.
We design to sample a convergence/shear map of $3\times 3\deg^2$ on $1024\times 1024$ pixels
corresponding to $4096\times 4096$ pixels over the total $16$ maps.
For the purpose of numerical accuracy, a finer mesh for 2-d density and potential calculations is needed
as pointed out by \cite{Sato2009}. We thus choose to sample the 2-d density and potential fields of a lens plane of
$585.2\times 585.2h^{-2}\hbox{Mpc}^2$ on $16384\times 16384$ pixels. The resolution is then about $\sim 35.7$ $h^{-1}\mathrm{kpc}$. To suppress the Poisson noise,
we further smooth the potential field with a Gaussian window function with the smoothing scale $30h^{-1}\hbox{kpc}$ \citep[e.g,][]{WV2004}.

The deflection angle $\hat \alpha$ and the shear matrix $U_{ij}$ on the mesh are calculated by finite difference using the nearest
neighboring grids through

\begin{equation}
{\hat {\boldsymbol \alpha}}^{(k)} = \nabla \hat \psi^{(k)}
%\frac{\partial \psi^{(k)}}{\partial \theta_i},
\label{eq:alpha}
\end{equation}
and
\begin{equation}
U^{(k)}_{ij} = \partial^2_{ij}\hat \psi^{(k)}.
%U_{ij} = \frac{\partial^2 \psi}{\partial {\theta_i}\partial {\theta_j}}.
\label{eq:shearm}
\end{equation}

To calculate the light ray position at $k$th plane, we follow the method of \cite {Hilbert2009} to use the ray positions
at $k-2$ and $k-1$ planes. Specifically, we have

\begin{eqnarray}
\boldsymbol \theta^{(k)} =&&  \left( 1- \frac{f_K^{(k-1)}}{f_K^{(k)}}\frac{f_K^{(k-2,k)} }{ f_K^{(k-2,k-1)} }\right) \boldsymbol \theta^{(k-2)}+ \frac{f_K^{(k-1)}}{f_K^{(k)}}\frac{f_K^{(k-2,k)}}{f_K^{(k-2,k-1)}}\boldsymbol \theta^{(k-1)} \nonumber \\
&& - \frac{f_K^{(k-1,k)}}{f_K^{(k)}}{\hat {\boldsymbol \alpha}}^{(k-1)}(\boldsymbol \theta^{(k-1)}),
\label{eq:ray}
\end{eqnarray}
where the deflection angle ${\hat {\boldsymbol \alpha}}^{(k-1)}$ is calculated at the ray position $\boldsymbol \theta^{(k-1)}$
by interpolating from the values at grids on the mesh. We start with $\boldsymbol \theta^{(0)}=\boldsymbol \theta^{(1)}=\boldsymbol \theta$ with
$\boldsymbol \theta$ being the light ray direction received by the observer. Therefore the light ray propagation
can be computed iteratively.

Taking derivatives with respect to $\boldsymbol \theta^{(0)}$, we obtain the corresponding distortion matrix

\begin{eqnarray}
A^{(k)}_{ij} && = \left( 1- \frac{f_K^{(k-1)}}{f_K^{(k)}}\frac{f_K^{(k-2,k)} }{ f_K^{(k-2,k-1)} }\right) A^{(k-2)}_{ij}+ \frac{f_K^{(k-1)}}{f_K^{(k)}}\frac{f_K^{(k-2,k)}}{f_K^{(k-2,k-1)}}A^{(k-1)}_{ij} \nonumber \\
&& - \frac{f_K^{(k-1,k)}}{f_K^{(k)}}\frac{\partial {\hat \alpha_i}}{\partial \theta_q^{(k-1)}}\frac {\partial \theta_q^{(k-1)}}{\partial \theta_j^{(0)}}
\nonumber \\
%U^{(k-1)}_{ik}A^{(k-1)}_{kj}
&& =  \left( 1- \frac{f_K^{(k-1)}}{f_K^{(k)}}\frac{f_K^{(k-2,k)} }{ f_K^{(k-2,k-1)} }\right) A^{(k-2)}_{ij}+ \frac{f_K^{(k-1)}}{f_K^{(k)}}\frac{f_K^{(k-2,k)}}{f_K^{(k-2,k-1)}}A^{(k-1)}_{ij} \nonumber \\
&& - \frac{f_K^{(k-1,k)}}{f_K^{(k)}}U^{(k-1)}_{iq}A^{(k-1)}_{qj},
\label{eq:distm}
\end{eqnarray}
which can also be calculated iteratively. Here again $U^{(k-1)}_{iq}$ is calculated at the ray position
$\boldsymbol \theta^{(k-1)}$ by interpolating. With the final $A_{ij}$, we can extract the convergence $\kappa$
and the shear $\gamma_i$ by noting that there is an unobservable rotation angle involved in $A_{ij}$
obtained through multiple-plane ray tracing.

\begin{figure}
\centering
\includegraphics[width=1.02\columnwidth]{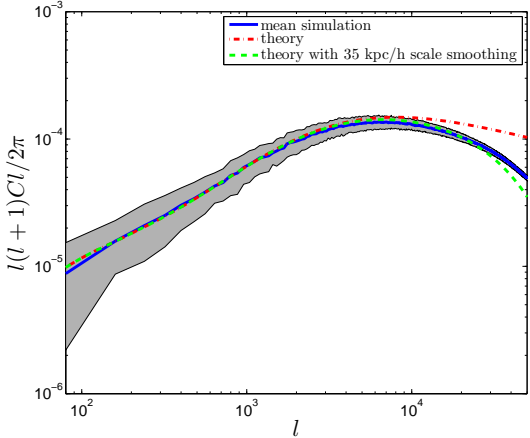}
\caption{The convergence power spectrum for the fiducial model. The blue solid line is for the simulation result obtained by averaging
over $128$ maps. The shaded region represents the $1 \sigma$ range for the variation from map to map. The red dash-dotted line is for the
theoretical result calculated from Eq.(\ref{eq:lp}). The green dashed line is for the smoothed theoretical result with the Gaussian smoothing scale of
$35 h^{-1} \mathrm{kpc}$.
%Correlation function of convergence map in fiducial case. The solid curve is from our lensing map, the dotted-dash curve
%denotes the theoretical prediction and dash curve is theory with smoothing scale of 35 $h^{-1} \mathrm{kpc}$. The shade
%region is the standard deviation from one $9~deg^2$ map.
\label{fig:PSfid}}
\end{figure}

Fig.~\ref{fig:PSfid} presents the power spectrum calculated from the the simulated convergence maps of the
fiducial model. The blue solid line is the mean result from the $128$ simulated maps of $3\times 3\deg^2$
with the shaded region showing the $1\sigma$ range of variation of the power spectrum from map to map.
The red dash-dotted line is for the theoretical result calculated using the Limber approximation
\citep{Limber1954, Kaiser1998} given by \citep[e.g.,][]{BS2001}

\begin{equation}
P_{\kappa}(l)=\bigg(\frac{ 9H_0^4\Omega_m^2}{4}\bigg)\int_0^{w_s} dw\frac{f_K^2(w_s-w)}{f_K^2(w_s)a^2(w)}P_{\delta}\bigg(\frac{l}{f_K(w)},w\bigg ),
\label{eq:lp}
\end{equation}
where $P_{\delta}$ is the power spectrum of the three-dimensional density fluctuations. We use the nonlinear $P_{\delta}$
calculated from CAMB updated according to the improved halofit model of \cite{Taka2012} \citep{CAMB}.
The green dashed line is the theoretical result smoothed with a Gaussian function with the
the smoothing scale of $35h^{-1}\hbox{kpc}$, approximately in accord with the simulation grid size for potential calculations.
We can see that up to $l\sim 10,000$, the result from the simulations agrees with the theoretical calculations
very well.

\subsection{Boundary problem}

As discussed in \cite{Hilbert2009}, a problem can rise if a fixed boundary is used to divide simulation particles into two different slices to construct the
density distribution on the corresponding lens planes. It leads to artificially cut particles of a
cross-boundary halo into two parts. This is particularly relevant to our studies on weak lensing peak statistics
in which halos are related directly to peaks in weak lensing maps. We follow the same procedures of \cite{Hilbert2009}
to deal with this boundary problem.

Specifically, in each of the $40$ snapshots, using FoF, we identify all the halos in the corresponding simulation boxes
and find cross-boundary halos that have member particles on either side of a boundary. For those halos, they are included as a whole in the slice
inside which their center of mass locates, and excluded completely from the other slice. Considering the possible cross-boundary motions of halos
that can lead to halo double counting or missing halos, a further step adopted from \cite{Hilbert2009} is taken to
avoid such a problem. For the two slices on the different sides of a boundary, if a halo is already included in the slice of the later snapshot
(closer to the observer) based on its position of the center of mass, it is excluded from the other slice of the earlier snapshot even its center of mass is inside
that slice at this earlier snapshot. If a halo is missed in both slices based on the position of its center of mass,
it indicates that the halo moves across the boundary in the direction
that is further away from the observer. In this case, we assign the halo to the slice of the earlier snapshot.

%In Fig.~\ref{fig:boundary}, we show a difference map between the convergence from the simple fixed boundary ray tracing calculation
%and the one that takes care of the boundary problem as presented above. It is seen that this boundary effect is not negligible.
%We find that $\sim 7\%$ of halos with a typical radius of $\sim h^{-1}\mathrm{Mpc}$ are involved in the cross-boundary problem.
%Fig.~\ref{fig:missing} shows an example of the convergence maps without (left) and with (right) the proper treatment of the boundary problem.
%The differences in the circled region are clearly seen and that can affect our peak identifications.

Detailed comparisons show that the differences between the convergence from the simple fixed boundary calculation and the adaptive one
described above can be as large as $\sim 0.5\sigma_0$ for $\sigma_0\approx 0.02$. For halos with a typical radius of $\sim h^{-1}\mathrm{Mpc}$,
about $7\%$ of them are involved in the cross-boundary problem.

In this paper, unless for comparison purposes as discussed in this subsection, all the analyses are based on the ray tracing simulations
including the proper treatment of the boundary problem.

\begin{figure*}[tbl]
%\plottwo{Fig4a.eps}{Fig4b.eps}
\begin{minipage}[c]{0.5\textwidth}
  \centering
  \includegraphics[width=0.95\textwidth]{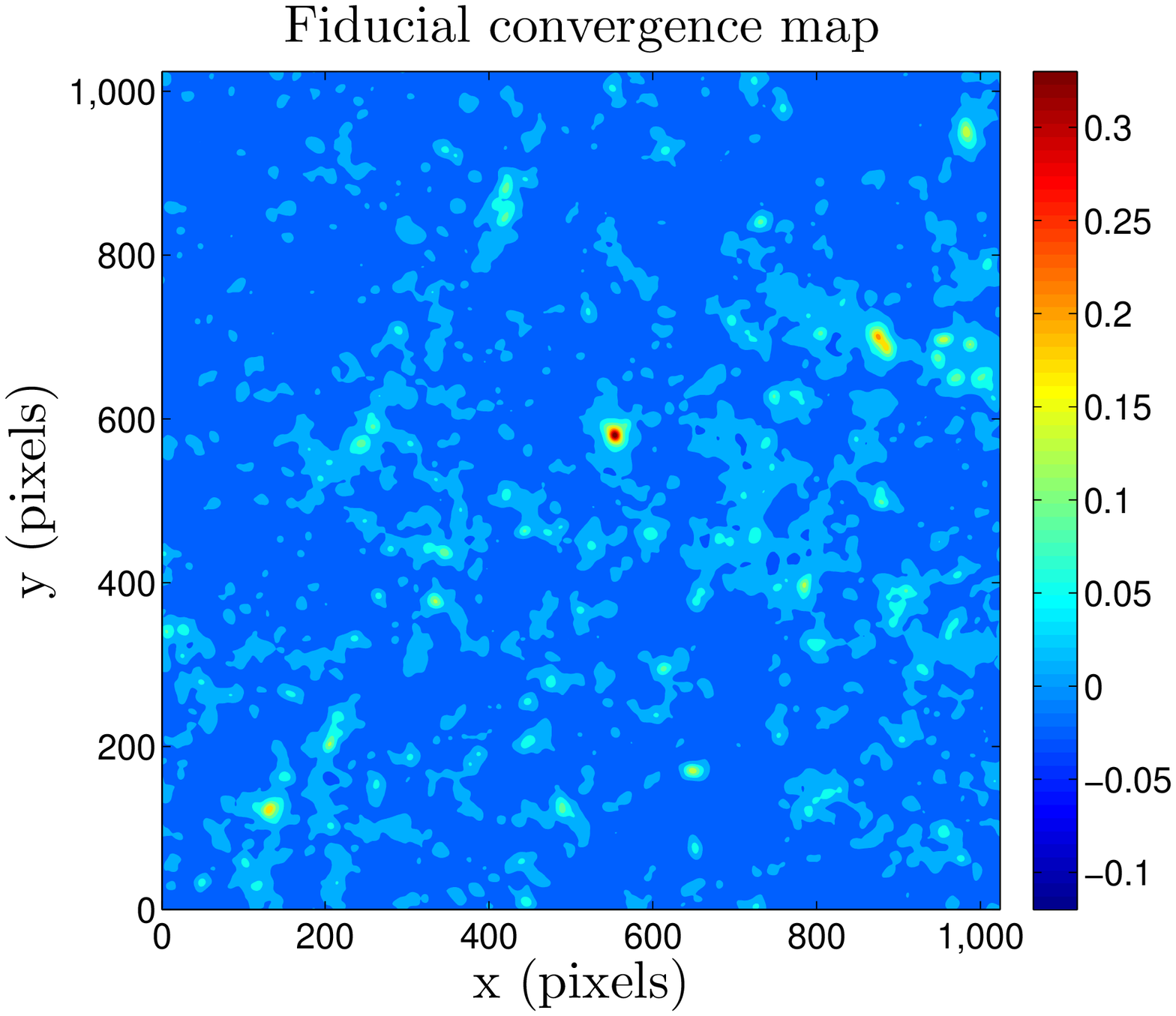}
\end{minipage}
\begin{minipage}[c]{0.5\textwidth}
  \centering
  \includegraphics[width=0.95\textwidth]{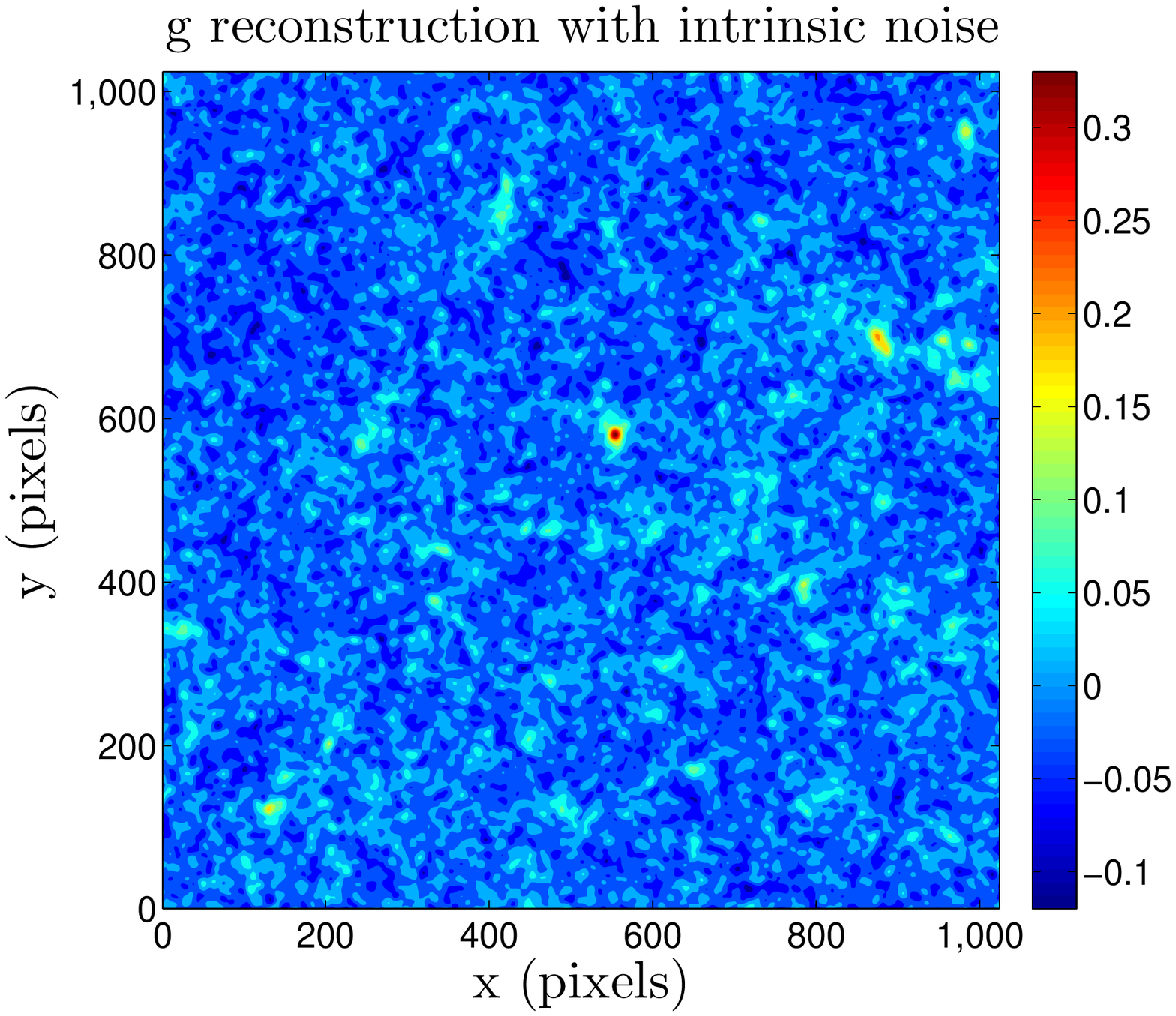}
\end{minipage}
\caption{Examples of convergence maps. The left one is from the base ray-tracing simulation smoothed with
$\theta_G=1\arcmin$. The right one is the corresponding `g reconstruction' convergence map from the populated galaxy catalog.}
%Left: One example of the smoothed base convergence map without noise;
%Right: The corresponding `$g$ reconstruction' map with intrinsic ellipticities of source galaxies included.}
%Lower-Left: The corresponding $g$ reconstructed convergence map;
%Lower-Right: The corresponding masked $g$ reconstructed convergence map, here the yellows are the masks in this case.
\label{fig:mapexample}
\end{figure*}

\begin{figure*}[tbl]
\plotone{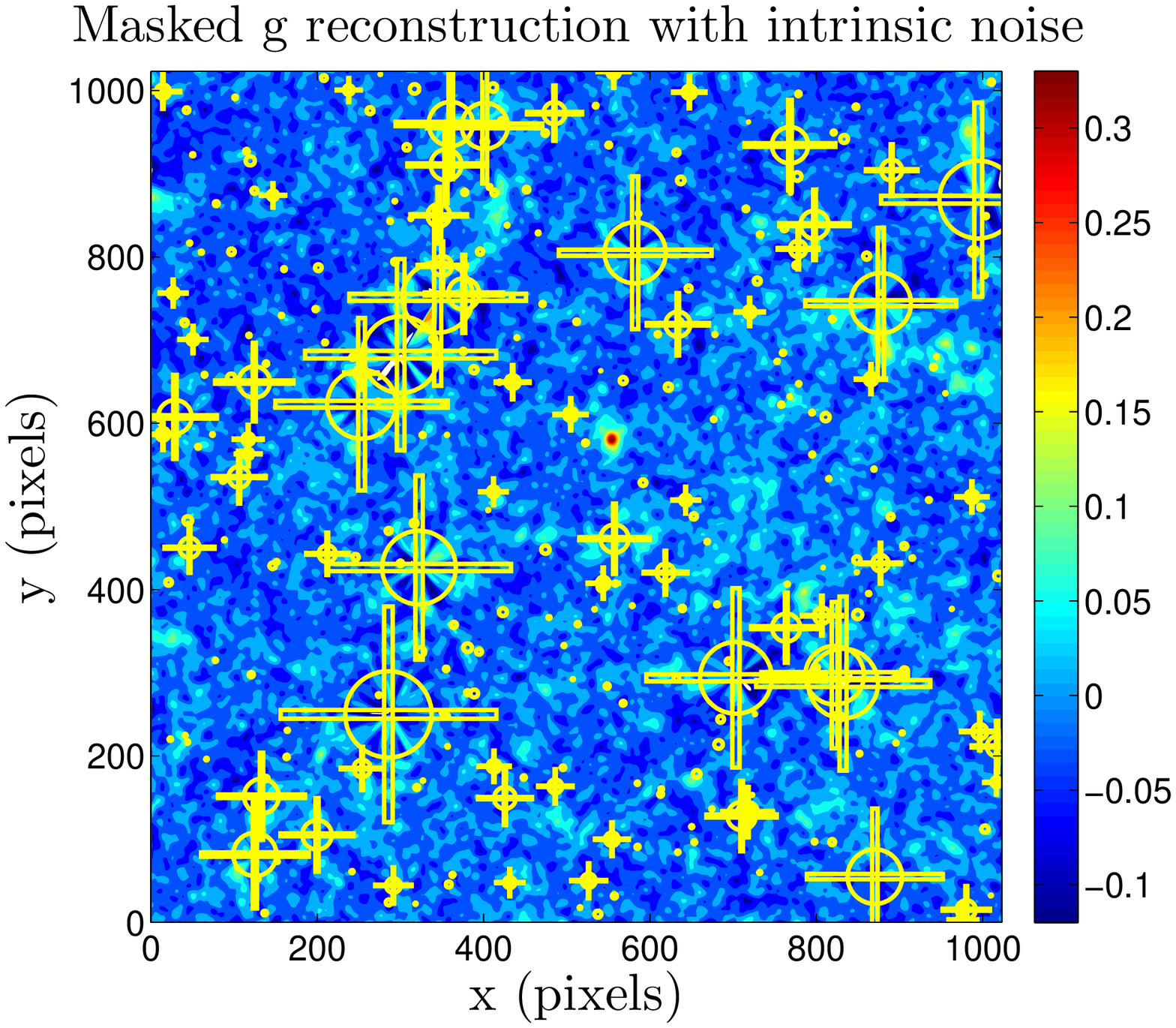}
%\begin{minipage}[c]{1\textwidth}
%  \centering
%  \includegraphics[width=1\textwidth]{Fig5.eps}
%\end{minipage}
\caption{ The masked `g reconstruction' convergence map corresponding to the right panel of Fig.~\ref{fig:mapexample},
here the yellow patterns are the masks occurred in this case.}
\label{fig:maskexample}
\end{figure*}

\subsection{Fiducial reconstructed convergence maps}

We follow the nonlinear reconstruction procedures described in \S 2 to derive the weak-lensing
convergence field from background galaxy ellipticities for peak analyses.

To generate source galaxy data, for each of the $128$ simulated fields for the fiducial model,
we randomly populate galaxies in angular positions at $z_s=1$ and assign them intrinsic ellipticities according to the
following probability distribution \citep[e.g.,][]{Bar1995}

%To analyze weak lensing convergence peaks, previous studies often add a Gaussian noise arising from intrinsic
%ellipticities directly to convergence maps from N-body simulations to construct `observed' maps
%\citep[e.g.,][]{HTY2004,TF2005,Yangxj2013}. Observationally however, a convergence map has to be
%reconstructed from the measured galaxy ellipticities. To mimic observations, we thus also
%perform the convergence reconstruction. We generate source galaxy catalogs by randomly populating them
%in angular positions at $z_s=1$ and assigning them intrinsic ellipticities according to the following probability distribution

\begin{equation}
 p_s(\epsilon_{s1},\epsilon_{s2})=\frac{\exp[-(\epsilon_{s1}^{2}+\epsilon_{s2}^{2})/\sigma_\epsilon^{2}]}{\pi\sigma_\epsilon^2[1-\exp(-1/\sigma_\epsilon^{2})]},
\,\,|\epsilon_s|\in[0,1]
\label{eint}
\end{equation}
where $\epsilon_{s1}$ and $\epsilon_{s2}$ are the two components of the intrinsic ellipticities, $|\epsilon_s|=\sqrt{\epsilon_{s1}^{2}+\epsilon_{s2}^{2}}$ and the rms dispersion of $|\epsilon_s|$ is taken to be $\sigma_\epsilon=0.4$.
We assume the number density of source galaxies to be $n_g=30\hbox{ arcmin}^{-2}$.
The spatial clustering and the intrinsic alignment of source galaxies are not considered here.

The reduced shear signal $\boldsymbol g$ for each source galaxy is calculated
from the simulated shear and convergence maps by interpolating the values on regular grids to the galaxy position.
The `observed' galaxy ellipticity $\boldsymbol \epsilon$ is then constructed according to Eq.(\ref{eobs}).

With these galaxy data in each $3\times 3\deg^2$ field, we first obtain a smoothed field of $\boldsymbol \epsilon$ on
a regular mesh of $1024\times 1024$ pixels by
%We calculate the average $\langle\boldsymbol \epsilon\rangle$ on positions of regular grids as follows
\begin{equation}
\langle\boldsymbol \epsilon\rangle(\boldsymbol \theta)=\frac{\sum_i W(\boldsymbol \theta_i - \boldsymbol \theta)
\boldsymbol \epsilon (\boldsymbol \theta_i)}{\sum_i W(\boldsymbol \theta_i - \boldsymbol \theta)},
\label{smoothg}
\end{equation}
where $\boldsymbol \theta$ here is for pixel position and $\boldsymbol \theta_i$ is
for galaxy position. The summation is over galaxy positions. The window function $W$ is taken to be Gaussian
given by
\begin{equation}
W(\boldsymbol x) = \frac{1}{\pi \theta_{G}^2}\exp{\left(-\frac{|\boldsymbol x|^2}{\theta_{G}^2}\right)}.
\label{windowf}
\end{equation}
Because we are interested in high peaks that are related to massive halos,
we take the smoothing scale $\theta_G$ to be $\theta_G=1\arcmin$,
suitable for halos with mass about $10^{14}\hbox{ M}_{\odot}$ and above \citep[e.g,][]{HTY2004}.
From the smoothed field $\langle\boldsymbol \epsilon\rangle$, the convergence reconstruction is done iteratively as described in \S2. The results
converge quickly with about eight iterations for the converging accuracy of $10^{-6}$,
defined to be the maximum difference between the corresponding reconstructed convergence maps from two consecutive iterations.
We then obtain $128$ reconstructed convergence maps, and the total area is $128\times 9=1152\deg^2$.
We refer such maps as `g reconstruction' maps.
It is emphasized again that the smoothing procedure is applied directly to $\boldsymbol \epsilon$.

%\textbf{For each base map, we generate $1$ sets of galaxy catalogs with different realizations of spatial positions and intrinsic ellipticities.}
%\textbf{We thus obtain total of $128\times 1=128$ reconstructed $3\times 3\hbox{ deg}^2$ convergence maps.}
%We refer such maps as `g reconstruction' maps.
%For comparison, we also generate sets of `observed' ellipticities by
%$\boldsymbol \epsilon=\boldsymbol \epsilon_s-\boldsymbol \gamma$ and perform linear reconstruction to get
%convergence maps, which are referred to as `shear reconstruction' maps.

Fig.~\ref{fig:mapexample} shows a set of convergence maps. The left one is the pure convergence map from ray tracing simulations smoothed with
a Gaussian window function with $\theta_G=1\hbox{ arcmin}$. The right one is the `g reconstruction' map.
%and lower left are the reconstructed map from shear and from the reduce shear $g$, respectively.
We can see that most of the high peaks in the left map are still apparent in the right reconstructed map.
However, the right one is noisy comparing to the left one due to the intrinsic ellipticities of source galaxies.
%We can see that the reconstructed map is noisy comparing to the base one, but most of the high peaks are still apparent.
The noise can affect the height of true peaks. It also generates pure noise peaks and their distribution is biased by the true
mass distribution. These two noise effects have to be taken into account properly in modeling the weak lensing peak statistics (F10).
%The two reconstructed maps from shear and from $g$ are very similar, but there are subtle systematic differences.
%The lower right map is the reconstruction with masks to be described in the next subsection.

In our peak statistics analyses, we identify peaks from the reconstructed convergence maps as follows. Considering
a pixel on a map of $3\times 3\deg^2$ ($1024\times 1024$ pixels), if its reconstructed convergence value is the highest among its
nearest $8$ neighboring pixels, it is identified as a peak. To reduce the map boundary effects, we exclude the outer most $10$ pixels
in each of the four sides of a map in our analyses.
%To do the peak analysis further, we need to identify peaks from the reconstructed convergence maps.
%It is identified as a peak if the convergence at this pixel is the largest among the surrounding $8$ pixels, and
The signal-to-noise ratio of a peak is defined by
\begin{equation}
 \nu=\frac{K}{\sigma_0}
 \label{SN}
\end{equation}
where $K$ is the reconstructed convergence value of the peak, and $\sigma_0$ is the root-mean-square of the noise that
depends on the number density of source galaxies and the smoothing scale of the window function used in obtaining the
smoothed ellipticity field $\langle\boldsymbol \epsilon\rangle$. For a Gaussian window function used in our studies,
%In this case, as we assume a Gaussian noise field, the variance is specified by the number of galaxies contained within a smoothing aperture
we have (Kaiser \& Squires 1993; Van Waerbeke 2000)
\begin{equation}
 \sigma_0^2=\frac{\sigma_\epsilon^2}{2}\frac{1}{2\pi\theta_G^2n_g}.
 \label{var}
\end{equation}
For $\sigma_\epsilon=0.4$, $n_g=30\hbox{ arcmin}^{-2}$ and $\theta_G=1\arcmin$, $\sigma_0\approx 0.02$.
In our analyses here, we consider high peaks with $\nu\ge 4$. We count peaks in $11$ bins in the range of
$4.25\le \nu \le 9.75$ with the bin width of $0.5$.  It is noted that different binnings can affect the peak abundance analyses
quantitatively. Because our main focus in this paper is on the mask effects, we do not discuss the binning optimization here.
We will see later that the existence of masks enhances systematically the weak-lensing peak counts in our considered signal-to-noise ratio
range. This should not be changed qualitatively by different choices of peak binning. On the other hand, careful and
quantitative comparisons of different binning methods for weak lensing peak analyses are desired, and will be explored in our future studies.

\subsection{Mask model and convergence reconstruction with masks}

Removing bad/low quality imaging data is essential in weak lensing observational analyses. This leaves holes
in the source galaxy distribution, which in turn affects the convergence reconstruction and the subsequent cosmological
studies. To investigate the mask effects on weak lensing peak counts statistically, we generate mock masks by modeling
the basic masks for point sources, bright saturated stars and bad pixels being circular in shape.
The mask size distribution is in accord with that of CFHTLS used in \cite{Shan2012}.
We also add rectangle-shaped masks in both x and y directions
to those circular ones with radius larger than $1\arcmin$ to mask out saturation spikes.
These extra masks have a size of $0.2r\times 5r$ with $r$ being the radius of the circular mask to be added on.
We populate masks randomly in each of the considered $3\times 3\hbox{ deg}^2$ fields. With the size distribution of
\cite{Shan2012}, the total number of masks in each field is set to be $N_{mask}$. We consider three cases with
$N_{mask}=140, 280$, and $420$, corresponding to the total masked area fraction of $\sim 7\%$, $\sim13\%$ and $\sim 19\%$,
respectively. We then remove galaxies within masks from the source galaxy catalogs generated in \S 3.4.
With the remaining galaxies, following the reconstruction procedures we first smooth the galaxy ellipticities
from Eq.~(\ref{smoothg}) to get the smoothed $\langle\boldsymbol \epsilon\rangle$ where the summation is over the remaining galaxies,
and then perform the nonlinear reconstruction to obtain the reconstructed convergence maps. Because of the removal of galaxies
in masked regions, the effective number of usable galaxies in obtaining the smoothed $\langle\boldsymbol \epsilon\rangle$
around those area is less than the other places, causing higher noise levels. We will show later in our analyses that
this non-uniform noise is mainly responsible for the mask effects on weak-lensing peak count statistics.
An example of the reconstructed convergence map with masks is presented in Fig.\ref{fig:maskexample}. The mask regions are shown in yellow.
%and reconstruct convergence maps from the remaining source galaxies.
%An example is presented in Fig.\ref{fig:maskexample}. The mask regions are shown in yellow.

For the fiducial model, we then have two separate sets of convergence maps reconstructed from `observed' ellipticities without and with masks, respectively.
Each set contains totally $128$ of $3\times 3\hbox{ deg}^2$ convergence maps for peak analyses.

\section{The peak abundance}

Our studies aim to understand the mask effects on weak-lensing peak abundances
and the consequent biases on cosmological parameter constraints derived from the peak counts.
To constrain cosmological parameters from weak lensing peak abundances, we need to calculate the
expected peak numbers for different cosmological models. Because true high peaks in weak-lensing convergence maps
correspond well to massive halos along line of sights, it is natural to relate the peak counts to the mass function of
dark matter halos \citep[e.g,][]{HTY2004}.
%The most simple model assumes a known spherical mass distribution for dark matter halos,
%and therefore the peak signal from a halo of a given mass at a given redshift can be computed. With the halo mass function,
%the expected peak counts for a cosmological model can then be calculated \citep[e.g,][]{HTY2004}.
However, the non-spherical mass distribution of dark matter halos and the projection effects of large-scale structures
can complicate the lensing signal of a halo and therefore affect the predicted peak abundance \citep[e.g,][]{TF2005,HOSS2012}.
Also, the intrinsic ellipticities of source galaxies generate noise that leads to significant
effects on weak-lensing peak counts from the reconstructed convergence maps as seen in Fig.~\ref{fig:mapexample}.
The easily seen noise effect is the occurrence of false peaks resulting from the chance alignments of the intrinsic
ellipticities of source galaxies. Different peak identification methods have been proposed
to suppress the contribution from noise peaks, such as the tomographic method, the optimal filtering method, etc.
\citep[e.g,][]{HS2005,Marian12}. However, yet another effect of noise is its influence on the measured lensing signals of
true peaks \citep[e.g.,][F10]{HTY2004, Yangxj2011, HOSS2012}. Therefore even we can pick out true peaks,
we still need to consider the noise effect on them.

Given the complications, extensive simulation studies have been done to understand the cosmological model dependence
of weak-lensing peak counts \citep[e.g.,][]{DH2010,Yangxj2011,Marian12}. Different phenomenological models derived from simulations
have also been proposed \citep[e.g.,][]{Marian09,HOSS2012}. Based on the theory of Gaussian random fields, \cite{Maturi10}
present an analytical model to predict the weak-lensing peak counts with relatively low signal-to-noise ratios where
peaks are dominantly due to the noise from galaxy intrinsic ellipticities and the line-of-sight projection effects from large-scale structures.
In F10, we develop a model for high signal-to-noise peak counts by taking into account the noise effects on the peak heights of
true halos and the biased spatial distribution of noise peaks around dark matter halos.

For the analyses here, we adopt the model of F10. In \S 4.1, we describe the basic ingredients of the model.
In \S 4.2, we show the model applicability by comparing with numerical simulations.

\subsection{Theoretical model}

%To constrain cosmological parameters from weak lensing peak abundances, we need to calculate the
%expected peak numbers for different cosmological models.
%From previous section, we see that the intrinsic ellipticities of lensed galaxies are the dominant source of
%errors in weak lensing analyses. With the process of averaging over a number of galaxies to obtain the estimate of
%the lensing signal, the residual noise is on the order of $\sigma_{\epsilon}/\sqrt{n_g\theta_G^2}$, where
%$\sigma_{\epsilon}$ is the rms of the intrinsic ellipticities,  $n_g$ is the surface number of lensed galaxies
%in the weak lensing analyses, and $\theta_G$ is the smoothing scale used in the averaging.
%Such a noise can affect significantly the weak lensing peak statistics. In our studies here, we employ
%the theoretical model of Fan et al. (2010) to calculate peak numbers for relatively high peaks in
%cosmological parameter fittings. The model takes into account the effects of noise
%by considering the noise-induced bias and the dispersion on the height of the true convergence peaks from massive halos
%and the enhancement of the noise peak abundances due to the existence of the true mass distribution properly.
Considering high peaks, the model of F10 takes into account the effects of noise from
intrinsic ellipticities of source galaxies, including the noise-induced bias and the dispersion on the heights of
true convergence peaks from massive halos, and the enhancement of the pure noise peak abundances due to the existence of the true mass distribution.

\begin{figure*}[tbl]
%\plottwo{Fig8a_dependence.eps}{Fig8b_dependence.eps}
 \begin{minipage}[c]{0.5\textwidth}
  \centering
  \includegraphics[width=1\textwidth,height=0.8\textwidth]{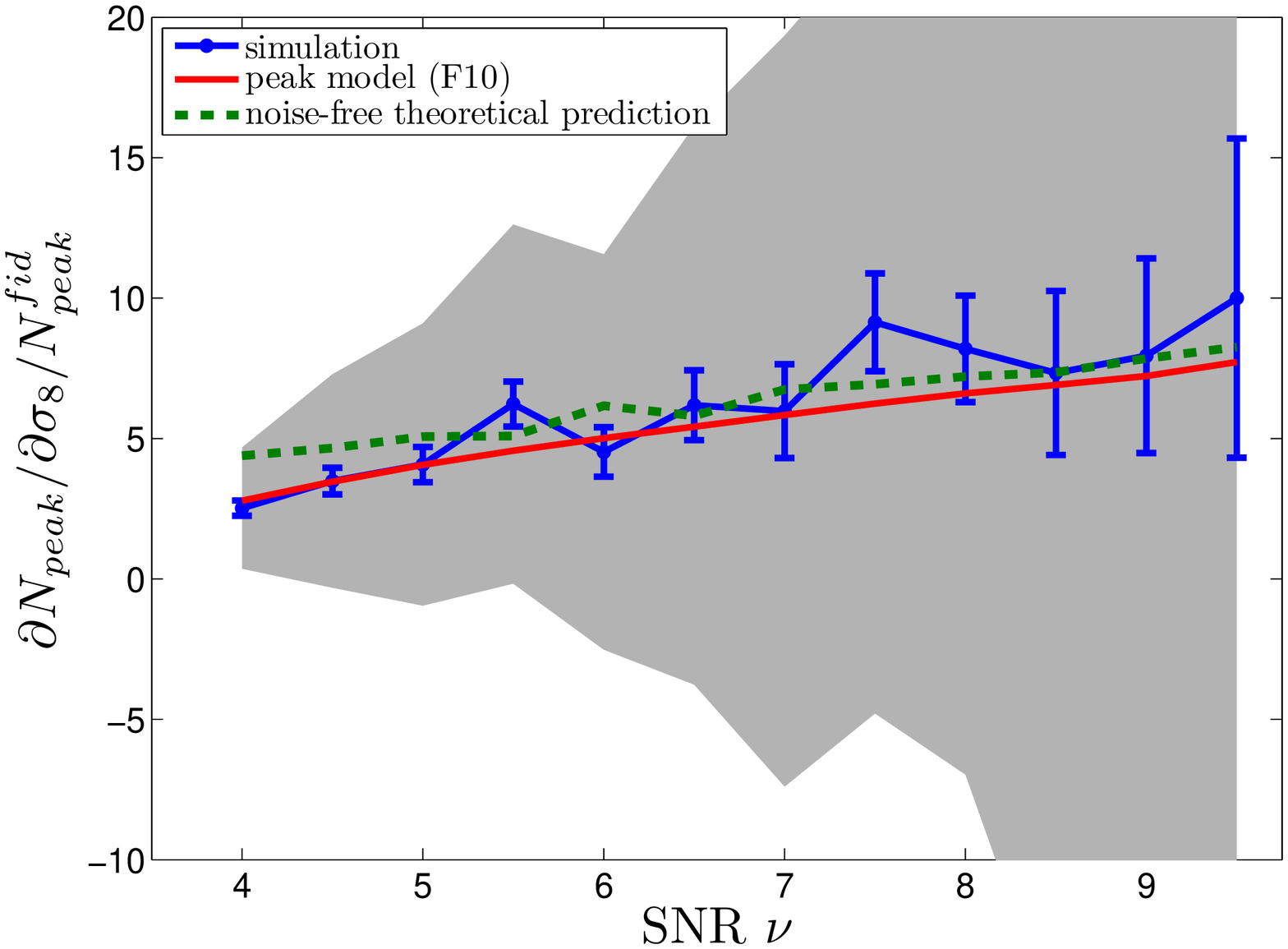}
\end{minipage}
\begin{minipage}[c]{0.5\textwidth}
  \centering
  \includegraphics[width=1\textwidth,height=0.8\textwidth]{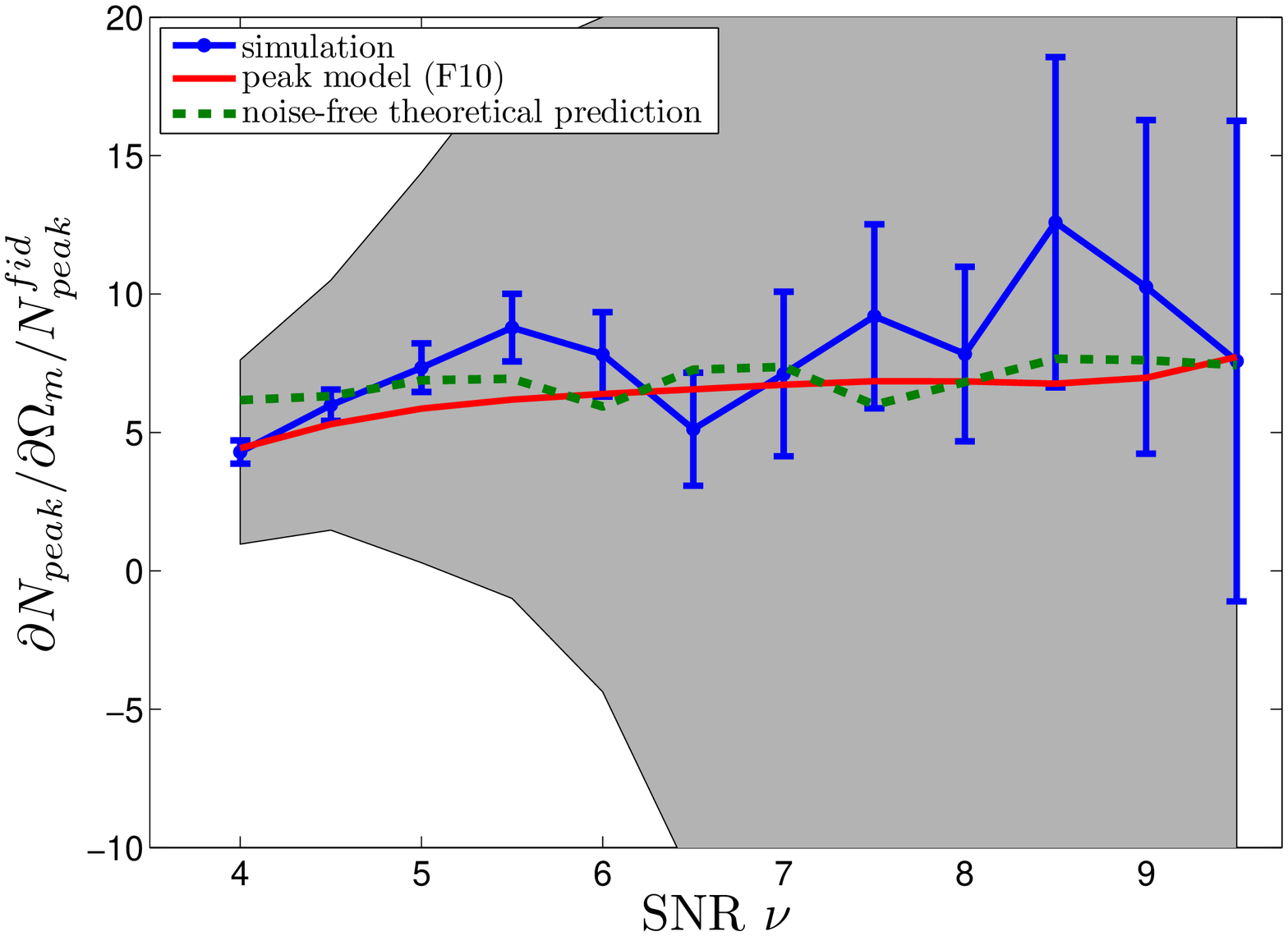}
\end{minipage}
\caption{Derivatives of the peak counts with respect to $\sigma_8$ (left) and $\Omega_m$ (right), respectively.
The blue symbols with error bars are for the average results from $64$ pairs of maps. The shaded regions indicate the $1\sigma$ variation
from pair to pair. The error bars show the $1\sigma$ range for the average derivatives over the $64$ pairs.
The red solid line is each panel is for the result predicted from the model of F10 including the noise effect.
The green lines are for the results from the theoretical model without noise.}
\label{fig:dependence}
\end{figure*}

The model assumes that the reconstructed smoothed convergence field can be written as
$K_N=K+N$, where $K$ represents the true lensing convergence, and $N$ is for the residual noise from intrinsic ellipticities.
The noise field $N$ is modeled as a Gaussian random field from the central limit theorem (e.g., van Waerbeke 2000).
The model concentrates on high peaks and assumes that true peaks come from individual massive halos.
Thus a considered survey area is split into halo regions and field regions.
Within an individual halo region, the peak number distribution can be calculated from the Gaussian statistics of $K_N$ with known $K$ from the halo.
Then the total number of peaks in halo regions can be obtained from the summation of the peaks in individual halo regions weighted
by the halo mass function. In field regions, the number distribution of peaks is computed directly from the noise field $N$.
The total surface number density of peaks can then written as
\begin{equation}
n_{peak}(\nu)d\nu=n_{peak}^c(\nu)d\nu+n_{peak}^n(\nu)d\nu,
\label{npeak}
\end{equation}
where $\nu=K_N/\sigma_0$ is the signal-to-noise ratio of a peak.
The term $n_{peak}^c(\nu)$ is for peaks in halo regions including both true peaks
corresponding to real halos and the noise peaks within halo regions, and $n_{peak}^n(\nu)$ is for peaks in field regions with only noise peaks.

For $n_{peak}^c(\nu)$, the peak count in halo regions, it can be written as
\begin{equation}
 n_{peak}^c(\nu)=\int{dz\frac{dV(z)}{dzd\Omega}}\int{dMn(M,z)f(\nu,M,z)},
\label{npeakc}
\end{equation}
where $dV(z)$ is the cosmological volume element at redshift z, $d\Omega$ is the solid angle element,
$n(M,z)$ is the mass function of dark matter halos and,
\begin{equation}
f(\nu,M,z)=\int_{0}^{R_{vir}}dR(2\pi R)\hat {n}^c_{peak}(\nu,R,M,z)
\label{fnumz}
\end{equation}
gives rise to the number of peaks in the area within the virial radius of a halo of mass $M$ at redshift $z$.
Here $\hat {n}^c_{peak}(\nu,R,M,z)$ describes the surface number density of peaks at the location of $R$ from the center of
the halo, which depends on the projected density profile of dark matter halos.
To calculate $\hat {n}^c_{peak}(\nu,R,M,z)$ in a particular halo region, we start from
$K_N=K+N$ where $K$ is the smoothed convergence of the halo assumed to be known and to follow the Navarro-Frenk-White (NFW) mass distribution
\citep{NFW1996,NFW1997}. The noise field $N$ is taken to be a Gaussian random field. Therefore $K_N$ is also a Gaussian
random field. We are interested in maxima peaks of $K_N$. By definition, such a maxima peak occurs in the place
where the first derivatives $\partial_i K_N=0$ for $i=1,\hbox{ } 2$, and the second derivative tensor $\partial_{ij}K_N$ should be
negative definite. Thus to calculate statistically the peak abundance, we need the joint probability distribution of $K_N$, $\partial_i K_N$
and $\partial_{ij}K_N$ \citep[e.g.,][]{Bardeen86, BE87}, which, for a Gaussian field, is given by (F10)

\begin{eqnarray}
p(&& K_N,K^{11}_{N},K^{22}_{N},K^{12}_{N},K^1_{N},K^2_{N})\hbox{\ }
%\nonumber \\ &&
 dK_N\hbox{\ }dK^{11}_{N}\hbox{\ }dK^{22}_{N}\hbox{ }dK^{12}_{N}\hbox{ }dK^1_{N}\hbox{ }dK^2_{N}\nonumber \\
&& = {1\over [2\pi(1-\gamma_N^2)\sigma_0]^{1/2}}\nonumber \\
&& \times \exp\bigg \{-{\{(K_N-K)/\sigma_0+\gamma_N [(K^{11}_N-K^{11})
+(K^{22}_N-K^{22})]/\sigma_2\}^2\over 2(1-\gamma_N^2)}\bigg \} \nonumber \\
&& \times {1\over 2\pi\sigma_2^2} \exp\bigg \{-{[(K^{11}_N-K^{11})-(K^{22}_N-K^{22})]^2\over 2 \sigma_2^2}\nonumber \\
&& -{(K^{11}_N-K^{11})^2\over \sigma_2^2}-{(K^{22}_N-K^{22})^2\over \sigma_2^2}\bigg \} \nonumber \\
&& \times {8\over (2\pi)^{1/2}\sigma_2}\exp\bigg \{-{4(K^{12}_N-K^{12})^2\over \sigma_2^2}\bigg \}
\nonumber \\
&&\times {1\over \pi\sigma_1^2}\exp\bigg [-{(K^1_N-K^1)^2\over \sigma_1^2}-{(K^2_N-K^2)^2\over \sigma_1^2}\bigg ] \hbox{\ } \nonumber \\
&& \times dK_N\hbox{\ }dK^{11}_{N}\hbox{\ }dK^{22}_{N}\hbox{ }dK^{12}_{N}\hbox{ }dK^1_{N}\hbox{ }dK^2_{N},
\label{jprob}
\end{eqnarray}
where we denote $K_N^{i}=\partial_i K_N$ and $K_N^{ij}=\partial_{ij} K_N$, and similarly for $K^i$ and $K^{ij}$.
Here the quantities $\sigma_i$ are the moments of the noise field $N$ given by \citep[e.g.,][]{vW2000}
\begin{equation}
\sigma_i^2=\int {d\vec k}\hbox{ }k^{2i}\langle|N(k)|^2\rangle,
\label{noisesigma}
\end{equation}
where $N(k)$ is the Fourier transform of the noise field $N$. With the diagonalization of $(-K_N^{ij})$, we
obtain its two eigen values $\lambda_{N1}$ and $\lambda_{N2}$ ($\lambda_{N1}\ge \lambda_{N2}$)
and the rotation angle $\theta_N$ constrained in the range $[0,\pi]$. For maxima peaks, we require $\lambda_{N1}\ge 0$ and $\lambda_{N2}\ge 0$.
We further define $x_N=(\lambda_{N1}+\lambda_{N2})/\sigma_2$ and $e_N=(\lambda_{N1}-\lambda_{N2})/(2\sigma_2x_N)$,
then the average number density of maxima peaks with a given signal-to-noise ratio $K_N/\sigma_0=\nu$ can be expressed as \citep[e.g.,][]{BE87}
\begin{eqnarray}
\hat n^c_{peak}(\nu, R, M, z)=&& \langle\delta(K_N/\sigma_0-\nu) \delta(K^1_N)\delta(K^2_N)(\sigma_2^2/4) x_N^2(1-4e_N^2)\nonumber \\
&& \Theta(1-2e_N)\Theta(e_N)\rangle,
%&&
%<\delta(\nu_N-\nu_0)\delta(x_N-x_0)\delta(e_N-e_0)\delta(\theta_N-\theta_0) \nonumber \\
%&& \delta(K^1_N)\delta(K^1_N)(\sigma_2^2/4) x_N^2(1-4e_N^2)\Theta(1-2e_N)\Theta(e_N)>,
\label{avenumpeak}
\end{eqnarray}
where the average is calculated by the probability distribution function corresponding to Eq.~(\ref{jprob})
using the variables $x_N$, $e_N$ and $\theta_N$ instead of $K_N^{11}$, $K_N^{22}$ and $K_N^{12}$.
The dependence on $R$, $M$ and $z$ comes in through the halo quantities $K$, $K^i$ and $K^{ij}$.
The step functions $\Theta(1-2e_N)$ and $\Theta(e_N)$ occur due to the requirements for maxima peaks. Then explicitly, we have
\begin{eqnarray}
\hat n^c_{peak}&&(\nu,R, M, z)=\exp \bigg [-{(K^1)^2+(K^2)^2\over \sigma_1^2}\bigg ]\nonumber \\
&& \times \bigg [ {1\over 2\pi\theta_*^2}{1\over (2\pi)^{1/2}}\bigg ]
\exp\bigg [-{1\over 2}\bigg ( \nu-{K\over \sigma_0}\bigg )^2\bigg ] \nonumber \\
&&\times \int_0^{\infty} dx_N\bigg \{ {1\over [2\pi(1-\gamma_N^2)]^{1/2}}\nonumber \\
&& \times \exp\bigg [-{ [{x_N+(K^{11}+K^{22})/ \sigma_2
-\gamma_N(\nu_0-K/\sigma_0)}]^2\over 2(1-\gamma_N^2)}\bigg ] \nonumber \\
&& \times  F(x_N)\bigg \},
\label{hatnc}
\end{eqnarray}
where $\theta_*^2=2\sigma_1^2/\sigma_2^2$ and $\gamma_N=\sigma_1^2/(\sigma_0\sigma_2)$.
For $K$, $K^{i}$ and $K^{ij}$ of a halo with mass $M$ at redshift $z$, we assume the
spherical NFW profile for the halo and adopt the concentration-mass relation from \cite{Bha2013} given by
\begin{equation}
c_{vir}(M,z)=\tilde {D}(z)^{0.9}7.7\bigg [ \frac{\delta_c}{\sigma(M,z)}\bigg ]^{-0.29}.
\label{cmrelation}
\end{equation}
Here $\tilde {D}(z)$ is the linear growth factor normalized to $z=0$ calculated with the
fitting formula given by \cite{CPT1992}. The quantity $\delta_c$ is the linear collapse threshold at redshift $z$
computed according to \cite{Hen2000}. The quantity $\sigma(M,z)$ is the rms of the smoothed linear density fluctuations
at redshift $z$ over the top-hat scale corresponding to $M$, and is calculated with
the same linear power spectrum as that used in our N-body simulations from CAMB taking into account the linear growth factor at $z$.
%For $\delta_c$, we adopt the fitting formula
%$\delta_c(z)=[3(12\pi)^{2/3}/20][1-0.0123\log(1+x^3)]$ with $x\equiv (\Omega_m^{-1}-1)^{1/3}/(1+z)$
%\citep{Hen2000}

The function $F(x_N)$ in Eq.~(\ref{hatnc}) is given by (F10)
\begin{eqnarray}
F(x_N)=&& \exp\bigg [-{(K^{11}-K^{22})^2\over \sigma_2^2}\bigg ]
\times \nonumber \\
&& \int_0^{1/2}de_N \hbox{ }8(x_N^2e_N)x_N^2(1-4e_N^2) \exp(-4x_N^2e_N^2)
\times \nonumber \\
&& \int_0^{\pi} {d\theta_N\over \pi} \hbox{ }
\exp\bigg [-4x_Ne_N\cos (2\theta_N){(K^{11}-K^{22})\over \sigma_2}\bigg ]. \nonumber \\
\label{fxn}
\end{eqnarray}

With $\hat n^c_{peak}(\nu,R, M, z)$ in Eq.~(\ref{hatnc}), we can then calculate $f(\nu, M, z)$ by Eq.~(\ref{fnumz}),
and further $n^c_{peak}(\nu)$ by Eq.~(\ref{npeakc}) where we adopt the Sheth-Tormen mass function in the calculations \citep{ST1999} .

For the field term $n_{peak}^n(\nu)$ in Eq.~(\ref{npeak}), it is given by
\begin{equation}
 \begin{split}
n_{peak}^{n}(\nu)=\frac{1}{d\Omega}\Big\{n_{ran}(\nu)\Big[d\Omega-\int dz\frac{dV(z)}{dz}\\
   \times\int dM\,n(M,z)\,(\pi R_{vir}^{2})\Big]\Big\},
\end{split}
\label{npeakn}
\end{equation}
where $n_{ran}(\nu)$ is the surface number density of pure noise peaks without foreground halos.
It can be calculated by Eq.~(\ref{hatnc}) with $K=0$, $K^{i}=0$ and $K^{ij}=0$.

Further details of the model can be found in F10.

% and is given by Eq.(15) of Fan et al. (2010).
%We adopt the Sheth-Tormen mass function in our calculations \citep{ST1999}.
%The linear power spectrum used in computing the rms density perturbations is the same as that
%used in the simulations of \cite{WV2004} from \cite{EH1999}.
%\textbf{The NFW profile (Navarro et al. 1996, 1997) is applied to describe the density distribution of dark matter halos with
%the concentration-mass relation given by \cite{Bha2013}}
%\begin{equation}
%c_{vir}(M,z)=D(z)^{0.9}7.7\nu^{-0.29},
%\label{eq17}
%\end{equation}
%where $D(z)$ is the linear growth factor at redshift z, and the peak height parameter, $\nu=\delta_c(z)/\sigma(M,z)$,
%where $\delta_c(z)$ is the linear collapse threshold and $\sigma(M,z)$ specifies the variance of the matter fluctuations
%over the scale $\varpropto M^{1/3}$ at a redshift $z$. The field term $n_{peak}^n(\nu)$ is given by
%\begin{equation}
% \begin{split}
%n_{peak}^{n}(\nu)=\frac{1}{d\Omega}\Big\{n_{ran}(\nu)\Big[d\Omega-\int dz\frac{dV(z)}{dz}\\
%   \times\int dM\,n(M,z)\,(\pi R_{vir}^{2})\Big]\Big\},
%\end{split}
%\label{eq18}
%\end{equation}
%where $n_{ran}(\nu)$ is the surface number density of pure noise peaks (Fan et al. 2010).

\subsection{The comparison of the model with simulations}

To test the model applicability in cosmological studies, we compare the peak counts predicted from the model of
F10 with simulation results in terms of their cosmological dependence. Within the flat $\Lambda$CDM framework,
we concentrate on $\Omega_m$ and $\sigma_8$, the two most important parameters for weak-lensing analyses.
Therefore for comparison purposes, in addition to the fiducial model runs, we also perform ray-tracing simulations for
four other cosmological models with different $(\Omega_m, \sigma_8)$ around the fiducial values as shown in Table~\ref{tab:param}.

For each of the variational model, we run four sets of ray-tracing simulations and obtain totally $4\times16=64$ weak-lensing maps
each with $3\times 3\deg^2$. In order to suppress the cosmic variance so that to reveal the cosmological dependence of the peak counts clearly,
except with different $\Omega_m$ or $\sigma_8$, each set of the simulations is done in the identical way as that of the corresponding fiducial model
with matched initial conditions for each N-body run. For each of the maps, we perform the convergence reconstruction also
in a way that is identical to the corresponding fiducial one using the same background galaxy catalog.
With these matched reconstructed maps, the derivatives of the peak counts with respect to $\Omega_m$ and $\sigma_8$ are then analyzed separately
as follows using the double-sided derivative estimator \citep[e.g.,][]{Marian13}
\begin{equation}
\frac{{\partial {N}_{peak}(\nu_i)}}{\partial{p_\alpha}}|_{p_{\alpha}}=\frac{1}{M}\sum_{f=1}^{M}\frac{N_{peak}^f(\nu_i, p_\alpha+\Delta p_\alpha)
-N_{peak}^f(\nu_i, p_\alpha-\Delta p_\alpha)}{2\Delta p_\alpha},
\label{derivative}
\end{equation}
where $p_{\alpha}$ stands for the cosmological parameter we are interested in, and specifically $\Omega_m$ or $\sigma_8$
for the analyses here, $f$ is for different matched pairs of maps with the total number of pairs $M=64$,
$N_{peak}^f(\nu_i, p_\alpha\pm \Delta p_\alpha)$ is for the number of peaks in the signal-to-noise ratio bin
centered on $\nu_i$ with bin width of $0.5$ in the map of $3\times 3\deg^2$ with the cosmological parameter $p_\alpha\pm \Delta p_\alpha$ in pair $f$.
The derivatives are estimated at the fiducial value $p_{\alpha}$.

The results are shown in Fig.~\ref{fig:dependence}, where the left and right panels are for the derivatives with respect to $\sigma_8$
and $\Omega_m$, respectively, divided by the corresponding average peak number from the fiducial model.
The blue symbols with error bars are for the simulation results. The shaded regions indicate the $1\sigma$ ranges
of the variations of the derivatives estimated from single pairs. The error bars show the expected errors for the values averaged over the $64$ pairs of maps.
The red solid lines are the results calculated from our model F10 taking into account the noise effects,
and the green dash lines are for the theoretical results without
including the noise effects calculated from the halo mass function assuming spherical NFW halos \citep[e.g.,][F10]{HTY2004}.
It is seen that within the error ranges, our model predictions (red lines) agree with the simulations results very well.
Comparing the red and green lines, we can see that the two are in good accord with each other for peaks with $\nu\ge 6$.
On the other hand, for peaks with $\nu\sim 4-5$, the green lines are higher than the red lines, signifying more
cosmological information predicted by the green ones. This shows that the noise is important for peak counts with
$\nu\sim 4-5$. It is noted that in our model F10, we only include the noise effect from intrinsic ellipticities of
source galaxies, and do not consider the projection effect from line-of-sight large-scale structures and the
nonspherical mass distribution for dark matter halos. While the noise is indeed the dominant source of errors,
the latter two effects can also affect the peak counts to some extent, and they contain
cosmological information themselves \citep[e.g.,][]{TF2005, HOSS2012}. This may be related to the tendency seen in Fig.~\ref{fig:dependence} that the simulation results
are mildly higher than the red lines. We will explore the model improvements further
in our future studies. For the current analyses, we conclude that within the error ranges, the cosmological dependence predicted
by our model with the noise effect included are in very good agreements with the simulation results.

\begin{figure*}[tbl]
%\plottwo{Fig7a.eps}{Fig7b.eps}
 \begin{minipage}[c]{0.5\textwidth}
  \centering
  \includegraphics[width=1\textwidth,height=0.8\textwidth]{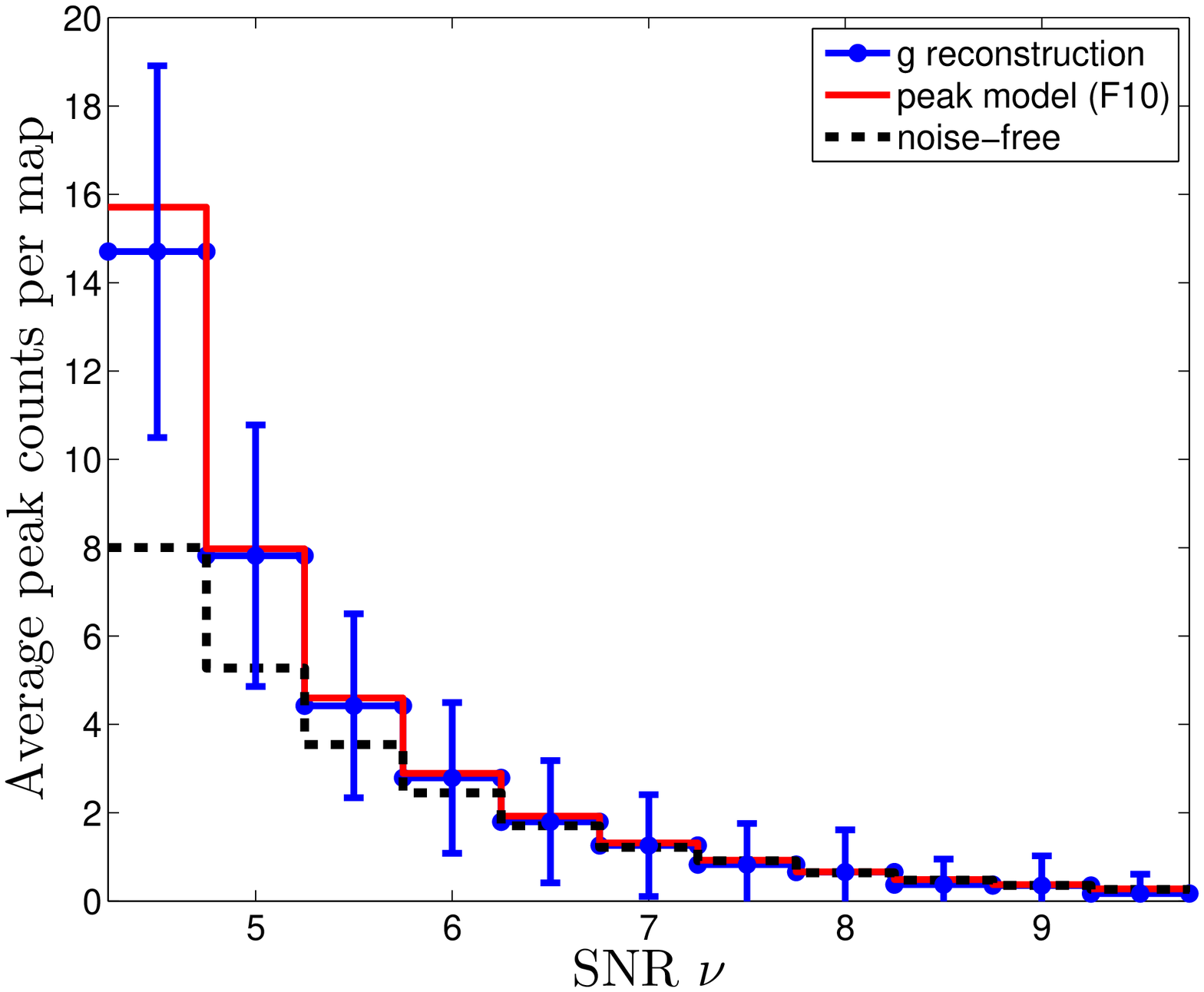}
\end{minipage}
\begin{minipage}[c]{0.5\textwidth}
  \centering
  \includegraphics[width=1\textwidth,height=0.8\textwidth]{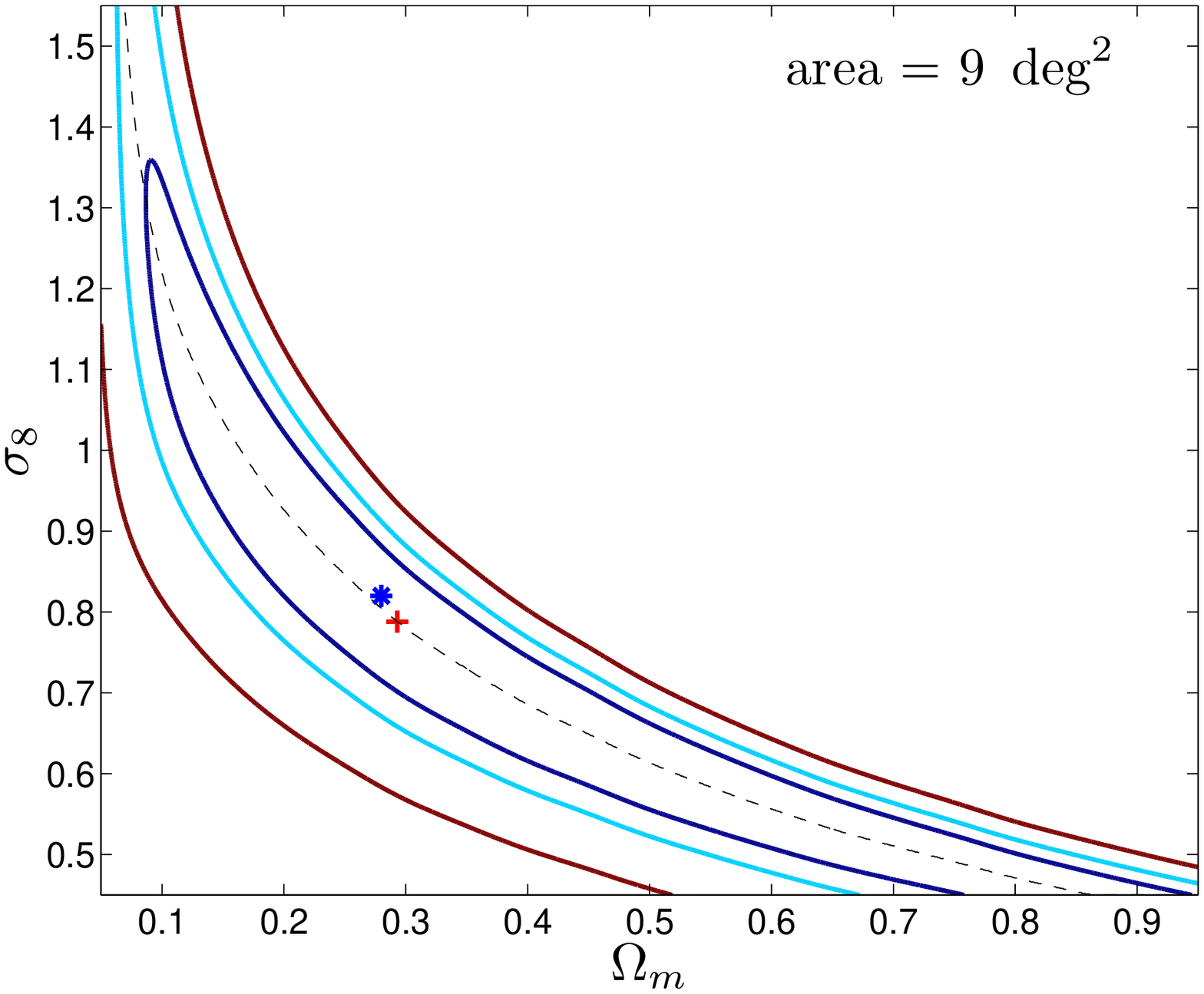}
\end{minipage}
\caption{Left: The average numbers of peak counts per map for the fiducial model.
The blue histograms with error bars are for the results from `g reconstruction' maps. The red histograms are the results
predicted from F10. The black histograms are for the results from the theoretical model without noise.
Right: Cosmological constraints on $(\Omega_m,\sigma_8)$ from $\chi^2$ fitting for a survey of $9 \deg^2$ from the `g reconstruction'
maps without masks. The contours from inside out show the $1\sigma$, $2\sigma$ and $3\sigma$ ranges, respectively. The dashed line
gives more or less the degeneracy direction between $\Omega_m$ and $\sigma_8$ in terms of weak-lensing peak counts considered here.
} \label{fig:grec}
\end{figure*}

Besides the derivatives with respect to cosmological parameters,
we also perform a direct comparison between peak counts from simulations
and our model prediction. The results are shown in the left panel of Fig.~\ref{fig:grec}. The blue histograms show the peak counts in
$3\times 3\deg^2$ averaged over the $128$ `g-reconstruction' maps for the fiducial model. The attached error bars are for the $1\sigma$ ranges
of the map-to-map variations. The red histograms are our model predictions, and the black ones are for the theoretical results without including
the noise effects. We can see that in the considered peak range, there is an excellent agreement between the results from our model prediction
and those from simulations. The black histograms are systematically lower than the simulation results for $\nu\sim 4-6$, again
demonstrating clearly the noise effect on peak counts. Therefore if this model without including the noise effect is used in
cosmological parameter fitting, a significant bias can arise. On the other hand, our model F10 can expectedly give better
constraints. We note again that here we use $11$ bins, linearly distributed in the considered signal-to-noise ratio range with
the bin width of $0.5$, in our peak counting. Different binning methods can give rise to specifically
different values of peak counts. However the systematic agreement of the trend between the blue and red histograms
and the systematic differences between them and the black ones
indicate that a different choice of binning should not change the results of their comparisons qualitatively.
Similarly for the mask effects to be discussed in the following.

Given the good agreements within error ranges shown above,
%seen in the comparisons both in terms of the derivatives of the peak counts with respect to $\sigma_8$ and
%$\Omega_m$, and the peak counts themselves,
in the studies for the mask effects on weak-lensing peak counts and
the consequent bias on the derived cosmological parameters, we adopt the model of F10 in the cosmological parameter fitting analyses.

\subsection{Cosmological parameter fitting from peak counts}
Shown in the previous subsection, our model including the noise effects agree well with simulation results.
We then use the peak counts identified directly from reconstructed maps for cosmological studies without the need to
distinguish true and false peaks. To derive cosmological parameter constraints from weak-lensing peak counts,
we minimize the $\chi^2$ defined as follows
\begin{equation}
 \chi_{p'}^{2}=\boldsymbol{dN}^{(p')}(\widehat{\boldsymbol{C}^{-1}})\boldsymbol{dN}^{(p')}
=\sum_{ij=1,...,11}dN_{i}^{(p')}(\widehat{C_{ij}^{-1}})dN_{j}^{(p')},
\label{chi2}
\end{equation}
where $dN_i^{(p')}=N_{peak}^{(p')}(\nu_i)- N_{peak}^{(d)}(\nu_i)$ with $N_{peak}^{(p')}(\nu_i)$
being the prediction for the cosmological model $p'$ from F10 and $N_{peak}^{(d)}(\nu_i)$ being the `observed' data
for the peak count in the signal-to-noise ratio bin centered on $\nu_i$,
and $C_{ij}$ is the covariance matrix of the peak counts including the error correlations between different
$\nu$ bins. It has been shown that the direct inversion of $C_{ij}$ estimated from simulated maps leads to an biased estimate of its inverse.
%However, the direct expectation value of $(\boldsymbol{C}^{-1})$ is not the inverse of the population covariance due to noise in the covariance matrix measurement.
An unbiased estimator of the inverse covariance matrix is given by \cite{HSS2007}
\begin{equation}
\widehat{\boldsymbol{C}^{-1}}=\frac{R-N_{bin}-2}{R-1}(\boldsymbol{C}^{-1}),~~N_{bin}<R-2
\label{nobiascov}
\end{equation}
where $N_{bin}$ is the number of bins used for peak counting, and R is the number of independent maps used in calculating $C_{ij}$.
In our case,  $N_{bin}=11$ and $R=128$. We adopt $\widehat{\boldsymbol{C}^{-1}}$ to evaluate the inverse covariance matrix during the whole analysis.

The `observed' data are constructed from the simulations for the fiducial model as follows.
For each of the $128$ reconstructed maps without or with masks, we identify peaks following the descriptions in \S 3.4.
To reduce the boundary effect on peak counts, we exclude the outermost $10$ pixels in each direction in peak counting.
Thus the effective area of each map is $[3(1-20/1023)]^2\approx 8.65\hbox{ deg}^2$.
For each map $r$, we count peaks in each of the $11$ signal-to-noise ratio bins of width $\Delta \nu=0.5$ in the range of $4.25\le \nu \le 9.75$.
We then calculate the mean number of peaks in each bin by averaging over the $128$ maps,
and scale it back to $9\hbox{ deg}^2$ by multiplying a factor of $9/8.65$.
These average peak counts form the `observed' data $N_{peak}^{(d)}(\nu_i)$ with $\nu_i=\{4.5,5.0,5.5,6.0,6.5,7.0,7.5,8.0,8.5,9.0,9.5\}$, respectively.

The covariance matrix $C_{ij}$ is also calculated from the $128$ simulated maps by
\begin{equation}
% C_{ij}^{(f)}=\frac{1}{R-1}\sum_{r=1}^{R}(N_{i}^{(f;r)}-\bar{N}_{i}^{(f)})(N_{j}^{(f;r)}-\bar{N}_{j}^{(f)})
 C_{ij}=\frac{1}{R-1}\sum_{r=1}^{R}[N^r_{peak}(\nu_i)-{N}_{peak}^{(d)}(\nu_i)][N^r_{peak}(\nu_j)-{N}_{peak}^{(d)}(\nu_j)]
%(N_{j}^{(r)}-\bar{N}_{j}),
\label{covar}
\end{equation}
where $r$ denotes for different maps with the total number of maps $R=128$, and $N^r_{peak}(\nu_i)$ is for the
peak count in the bin centered on $\nu_i$ from the map $r$ (scaled back to $9\deg^2$).

The right panel of Fig.~\ref{fig:grec} shows the fitting result of $(\Omega_m, \sigma_8)$ with the data obtained
from the reconstructed maps for the fiducial model without masks. Here the red symbol indicates the best fit values, and the
contours from inside out show the $1,2,3\sigma$ ranges, respectively. The blue symbol is for the input $(\Omega_m, \sigma_8)$
for the fiducial model. We see that the best fit result obtained using our model agrees with the fiducial input very well.
This further demonstrates the cosmological applicability of our model within error ranges in addition to the comparisons shown in \S 4.2.

We now proceed to analyze the mask effects on weak-lensing peak counts and consequently on cosmological studies.

\begin{figure*}[tbl]
% \plottwo{Fig8a.eps}{Fig8b.eps}
\begin{minipage}[c]{0.5\textwidth}
  \centering
  \includegraphics[width=1\textwidth]{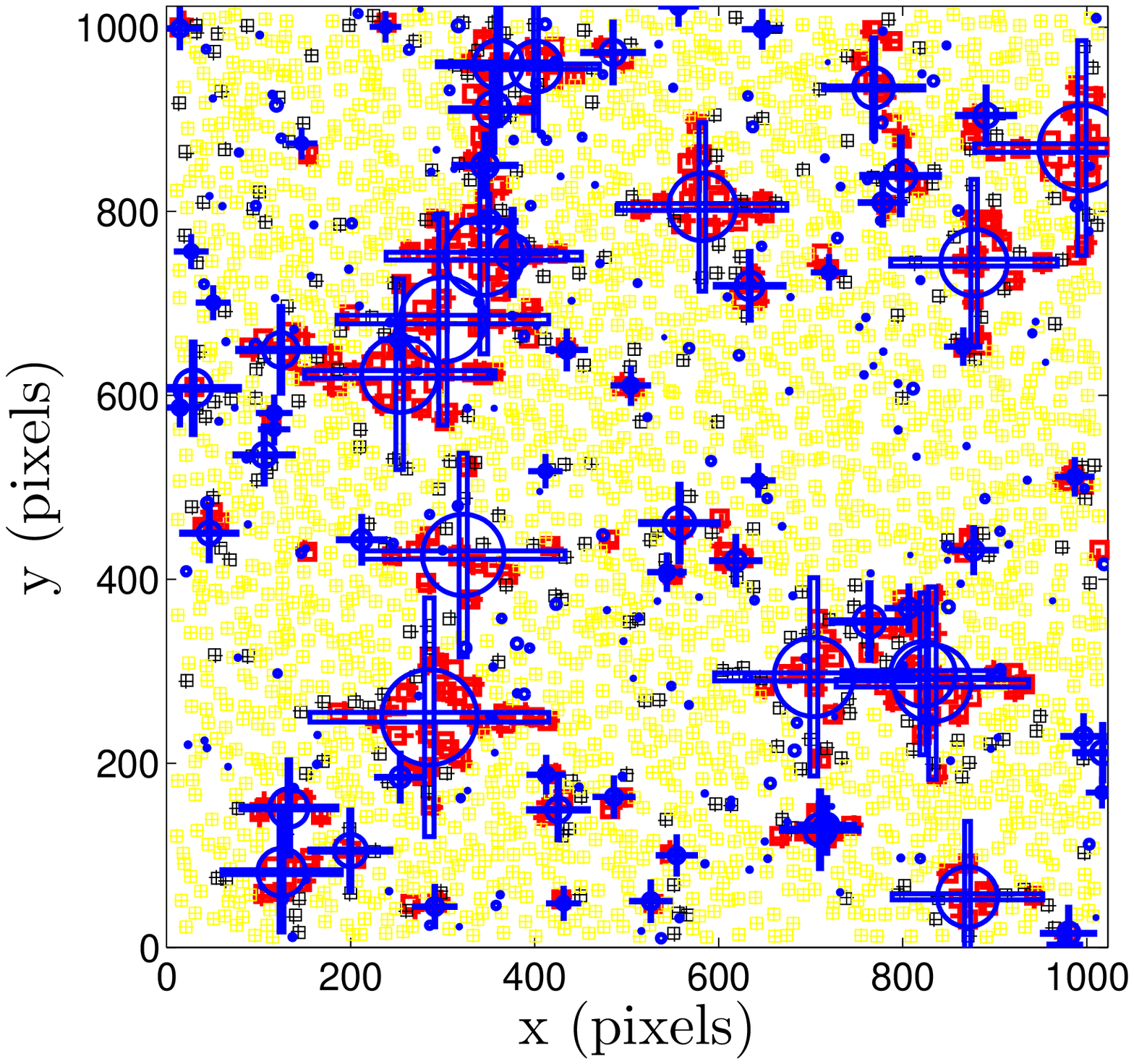}
\end{minipage}
\begin{minipage}[c]{0.5\textwidth}
  \centering
  \includegraphics[width=1\textwidth]{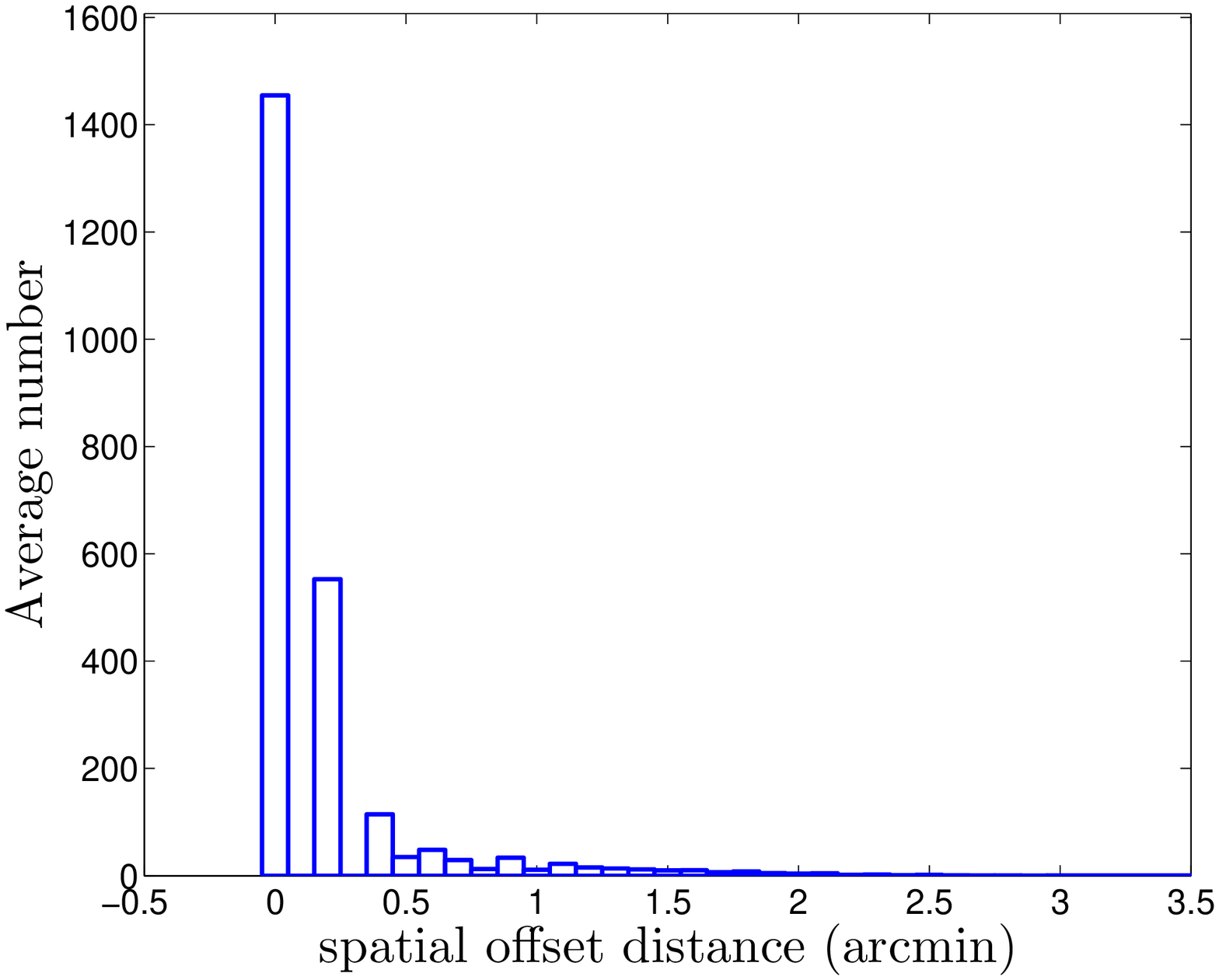}
\end{minipage}
\caption {The mask effects on peak positions. The left panel shows the spatial distribution of peaks in one map. The
square and plus symbols are for peaks in the case with and without masks, respectively. The red, black and yellows ones
are, respectively, for peaks with spatial offsets larger than $0.5\hbox{ arcmin}$, in the range of $[0.2\hbox{ arcmin},0.5\hbox{ arcmin}]$
and less than $0.2\hbox{ arcmin}$. The masks are shown in blue. The right panel is the statistical distribution
of the spatial offset obtained by averaging over the $128$ pairs of maps.
}\label{fig:spatial}
\end{figure*}

\begin{figure*}[tbl]
% \plottwo{Fig9a.eps}{Fig9b.eps}
\begin{minipage}[c]{0.5\textwidth}
  \centering
  \includegraphics[width=1\textwidth,height=0.8\textwidth]{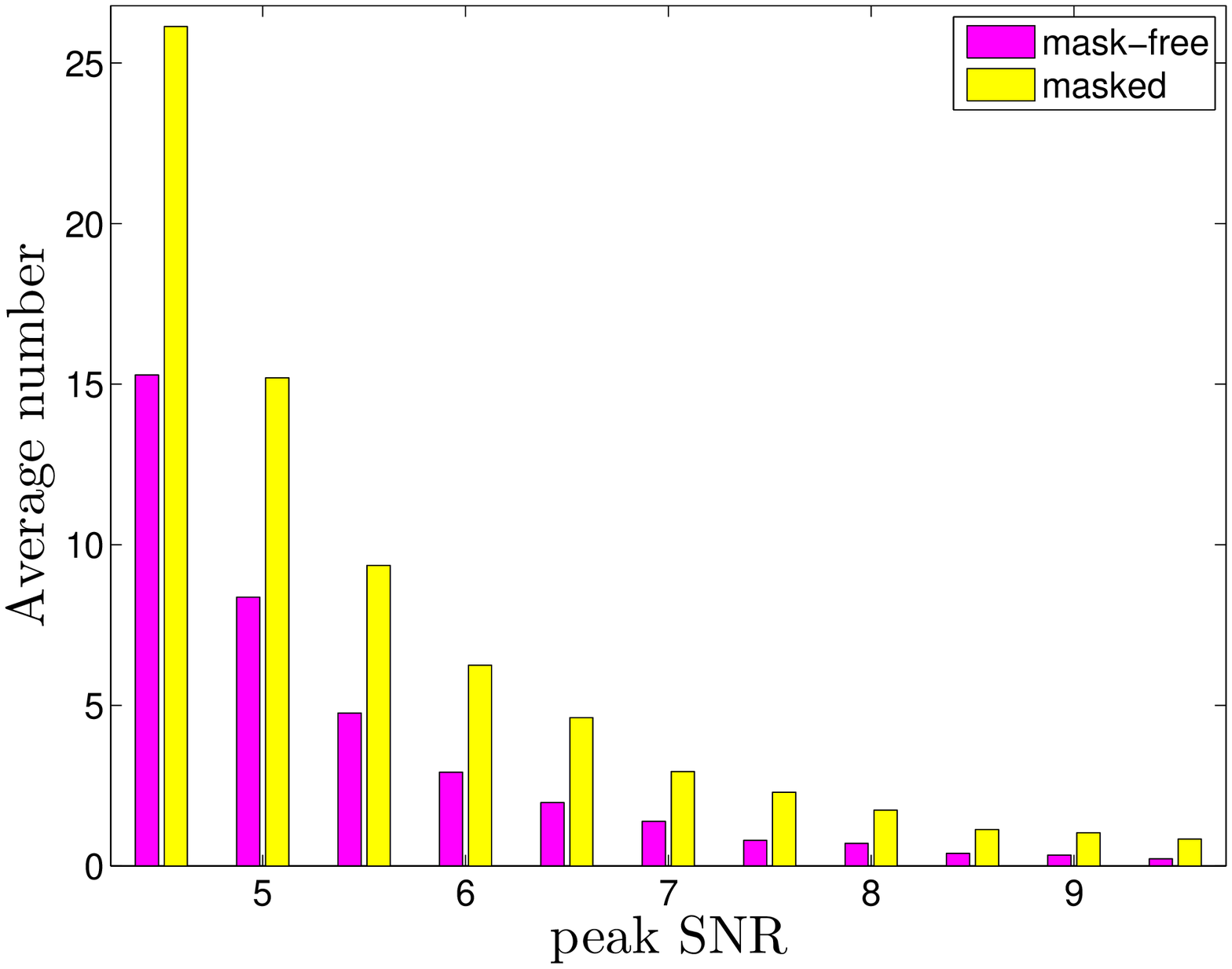}
\end{minipage}
\begin{minipage}[c]{0.5\textwidth}
  \centering
  \includegraphics[width=1\textwidth,height=0.8\textwidth]{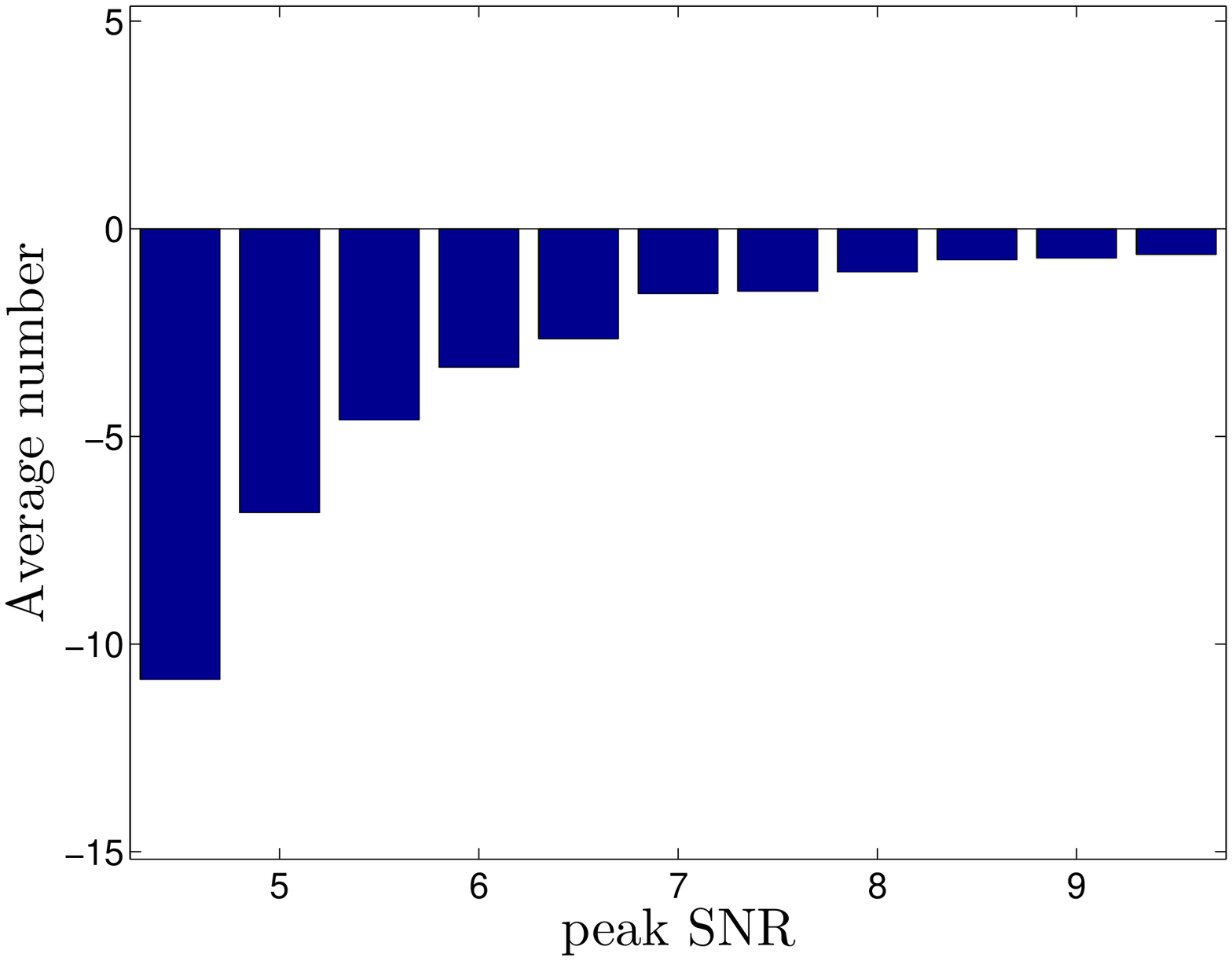}
\end{minipage}
\caption {The peak counts distribution. The left panel shows the peak number distributions for the cases without (purple)
and with (yellow) masks, respectively. The right panel shows the peak number difference between the two cases as
a function of $\nu=K_N/\sigma_0$ with $\sigma_0=0.02$.
}\label{fig:height}
\end{figure*}

\section{Results}

\subsection{Mask effects}

In this section, we discuss the mask effects on weak lensing peak analyses by comparing two sets of $g$-reconstructed convergence maps
with and without masks, respectively. There are totally $128$ maps for each set. For each map in the case without masks,
there is a corresponding map with masks that the source galaxies are exactly the same as the other one except that the galaxies
within the masked regions are discarded and the convergence reconstruction is done from the smoothed $\langle\boldsymbol \epsilon\rangle$ field obtained
from the remaining galaxies. We then have $128$ pairs of maps that allow us to do detailed comparisons.
The mask size distribution model is described in \S3.5. Three cases with the total number of masks $N_{mask}=140, 280$ and $420$ for each $9\hbox{ deg}^2$
are considered, which corresponds to the masked area fraction of $\sim 7\%, \sim 13\%$ and $\sim 19\%$, respectively. The case with $N_{mask}=280$
is taken to be our fiducial case for most of the results presented in the following.

To perform detailed comparisons for peaks in convergence maps with and without masks, we need to identify the peak correspondences between
each pair of maps. This is done by peak matching. For each peak in a map from one set, we check for peaks within $3.5\hbox{ arcmin}$ in each
dimension around it in its peer map from another set and define the nearest peak within this region as its partner peak. Only those pairs of peaks
that are partners to each other are identified as peaks with correspondences.

The existence of masks affects the convergence reconstruction and consequently the peak properties both in their spatial location and
peak height.

\begin{figure*}[tbl]
%\plottwo{Fig10a.eps}{Fig10b.eps}
 \begin{minipage}[c]{0.5\textwidth}
  \centering
  \includegraphics[width=1\textwidth]{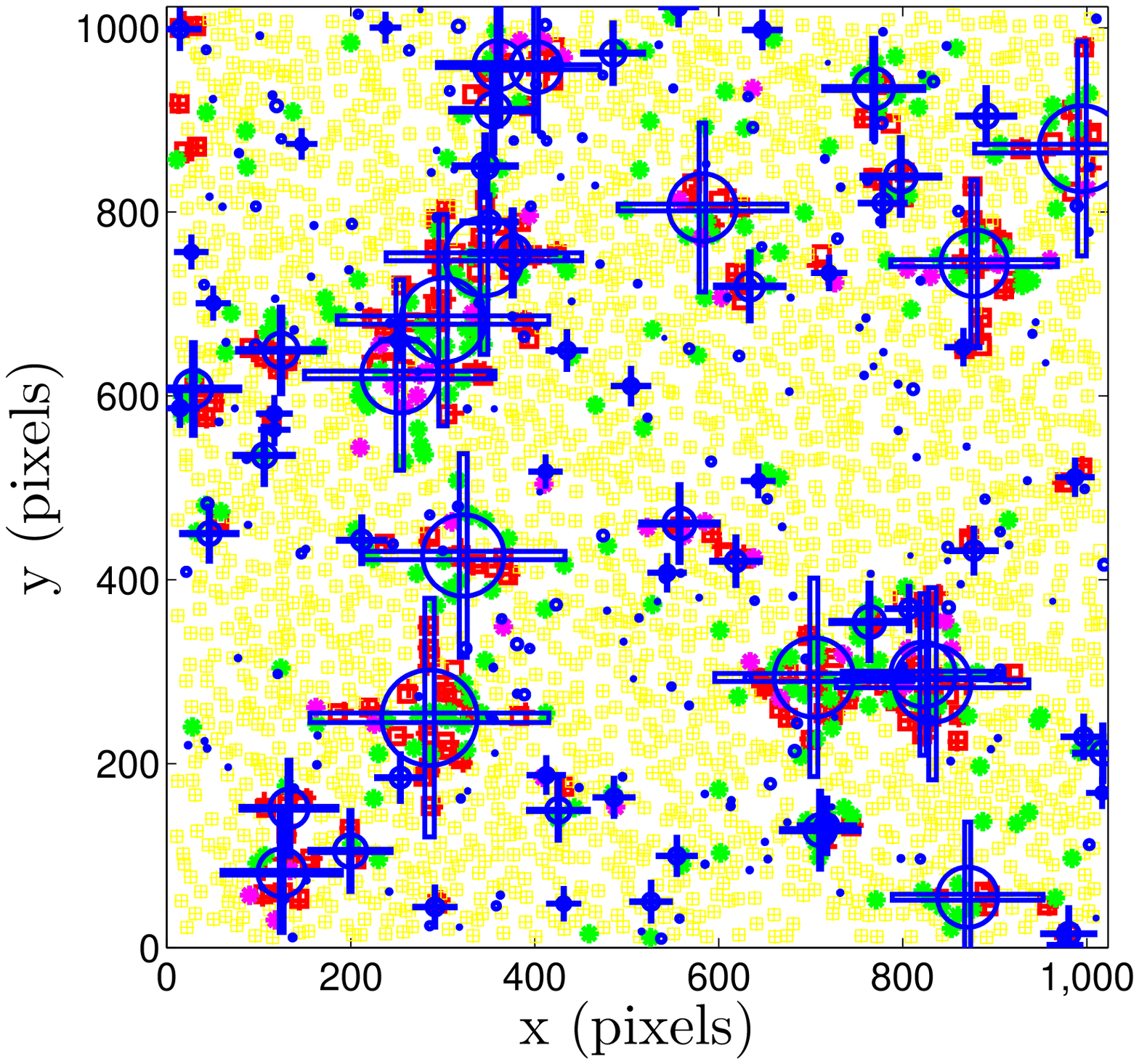}
\end{minipage}
\begin{minipage}[c]{0.5\textwidth}
  \centering
  \includegraphics[width=1\textwidth]{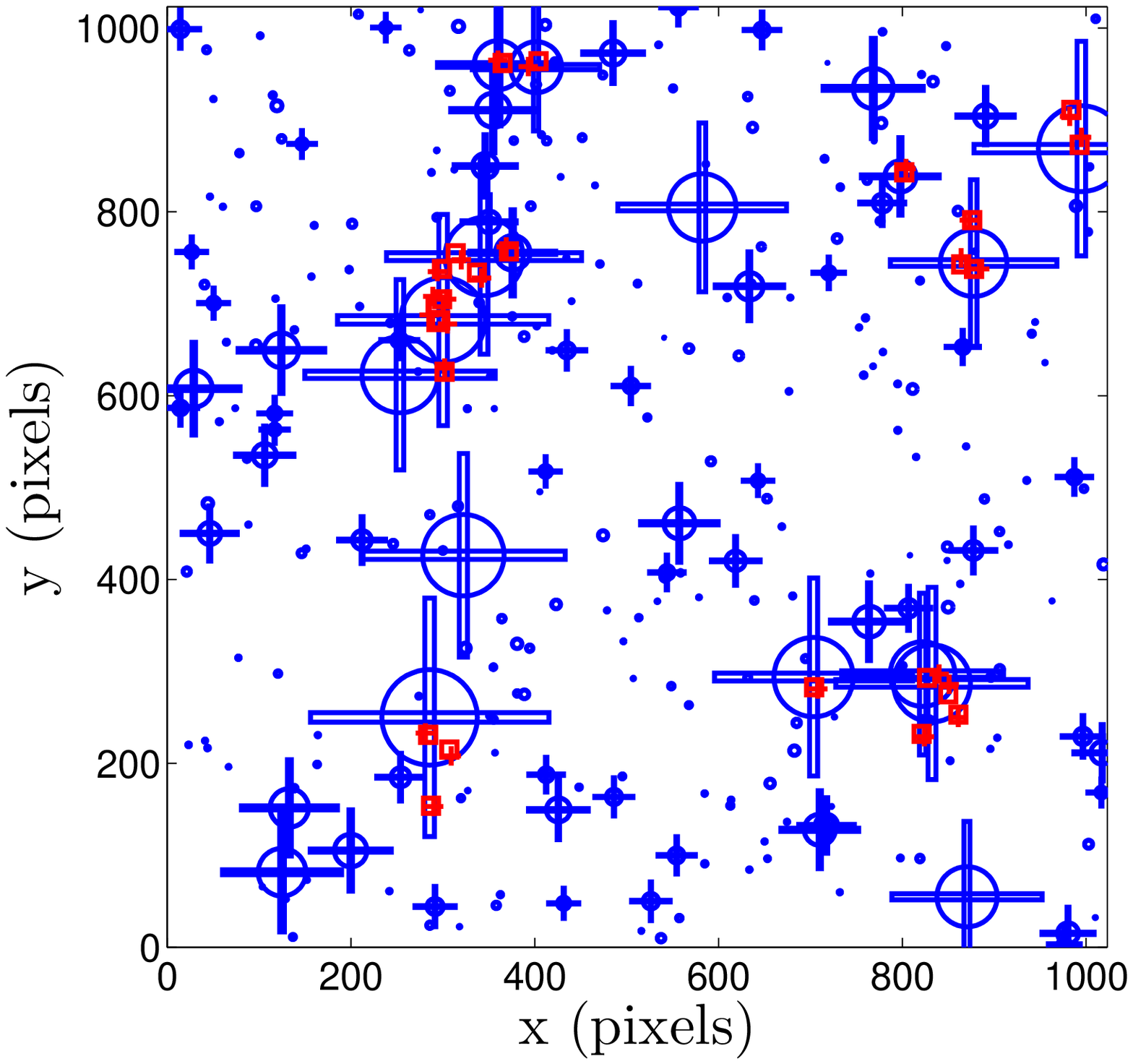}
\end{minipage}
\caption {The illustration of the spatial distribution of the affected peaks. In the left panel, the red, green and purple symbols are for
Type I affect peaks, Type II affected peaks and Type II with $\nu\ge 2$, respectively. The yellow symbols are for the rest of the
peaks with correspondences between the case with and without masks. The right panel shows the Type I peaks with $\nu< 3.25$
in the case without masks but with the corresponding peak height shifting to $\nu>4.25$ in the case with masks (red symbols).
}\label{fig:largeinfluence}
\end{figure*}

\begin{figure*}[tbl]
%\plotone{Fig11.eps}
\begin{minipage}[c]{0.95\textwidth}
  \centering
  \includegraphics[width=1\textwidth]{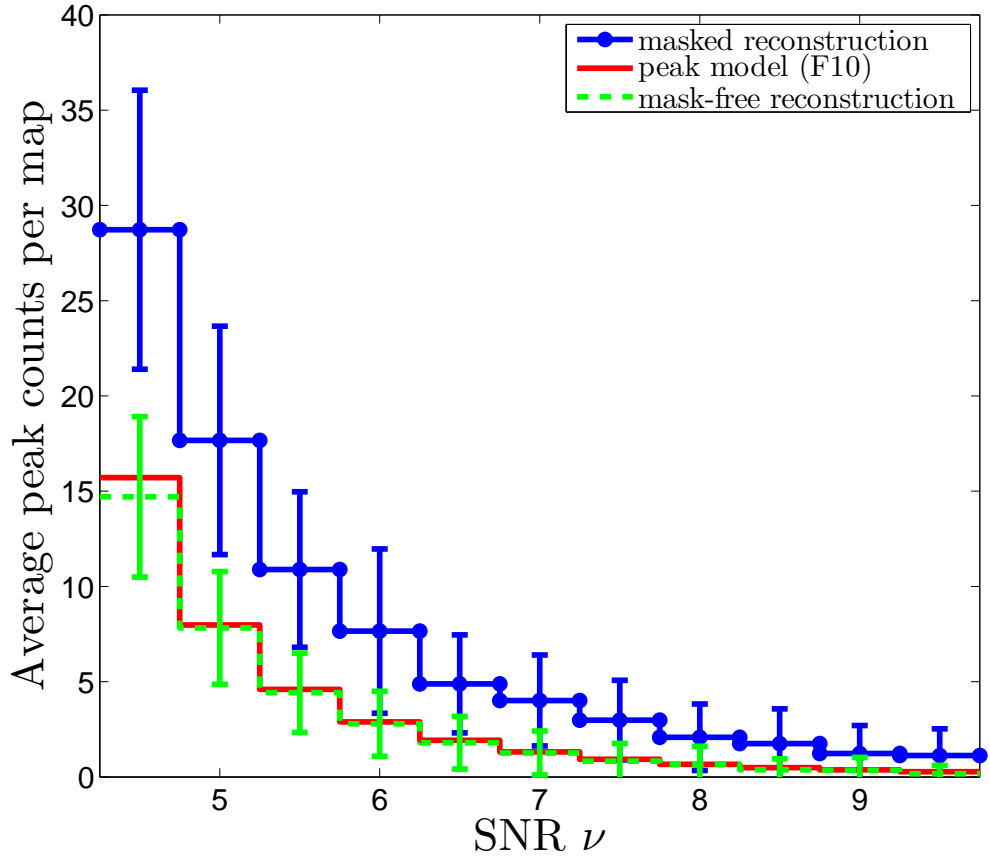}
\end{minipage}
\caption {The comparison of the peak counts between the cases with and without masks, where the blue, green and red histograms
are for the cases with masks, without masks and the model prediction of F10 with a uniform noise of $\sigma_0=0.02$.}
%The upper right panel shows the fitting results for the survey area of $9\hbox{ deg}^2$. The blue symbol is for the
%fiducial values and the red symbol is for the best fit with the peak counts in the case with masks as the 'observed data'
%and $\sigma_0=0.02$in the model of Fan et al. (2010). The lower left and right panels are for the survey area of $150\hbox{ deg}^2$
%and $1000\hbox{ deg}^2$,respectively.}
\label{fig:avepeakmask}
\end{figure*}

\begin{figure*}[tbl]
%\plottwo{Fig12a.eps}{Fig12b.eps}
\begin{minipage}[c]{0.5\textwidth}
  \centering
  \includegraphics[width=1\textwidth,height=0.8\textwidth]{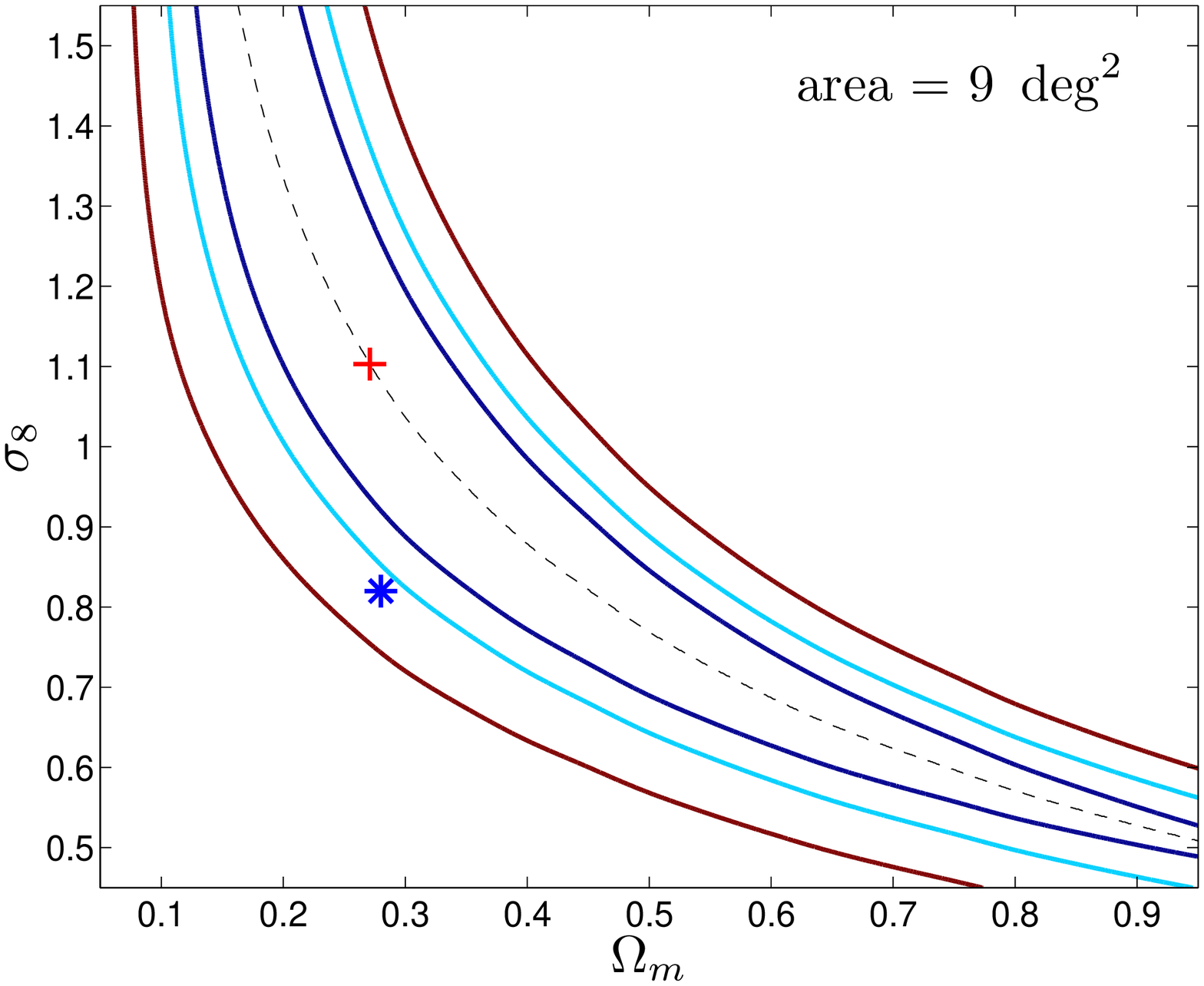}
\end{minipage}
\begin{minipage}[c]{0.5\textwidth}
  \centering
  \includegraphics[width=1\textwidth,height=0.8\textwidth]{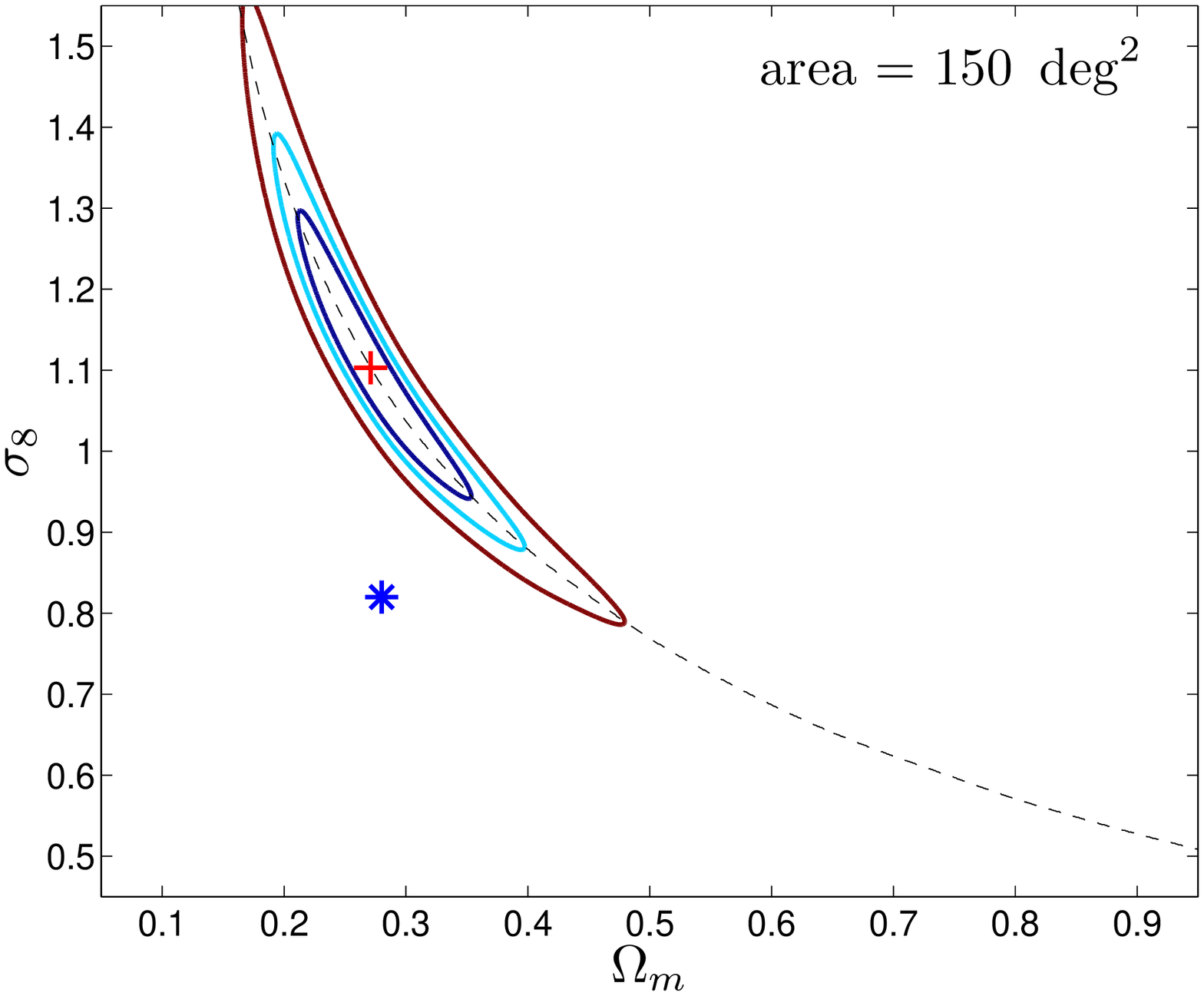}
\end{minipage}
\caption {The bias in cosmological parameter constraints due to the mask effects.
The left panel shows the fitting results for the survey area of $9\hbox{ deg}^2$. The blue symbol is for the
fiducial values and the red symbol is for the best fit with the peak counts in the case with masks as the `observed data' and $\sigma_0=0.02$
in the model of F10. The right panel is for the results with the survey area of $150\hbox{ deg}^2$.}
% and $1000\hbox{ deg}^2$, respectively.}}
\label{fig:maskfitting}
\end{figure*}

\subsubsection{Spatial location}

In Fig.~\ref{fig:spatial} we present the mask effect on spatial positions of peaks for $N_{mask}=280$. The left panel shows an example map of the spatial distribution of peaks with correspondences. The squared and plus symbols are respectively for peaks in the cases with and without masks. The red, black and yellow ones are for the pairs of peaks with their spatial offset larger than $0.5\hbox{ arcmin}$, in the range of $[0.2,0.5]\hbox{ arcmin}$ and less than $0.2\hbox{ arcmin}$, respectively. It is seen clearly that strongly affected peaks are almost all closely associated with masks, especially large masks. In the right panel of Fig.~\ref{fig:spatial}, the statistical offset distribution averaged over $128$ pairs of maps is shown. There are about $40\%$ of peaks with offset larger than $0.1\hbox{ arcmin}$. The fraction with offset larger than $0.5\hbox{ arcmin}$ is $\sim 11\%$. We also notice that lower peaks are more strongly affected by masks. This offset due to mask can have significant effects on weak lensing analyses for individual clusters. For a typical weak lensing observation targeting at a particular cluster, the observed size is about $20\hbox{ arcmin}$. If there happens to be a large mask close to the central region of the cluster, the weak lensing determined center for the cluster can be considerably offsetted from its true center, which in turn can lead to large errors in the weak lensing determination of the density profile for the cluster.

\subsubsection{Peak height}

We now discuss the mask effects on peak heights. Fig.~\ref{fig:height} shows the results,
where the left panels are the peak number distribution in $9\hbox{ deg}^2$ averaged over $128$ maps in each case and
the right panels are the peak number differences between the cases without and with masks, respectively, as a function of
S/N ($\nu=K_N/\sigma_0$ with $\sigma_0=0.02$).
It is clearly seen that the number of peaks in high signal-to-noise bins is systematically higher in the case with masks,
which can expectedly affect the cosmological parameter constraints with weak lensing peak counts significantly. We further
exam the correlation between the positions of the strongly affected peaks and the locations of masks. We define two types of
strongly affected peaks. Type I is for peaks with their peak height difference between the cases with and without masks
higher than $1\sigma$. Type II is for peaks without correspondences between the two cases. Fig.~\ref{fig:largeinfluence}
presents a typical map with masks. The left panel shows the spatial distribution of peaks with squares and pluses
for peaks in the cases with and without masks, respectively. The red symbols are for Type I peaks, the green symbols are for
Type II peaks with the purple ones for Type II peaks with $\nu\ge 2$, and the yellows ones are for the rest. The clustering of
the strongly affected peaks around large masks are apparent. The right panel shows particularly the Type I peaks with $\nu< 3.25$
in the case without masks but with the corresponding peak height shifting to $\nu>4.25$ in the case with masks. It is found
that they all trace large masks. It is these high Type I peaks that can affect profoundly the cosmological parameter constraints.

Fig.~\ref{fig:avepeakmask} shows the effects of masks on the peak counts, where the peak counts are calculated by averaging over $128$ maps in each case
and $\nu$ is computed with $\sigma_0=0.02$ in all cases. The blue, green and red histograms are for peak counts in the case with masks, without masks and
the theoretical prediction of F10 with a uniform noise with $\sigma_0=0.02$. It is seen clearly that the peak counts considered here are
systematically higher due to the presence of masks. We do not expect the results to change qualitatively with different choices of
peak binning. Fig.~\ref{fig:maskfitting} shows the corresponding fitting results with the survey area of $9\hbox{ deg}^2$
(left) and $150\hbox{ deg}^2$ (right), respectively. The fittings are done with the `observed data' being the peak counts for the masked case,
and the model of F10 with $\sigma_0=0.02$ uniformly. The covariance matrix is calculated from the $128$ reconstructed maps with masks.
The meanings of the lines and symbols are similar to those of Fig.~\ref{fig:grec}.
Clearly, the enhanced peak counts due to the occurrence of masks lead to a large bias in cosmological parameter fitting.
Even for a survey of $9\hbox{ deg}^2$, the true cosmological parameter values (blue symbols) lie outside the $2\sigma$ contour around the best fit (red symbol).
This demonstrates the significance of the mask effects, which must be taken into account carefully in cosmological parameter constraints
with weak lensing peak counts. For the results with the survey area of $150\hbox{ deg}^2$ (similar to the survey area of CFHTLenS \citep{Ecfhtlens13}),
we take a simple approach to rescale the covariance matrix calculated from $128$ masked convergence maps to that of the larger survey area
assuming a Poisson scaling relation to the survey area $S$ as $1/S$ \citep{K2010}. This may underestimate the covariance matrix
by a factor of $\sim 1.5$ given the existence of long-range correlations of the true peaks \citep{K2010}.

%The lower left and right panels show the results for the survey areas of $150\hbox{ deg}^2$
%(similar to the survey area of CFHTLenS \citep{Ecfhtlens}) and $1000\hbox{ deg}^2$, respectively. Here we take a simple approach
%to rescale the covariance matrix calculated from $128$ noisy convergence maps to the two larger surveys assuming a Poisson
%scaling relation to the survey area $S$ as $1/S$ \citep{K2010}. This may underestimate the covariance matrix by a factor of $\sim 1.5$
%given the existence of long-range correlations of the true peaks \citep{K2010}.}

\begin{figure*}[tbl]
 \begin{minipage}[c]{0.33\textwidth}
  \centering
  \includegraphics[width=1.1\textwidth]{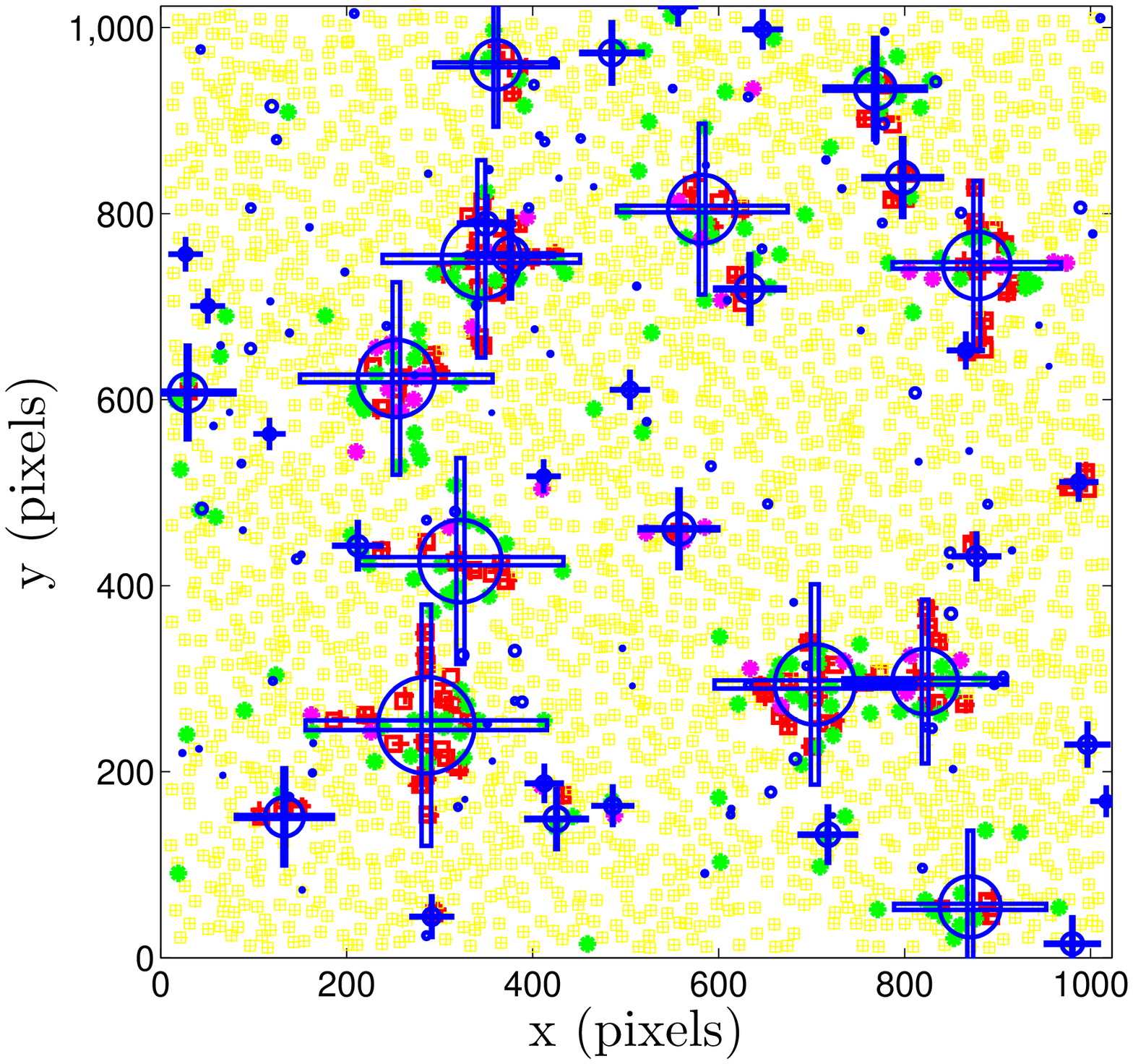}
\end{minipage}
\begin{minipage}[c]{0.33\textwidth}
  \centering
  \includegraphics[width=1.1\textwidth]{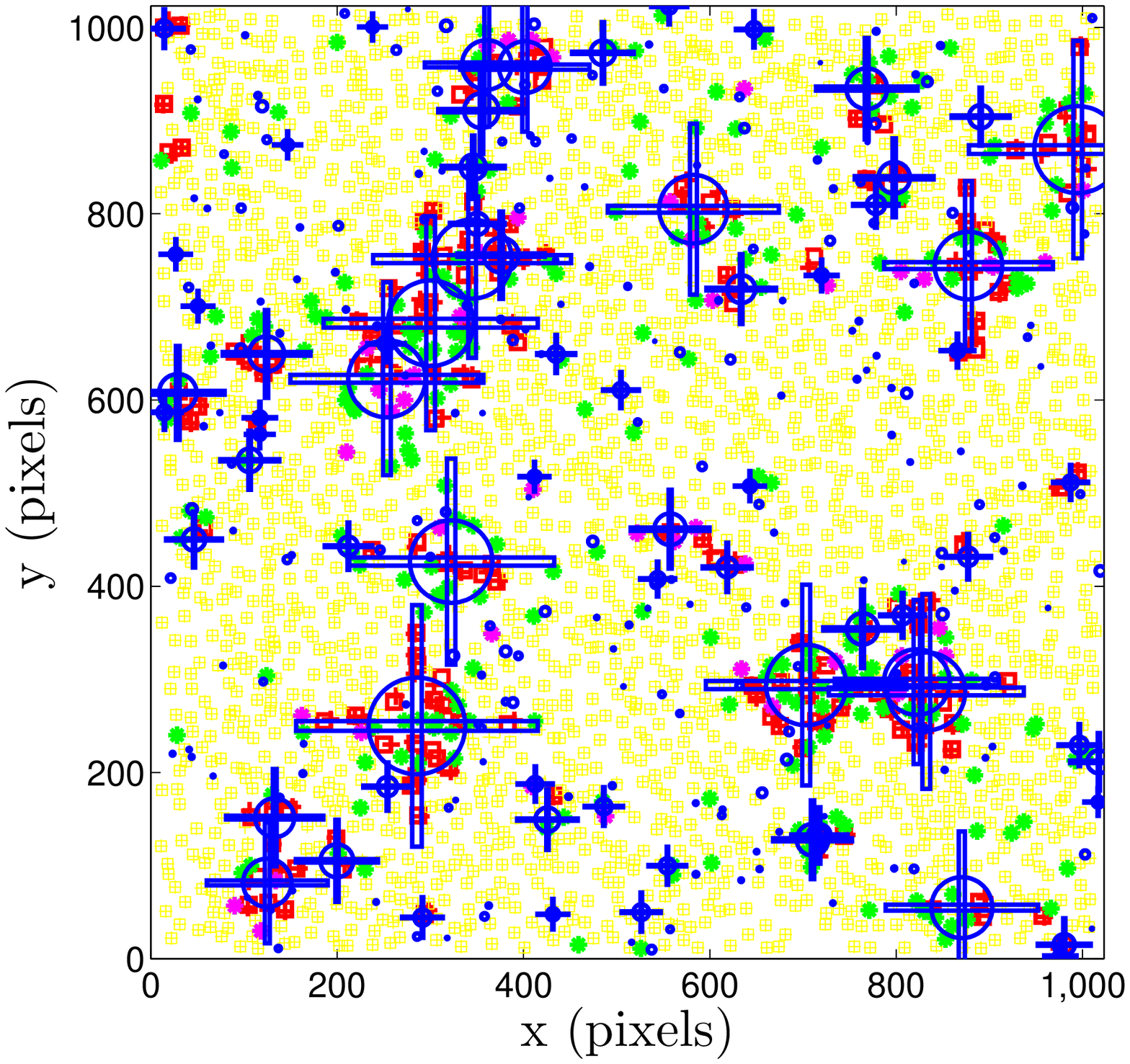}
\end{minipage}
\begin{minipage}[c]{0.33\textwidth}
  \centering
  \includegraphics[width=1.1\textwidth]{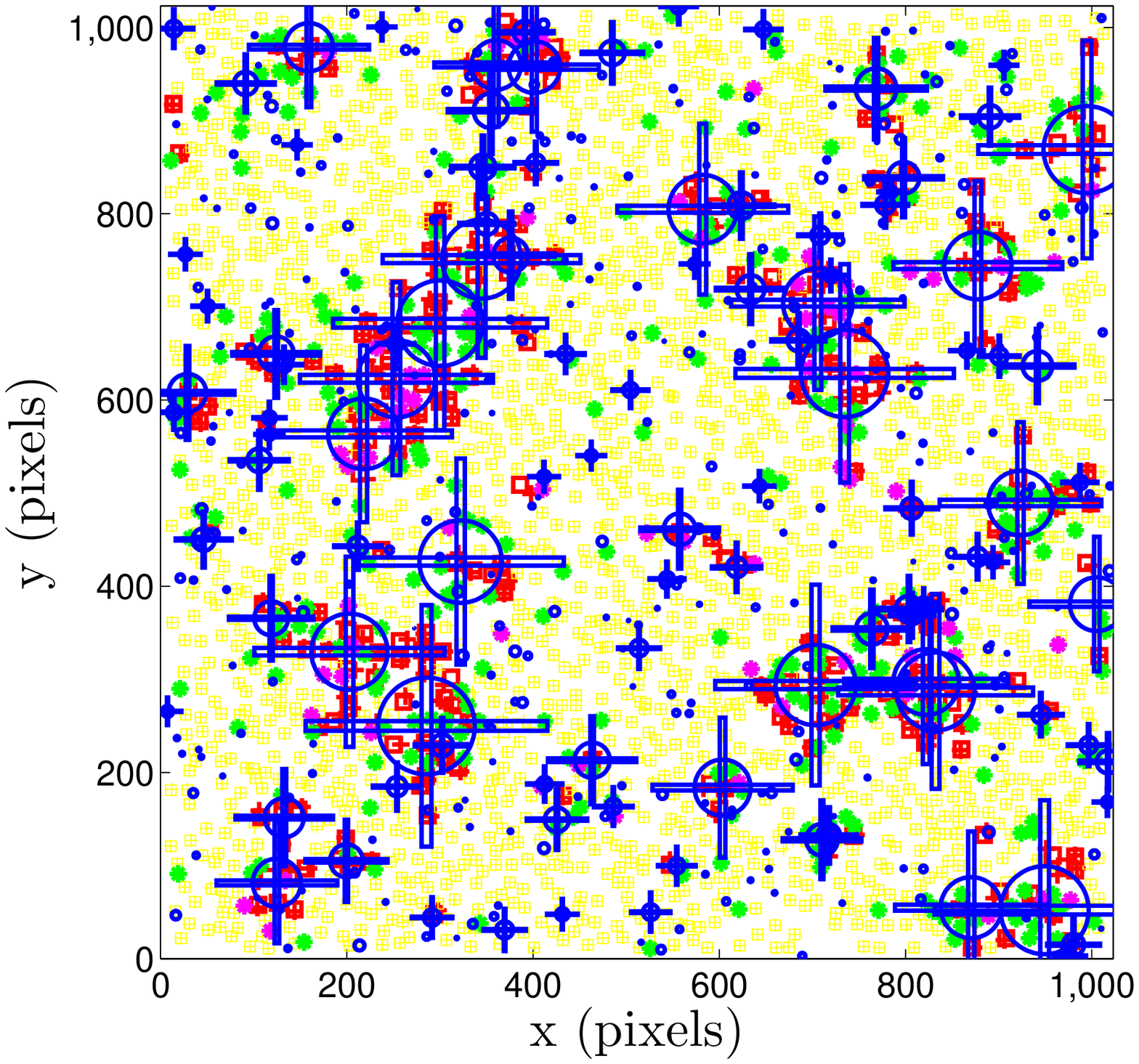}
\end{minipage}
\caption {The illustration of the dependence of the mask effects on the masked fraction.
From left to right, the masked area fraction is $\sim 7\%$, $\sim 13\%$, and $\sim 19\%$
($N_{mask}=140, 280$, and $420$ in $9\hbox{ deg}^2$), respectively. The meanings of the symbols are the same as those in the left
panel of Fig.~\ref{fig:largeinfluence}.
}\label{fig:fraction}
\end{figure*}

\begin{figure*}[tbl]
\begin{minipage}[c]{0.33\textwidth}
  \centering
  \includegraphics[width=1.1\textwidth]{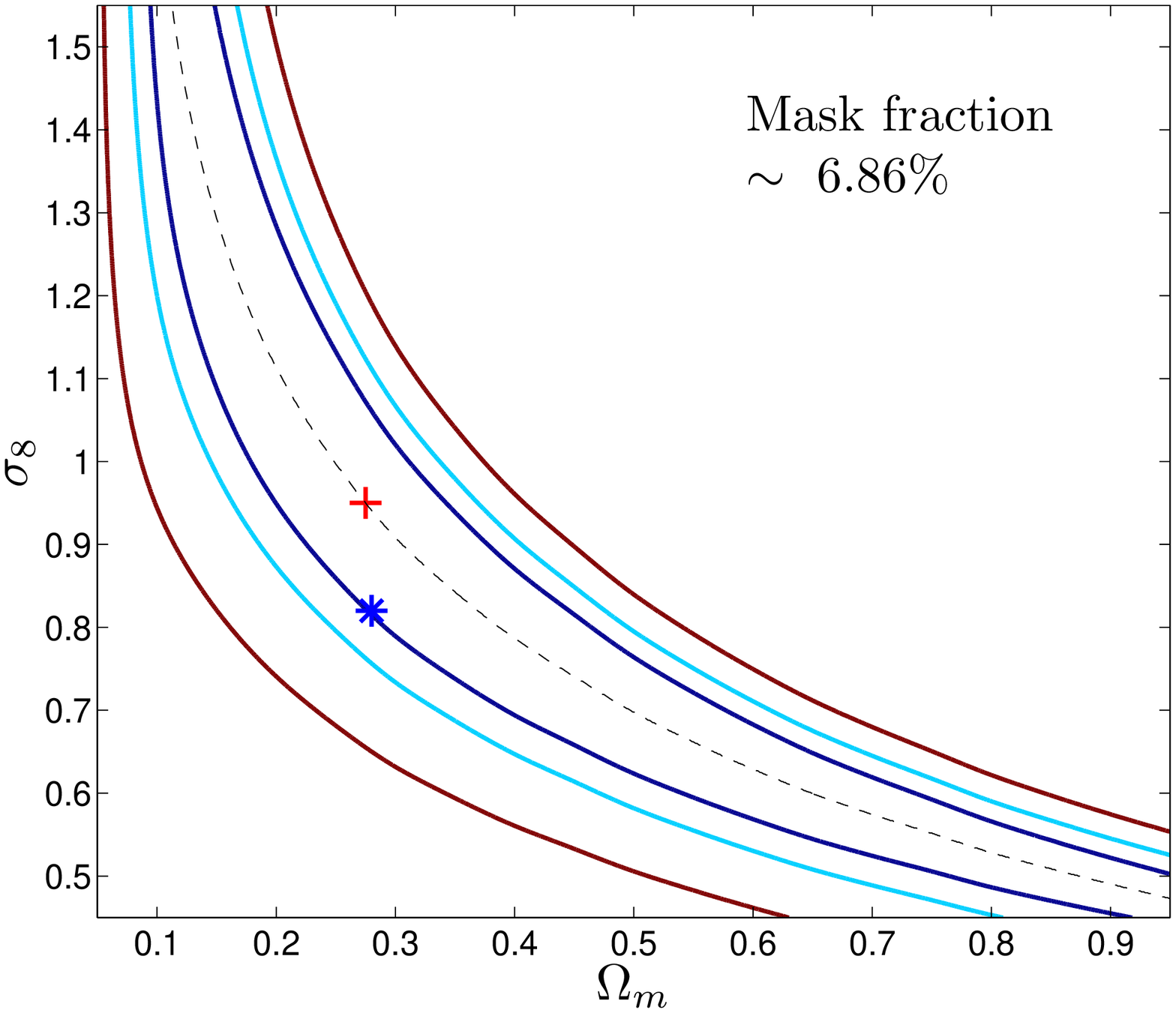}
\end{minipage}
\begin{minipage}[c]{0.33\textwidth}
  \centering
  \includegraphics[width=1.1\textwidth]{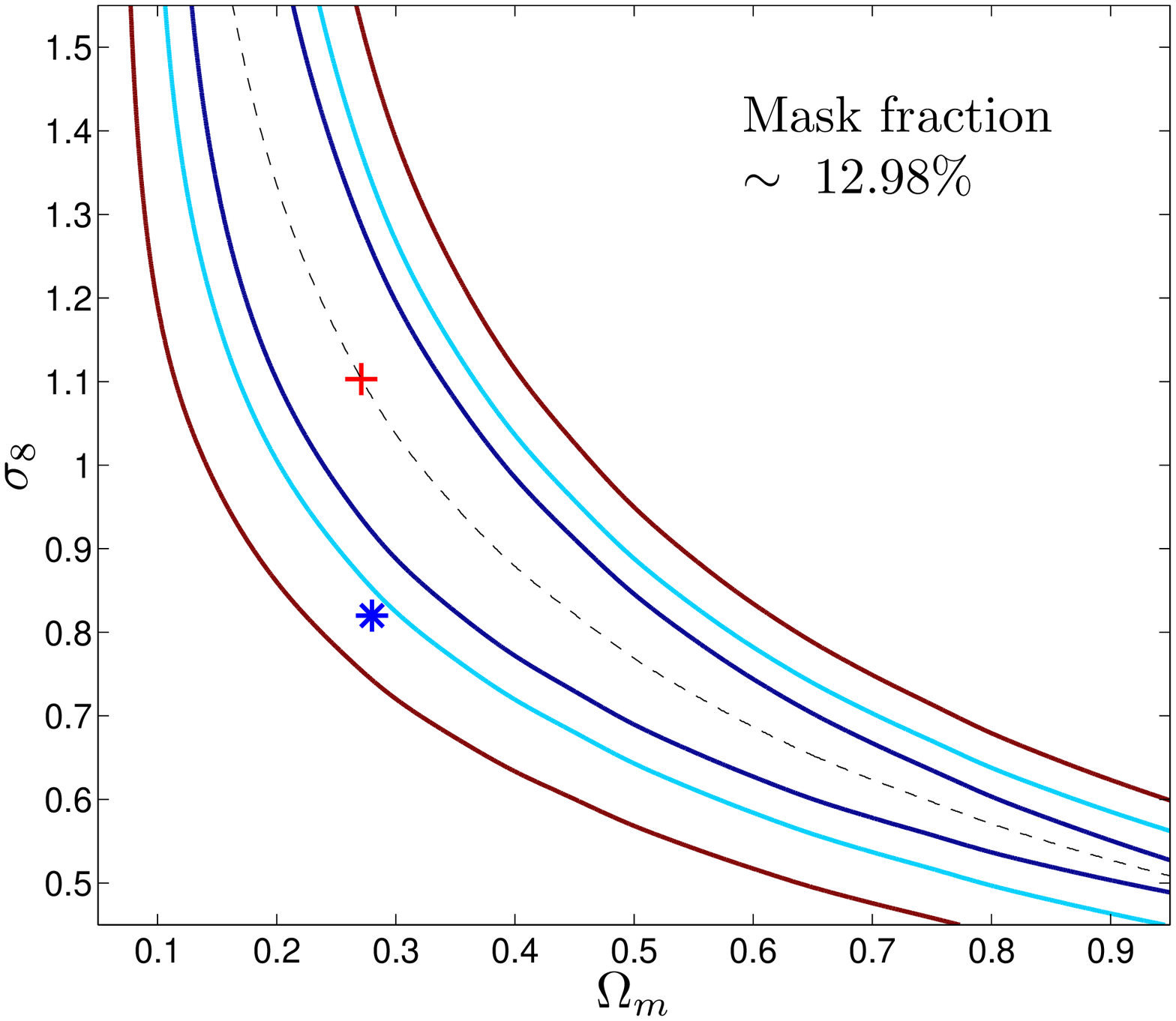}
\end{minipage}
\begin{minipage}[c]{0.33\textwidth}
  \centering
  \includegraphics[width=1.1\textwidth]{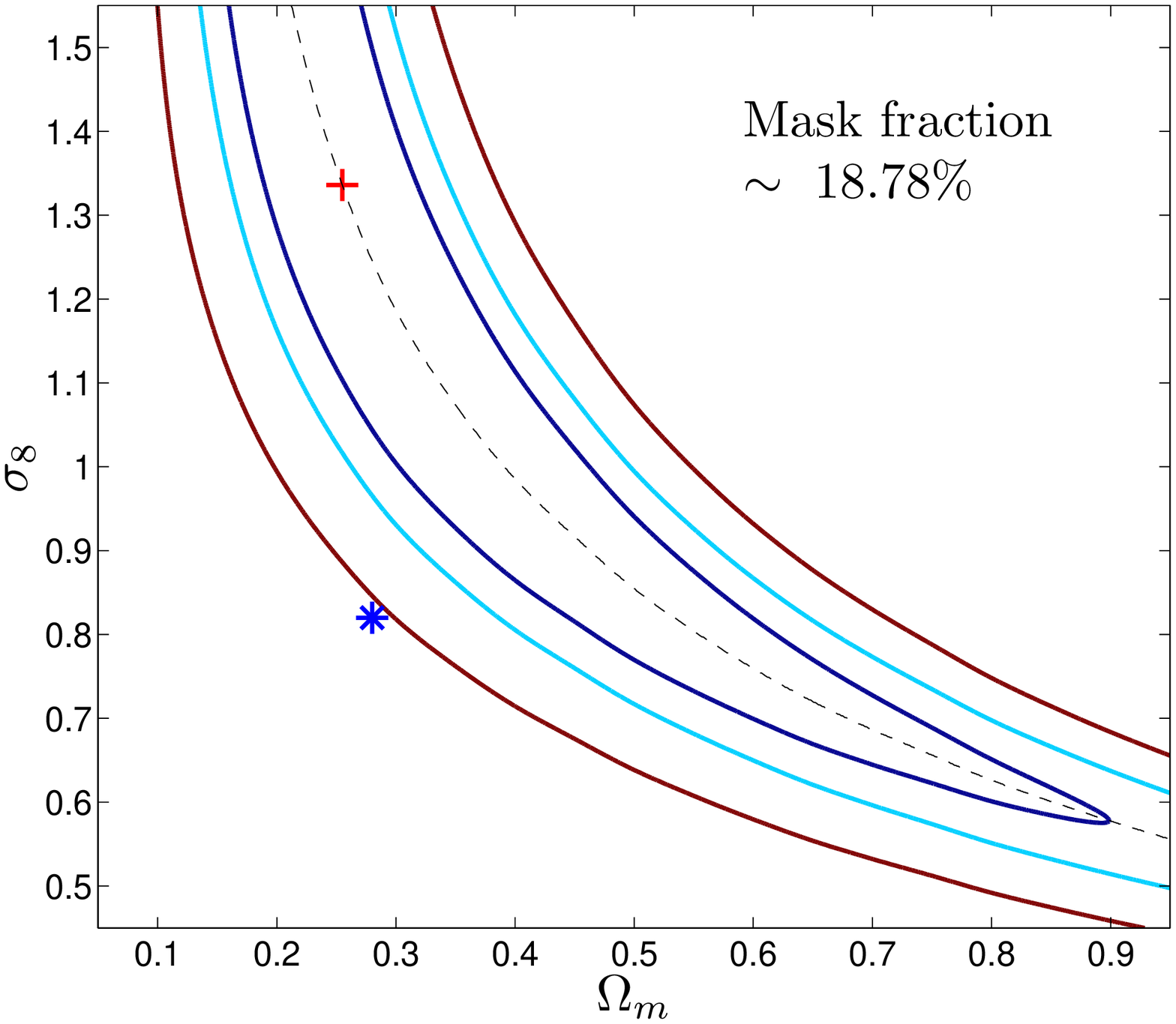}
\end{minipage}
\caption {The corresponding results of cosmological constraints for different masked fractions shown in Fig.~\ref{fig:fraction}.
}\label{fig:fractionfit}
\end{figure*}

The above results are shown for the average masked area fraction of $\sim 13\%$ with $N_{mask}=280$ in $9\hbox{ deg}^2$. We also analyze how the effects depend on the masked fraction. We consider three cases with the number of masks $N_{mask}=140, 280$ and $420$ in $9\hbox{ deg}^2$ and the corresponding masked fraction of $\sim 7\%$, $\sim 13\%$ and $\sim 19\%$, respectively. The peak statistics are listed in Table 2. The mask effects are clearly stronger for larger masked fraction. The fraction of peaks with $\nu> 3$ is about $7\%$ in the case without masks. This fraction increases to $\sim 9\%$, $\sim 11\%$ and $\sim 13\%$ for $N_{mask}=140, 280$ and $420$, respectively. More than $90\%$ and $70\%$ of Type I and Type II affected peaks, respectively, are within the regions around masks with a size of twice the mask radius. The results are further visually illustrated in Fig.~\ref{fig:fraction} with all the symbols the same as those shown in left panel of Fig.~\ref{fig:largeinfluence}. The corresponding fitting results for the survey area of $9\hbox{ deg}^2$ are shown in Fig.~\ref{fig:fractionfit}. We see that with the increase of the masked fraction, the effects become larger. For the masked fraction of $\sim 19\%$, the bias for $(\Omega_m, \sigma_8)$ is already larger than $3\sigma$ for a $9\hbox{ deg}^2$ survey.

\subsection{Mask effects correction}

We have demonstrated in \S5.1 that the mask effects on weak lensing peak counts are significant. The subsequent cosmological parameter
constraints are largely biased if they are not taken into account properly. We therefore need to explore ways to control the mask effects
on cosmological applications with weak lensing peak accounts.

\begin{figure*}[tbl]
% \plottwo{Fig15a.eps}{Fig15b.eps}
\begin{minipage}[c]{0.5\textwidth}
  \centering
  \includegraphics[width=1\textwidth,height=0.8\textwidth]{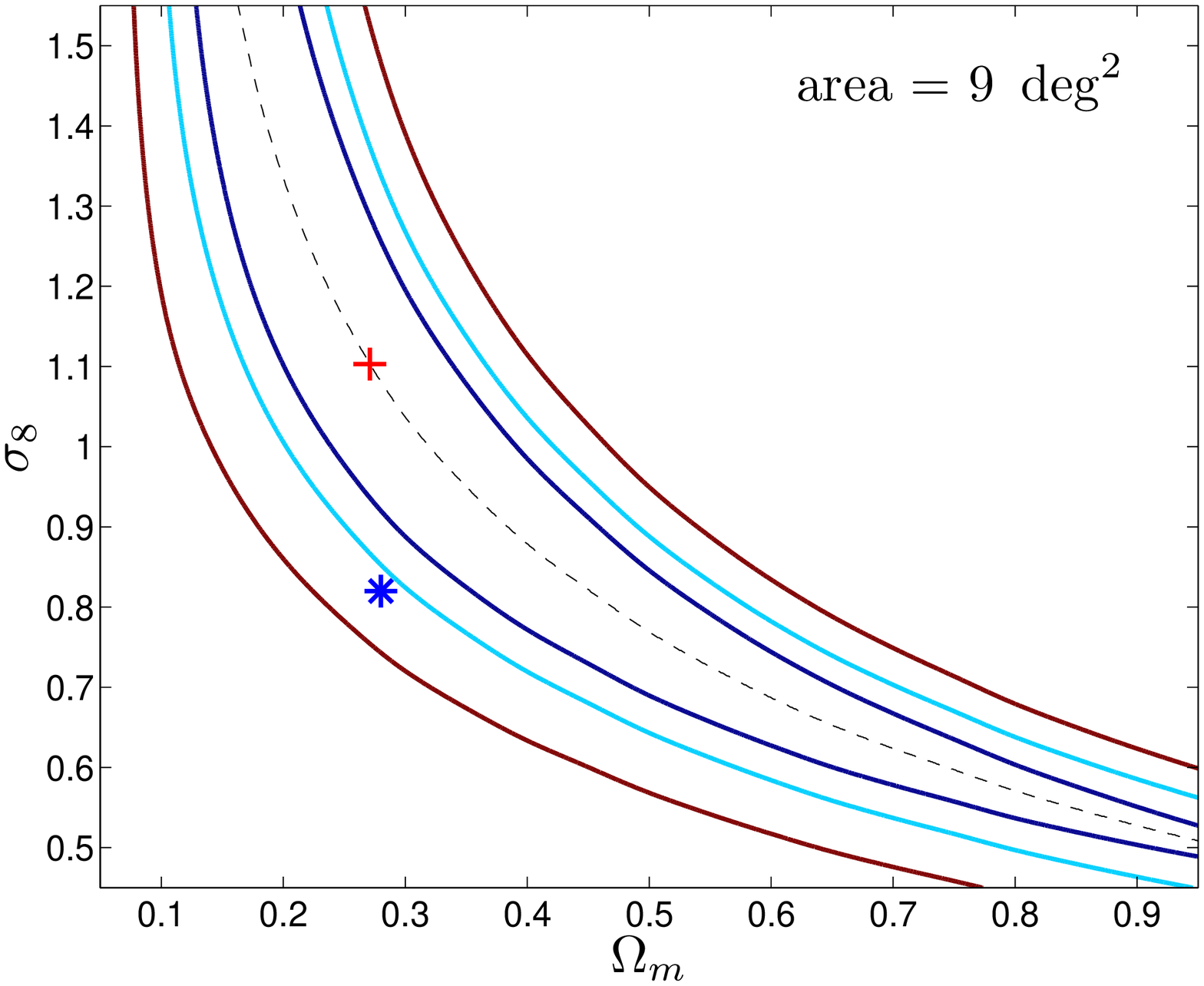}
\end{minipage}
\begin{minipage}[c]{0.5\textwidth}
  \centering
  \includegraphics[width=1\textwidth,height=0.8\textwidth]{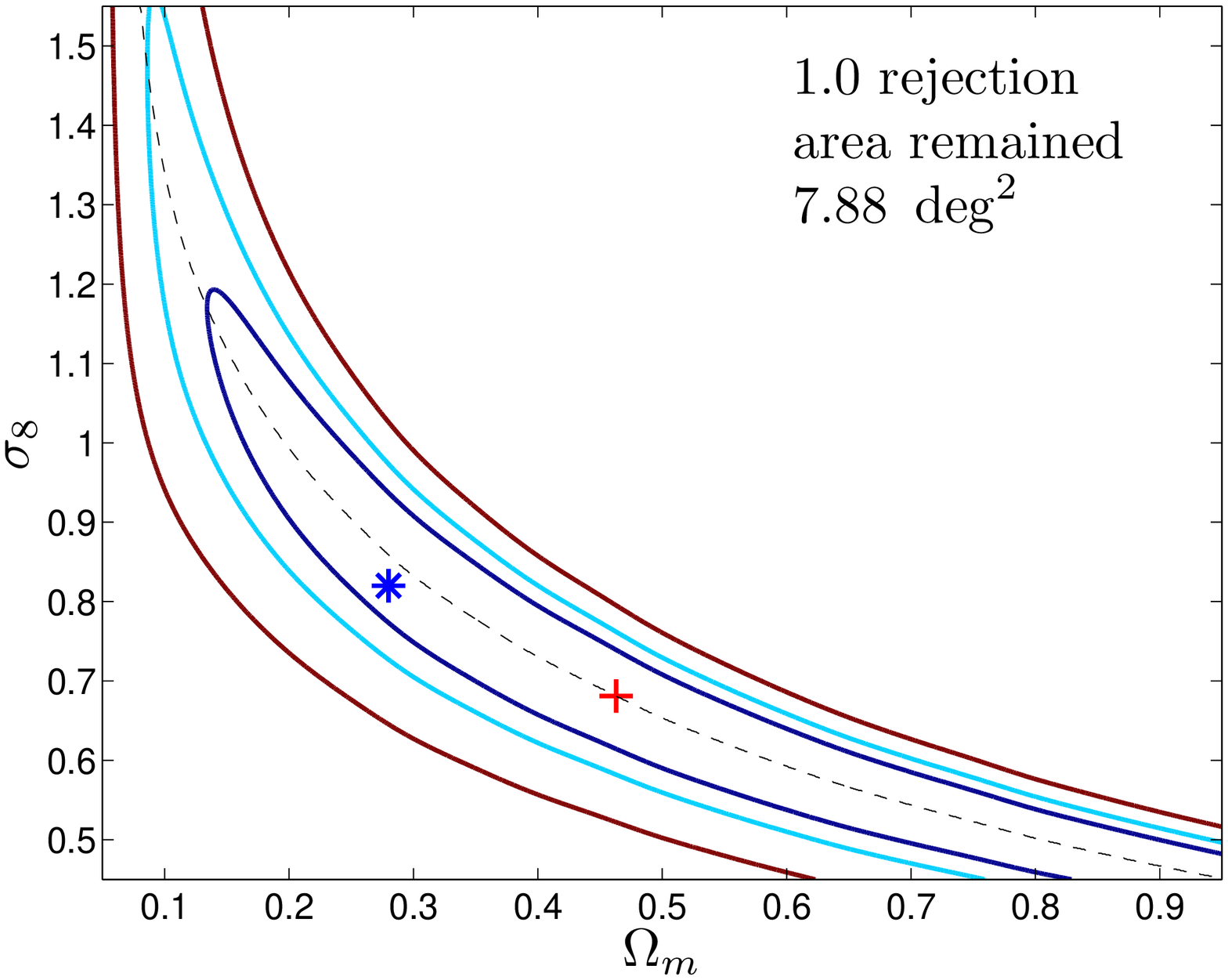}
\end{minipage}
% \plottwo{Fig15c.eps}{Fig15d.eps}
\begin{minipage}[c]{0.5\textwidth}
  \centering
  \includegraphics[width=1\textwidth,height=0.8\textwidth]{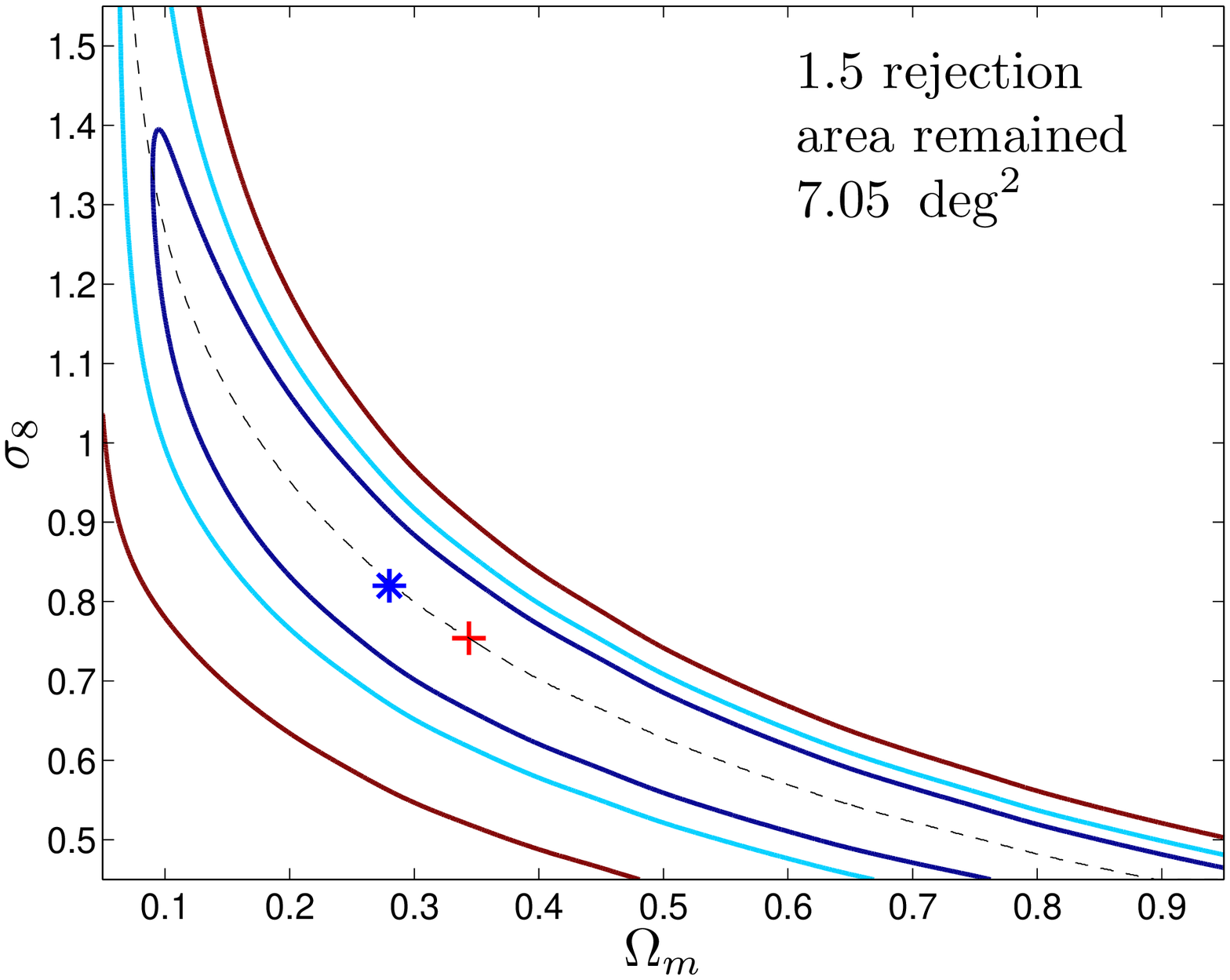}
\end{minipage}
\begin{minipage}[c]{0.5\textwidth}
  \centering
  \includegraphics[width=1\textwidth,height=0.8\textwidth]{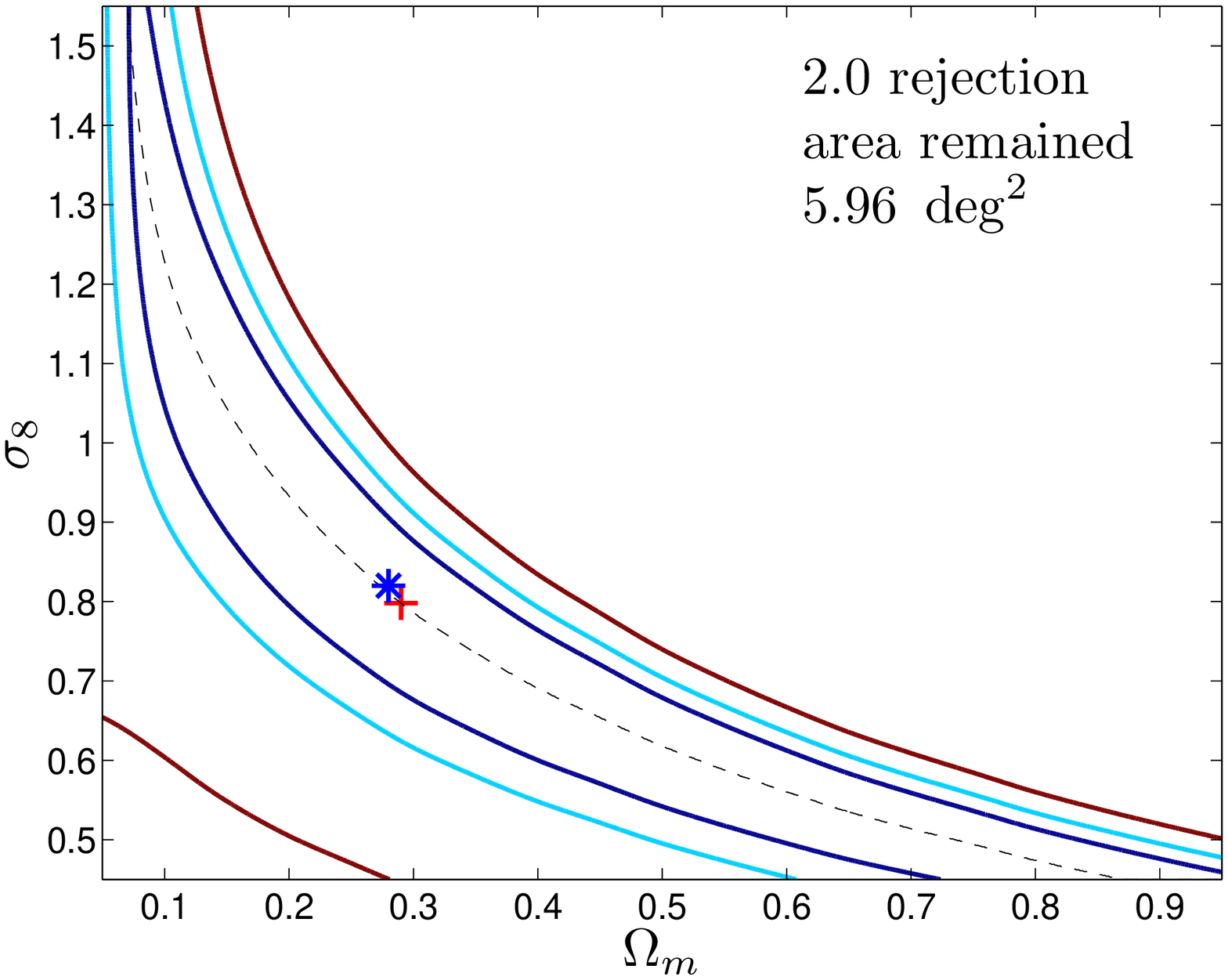}
\end{minipage}
\caption {Results after rejection of regions around masks. The upper left panel is the same as the left panel of Fig.~\ref{fig:maskfitting},
showing the fitting result without any rejections.
%shows the peak number distribution per map with the blue, black and red histograms for peak counts without rejection, with
%rejection of regions of $1.5$ times the mask size around masks and the model prediction of Fan et al. (2010).
The upper right panel shows the fitting results with the rejected regions the same as the masked regions.
The lower left and right panels are the results for rejections of regions with $1.5$ and $2$ times of mask size, respectively.
Here the number of masks is $N_{peak}=280$ in $9\hbox{ deg}^2$ and the corresponding masked area fraction is $13\%$. }
\label{fig:rejection}
\end{figure*}

From Table 2 and Fig.~\ref{fig:largeinfluence}, we see that the strongly affected peaks are mostly clustered around masks.
Thus the first method we use to suppress the mask effects is to exclude the severely affected regions around masks when preforming the peak counting.
It is expected that the bias on cosmological parameters can be considerably removed but inevitably at the expense of losing effective survey areas
and therefore enlarging the statistical error contours. We name this method as the rejection method. We consider three cases with the rejection
regions of $1$, $1.5$ and $2$ times of the mask size around each mask. We pay attention to the mask overlaps. Fig.~\ref{fig:rejection} shows the results,
where the model of F10 with a uniform noise of $\sigma_0=0.02$ is used in the fitting.
%The upper left panel shows the peak number distributions. The blue histogram is for the \textbf{masked peak counts simply
%scaled to the residual area $7.05~ deg^2$ after rejecting regions of $1.5$ times of the mask size}.
%The green one is for the \textbf{real} peak counts after rejecting regions of $1.5$ times of the mask size. The red one is the model prediction of Fan et al. (2010)
%with a uniform noise level of $\sigma_0=0.02$.
The upper left panel shows the result without any rejections, which is the same as the left panel of Fig.~\ref{fig:maskfitting}.
The upper right, lower left and lower right panels show the fitting results for the three considered rejections,
from the smallest to the largest rejections, respectively. Note that in each case, the covariance matrix used in the fitting is recalculated
with the peak counts from the $128$ maps with the corresponding rejections. It is seen that while it is reduced significantly, the bias is
still apparent with the rejection of only the masked areas in peak counting (upper right panel).
By rejecting regions of $1.5$ times of the mask size around masks, the bias is suppressed to an insignificant level noting the
degeneracy direction between the two parameters (lower left panel). To increase the rejection areas further leads to a mild improvement of the fitting result
(lower right panel). On the other hand, we see that with the increase of the rejection areas, the confidence contours become larger as expected.
We thus conclude that rejecting regions of $1.5\sim 2$ times of mask size around masks is an optimal choice
for controlling the bias in cosmological parameter constraints without loosing statistics significantly.

\begin{table*}[tbl]
\centering
\caption{Mask effects on peak statistics with different mask fractions.\label{Table 2}}
{
\begin{tabular}{ccccccccc}
\hline \hline
$f_{mask}$\footnotemark[1] \,    &    $N_{mask}$\footnotemark[2]  \,   &    $f_{nocorr}$\footnotemark[3]  \,   &  $f_{offset}$\footnotemark[4]  \,  &   $f_{\nu\,>3}$\footnotemark[5]  \, &   $f_{m,\nu\,>3}$\footnotemark[6]  \,  &  $f_{LPin}$\footnotemark[7]  \,  &  $f_{NCin}$\footnotemark[8] \,  &  $f_{LIinALL}$\footnotemark[9] \, \\
\hline
$\sim\,7\%$       &       $140$         &   $6.29\%$        &   $24.05\%$    &   $6.67\%$        &    $8.53\%$       &    $92.67\%$  &   $72.26\%$        &  $40.54\%$     \\
$\sim\,13\%$      &       $280$         &   $10.97\%$        &   $40.03\%$    &   $6.67\%$        &    $10.93\%$       &    $93.59\%$  &   $77.83\%$        &  $44.13\%$     \\
$\sim\,19\%$      &       $420$         &   $15.00\%$        &   $52.11\%$    &   $6.67\%$        &    $12.53\%$        &   $94.01\%$  &   $82.24\%$        &  $47.45\%$     \\
\hline \hline
\end{tabular}
}
\footnotesize
\footnotetext[1]{Masked area fraction}
\footnotetext[2]{Number of masks in $9\hbox{ deg}^2$}
\footnotetext[3]{Fraction of no-correspondence peaks among the total number of peaks}
\footnotetext[4]{Fraction of peaks with spatial offset larger than $0.1\,arcmin$.}
\footnotetext[5]{Fraction of peaks with $S/N\,>3$ in the mask-free case}
\footnotetext[6]{Fraction of peaks with $S/N\,>3$ in the case with masks}
\footnotetext[7]{Fraction of Type I affected peaks within regions around masks with a size of twice the corresponding masks among the
total number of Type I affected peaks}
\footnotetext[8]{Fraction of no-correspondence peaks within regions of twice the size of masks among the total number of no-correspondence peaks}
\footnotetext[9]{Fraction of (Type I+Type II) affected peaks within regions of twice the size of masks among the total number of
peaks within the regions}
\end{table*}

%\textbf{However, although the rejection method can be an optimal choice to correct the bias, it can loose statistics significantly
%when the survey area become larger and larger. Thus, we now consider another approach in which we can keep as much survey area as we can in the peak counting.}

We also explore ways to improve our theoretical modeling to take into account the mask effects properly.
From Fig.~\ref{fig:largeinfluence}, we see that the significantly affected peaks are closely associated with masked regions,
especially those of large masks. Therefore for theoretical modeling, we need to treat these masked regions separately
from the rest part of the survey area. For our peak abundance analyses, the presence of masks mainly affects
the number of galaxies that are usable in obtaining the smoothed ellipticity field $\langle\boldsymbol \epsilon\rangle$
around the masked regions. This in turn leads to non-uniform noises in the convergence field reconstructed from $\langle\boldsymbol \epsilon\rangle$
with higher noises near masks than that for the area away from them.
These higher noises affect the peak counts in regions around masks in two ways. One is that the systematic
peak height shift for true peaks is larger in these regions (F10). The other is the enhancement of the number of
noise peaks given their peak heights measured in signal-to-noise ratio with $\sigma_0$ still taken to be $0.02$, the value in the mask-free case.
The latter can be understood by noting that for different two-dimensional Gaussian random fields, their peak number density distributions
are the same if the peak heights in each field are measured in the signal-to-noise ratio defined with the noise level $\sigma_0$
of the field itself \citep[e.g.,][]{vW2000}. Thus if we use a fixed $\sigma_0$ to define the signal-to-noise ratio for peaks in different Gaussian random fields,
the peak number distribution would be different for different Gaussian random field. Here due to the presence of masks,
the true noise levels around the masks are higher. When counting peaks, however,
we use the fixed value $\sigma_0=0.02$ (corresponding to $n_g=30\hbox{ arcmin}^{-2}$) uniformly to define their signal-to-noise ratios.
Therefore the number density of peaks with high $\nu =K_N/\sigma_0\ge 4$ ($\sigma_0=0.02$) is higher than that of the mask-free case.

%\textbf{To begin with, we notice that regions with masks must be theoretically considered differently. In other words, we need to
%revise our theoretical model. Due to the presence of the masks, fewer galaxies are usable. Therefore the noise level in mask regions
%is systematically higher. This higher noise affects the peak counts in regions around masks in two ways. One is that the systematic
%peak height shift for true peaks is larger than regions ways from masks \citep{F2010}. The other is the enhancement of the number of
%noise peaks given their peak heights measured in signal-to-noise ratio with $\sigma_0$ still be $0.02$, the value in mask-free case.
%The latter can be understood by noting that for any two-dimensional Gaussian random fields, their peak number density distributions
%are the same if the peak height is measured in signal-to-noise ratio with the noise level $\sigma_0$ different for different
%noise field \citep[e.g.,][]{vW2000}. Here due to the presence of masks, the true noise level is higher. When counting peaks, however,
%we still use $\sigma_0=0.02$ (corresponding to $n_g=30\hbox{ arcmin}^{-2}$) uniformly to define the signal-to-noise ratio of peaks.
%Therefore the number density of peaks with high $\nu=K_N/\sigma_0$ ($\sigma_0=0.02$) is higher than the mask-free case.}

For further quantitative analyses of the non-uniform noise, we calculate the contribution of source galaxies
to each grid point in constructing the smoothed field of $\langle\boldsymbol \epsilon\rangle$ by

\begin{equation}
% \begin{split}
 R^{e}(\boldsymbol \theta) = \frac{\sum_{k=1}^{N_{gal}}R_k(\boldsymbol \theta_k)W(\boldsymbol \theta_k-\boldsymbol \theta)}
{\sum_{k=1}^{N_{gal}}W(\boldsymbol \theta_k-\boldsymbol \theta)},
% n_g^{e}=R^{e}n_g
% \end{split}
\label{effectw}
\end{equation}
where the summation is over all galaxies with $R_k=1$ for galaxies outside masks and $R_k=0$ for galaxies inside masks.
The kernel $W$ is taken to be the Gaussian smoothing function with $\theta_G=1\arcmin$ consistently.
The effective number density of source galaxies at each grid point
is then estimated by $n_g^{e}(\boldsymbol \theta)=R^{e}(\boldsymbol \theta)n_g$, where $n_g=30\hbox{ arcmin}^{-2}$ here.
We find that $R^e$ can be significantly smaller than $1$ in large mask regions.

Considering the fact that the noise cannot be suitably modeled as a Gaussian random field in regions with $R^e\ll1$, we exclude the circular masked regions with
the mask radius larger than $3\arcmin$ from our peak counting analyses. We name these regions as Part I regions which on average occupy
about $1\deg^2$ over the total $9\deg^2$ survey area in our studies. For the remaining $\sim 8\deg^2$ area, we develop a two-noise-level model
to include the non-uniform noise in our theoretical considerations. Specifically, for each of the $128$ reconstructed convergence maps,
we first exclude Part I regions. We then divide the remaining area of each map further into two parts, with
Part II being the left over spiked mask regions around the excluded circular regions of Part I,
and Part III for the rest of the area. We then calculate the effective number density of source galaxies for Part II and Part III
separately by averaging $n_g^e$ over the grid points inside the corresponding regions and over all the $128$ maps. For Part II, we obtain
$n_g^{eII}\approx 11.4 \hbox{ arcmin}^{-2}$, considerably smaller than $n_g=30\hbox{ arcmin}^{-2}$.
For Part III, $n_g^{eIII}\approx 28.4 \hbox{ arcmin}^{-2}$ which is close to $30\hbox{ arcmin}^{-2}$ as expected.

%\begin{equation}\label{51}
%% \begin{split}
%% R^{e}= \frac{\sum_{k=1}^{N_{gal}}Ru_k}{\sum_{k=1}^{N_{gal}}u_k},\\
% n_g^{e}=R^{e}n_g.
%% \end{split1}
%\end{equation}
%where $n_g=30\hbox{ arcmin}^{-2}$ here.

%\textbf{We notice that the effective number density of background galaxies are much more suppressed within large masks
%than the regions away from the masks. Within the large masks, the effective number density of background galaxies in the
%center circle regions is much less than the remained large mask regions, which can even close to 0. Meanwhile,
%from Fig.~\ref{fig:largeinfluence}, we also notice that most abnormal peaks are within large masks, i.e. masks with radius
%larger than 3 arcmins, especially within those center circle regions. Thus, based on these considerations, we consider
%separately about the large mask regions and other part regions and divide the whole survey area into three parts: Center
%circle regions of those masks whose radius larger than 3 arcmins (PartI), rest parts of those large masks (PartII) and
%other survey regions (PartIII). Consider that the effective number density with PartI is largely decreased due to the presence
%of these large masks, so that the shear information is mostly lost and the peaks within this region are mostly false peaks,
%from numerical reconstruction, we consider to reject PartI and remain the other two parts. Hence, the average remained area is $\sim 8~deg^2$ for each map.}

\begin{figure*}[tbl]
% \plottwo{Fig16a.eps}{Fig16b.eps}
\begin{minipage}[c]{0.5\textwidth}
  \centering
  \includegraphics[width=1\textwidth,height=0.8\textwidth]{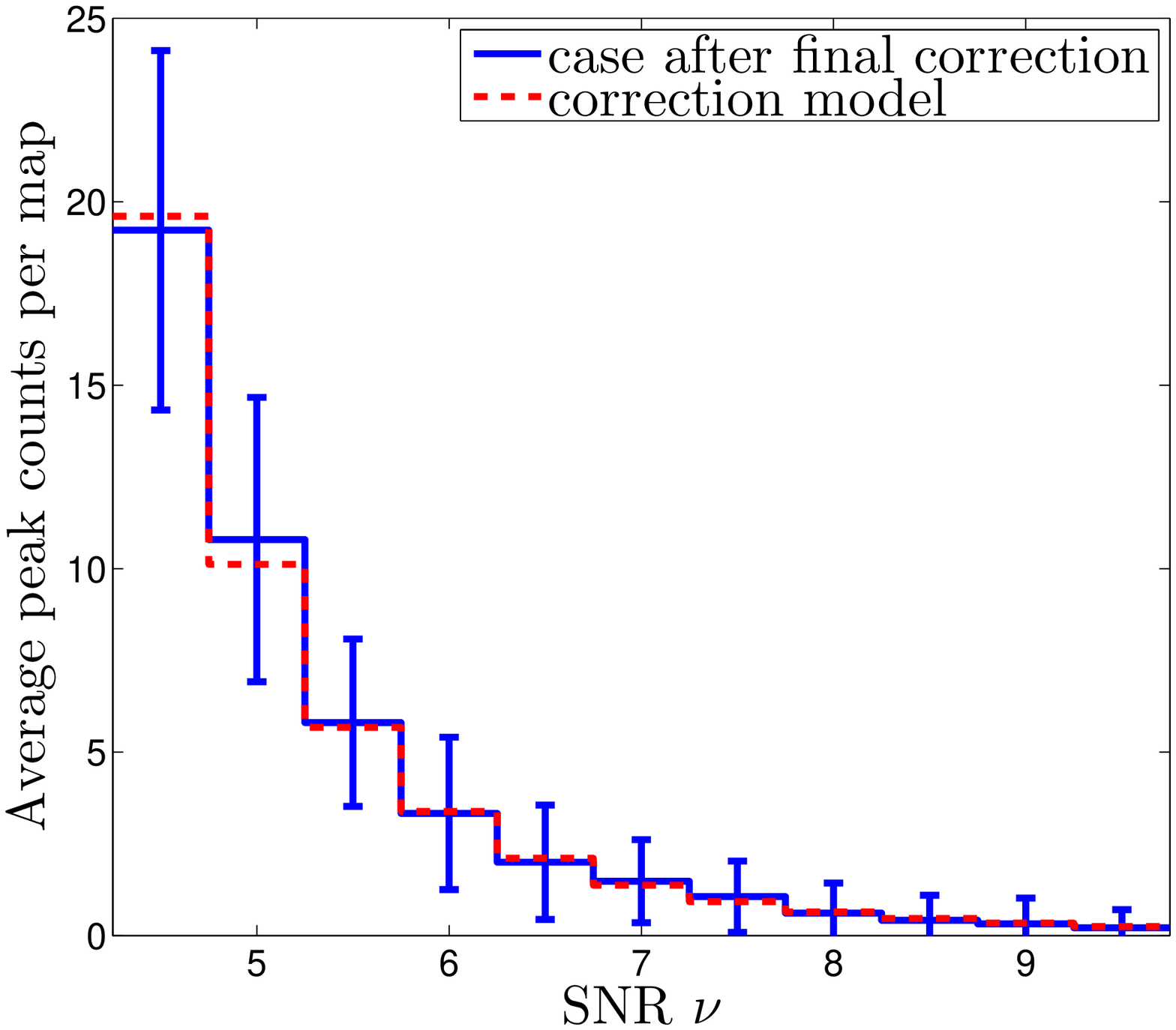}
\end{minipage}
\begin{minipage}[c]{0.5\textwidth}
  \centering
  \includegraphics[width=1\textwidth,height=0.8\textwidth]{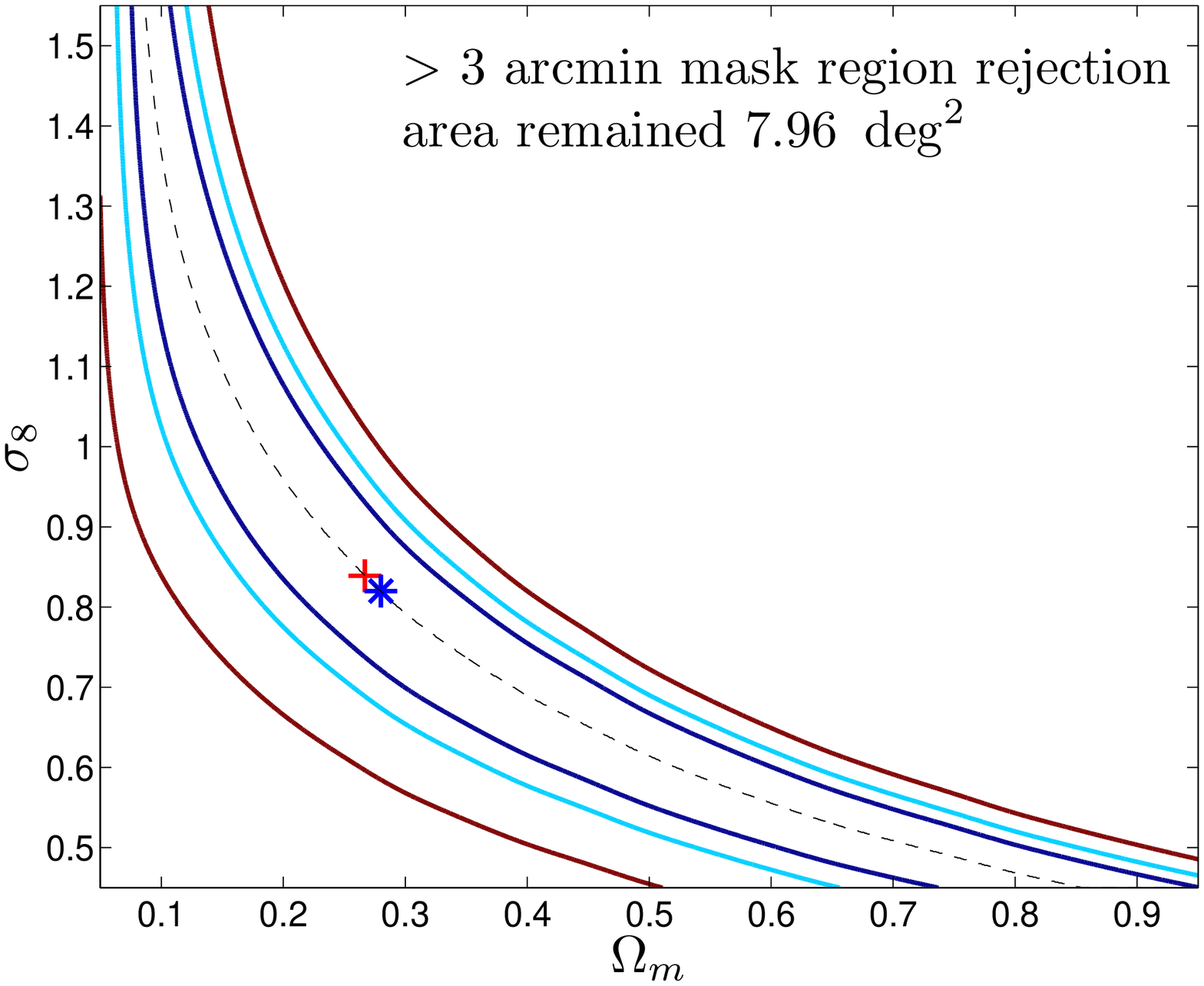}
\end{minipage}
\caption { Left: Average peak counts. The blue histograms with error bars are for the results from masked reconstructed maps excluding Part I regions.
The red histograms are calculated with our two-noise-level model. Right: The corresponding cosmological constraints. Here $N_{peak}=280$.
}
\label{fig:final}
\end{figure*}

%\textbf{For the remained two parts (PartII and Part III), we develop a two-noise-level model that takes into account the
%lower number density of usable galaxies in these two regions. According to Eqn.~\ref{50} and Eqn.~\ref{51}, We calculate
%the effective number density of usable galaxies for PartII $n_g^{ePartII}$ by averaging $n_g^e$ of all the grid points
%within PartII. We obtain $n_g^{ePartII}\approx 11.4\hbox{ arcmin}^{-2}$ for the case with the masked
%fraction of $13\%$ considered here. For the PartIII regions, the effective number density $n_g^{ePartIII}$ is calculated
%similarly by averaging all the grid points in the PartIII regions. We have $n_g^{ePartIII}\approx 28.4\hbox{ arcmin}^{-2}$,
%close to $n_g=30\hbox{ arcmin}^{-2}$ as expected.}

With $n_g^{eII}$ and $n_g^{eIII}$, we calculate the expected number density of peaks
with the model of F10 separately for Part II and Part III regions. We then rescale the signal-to-noise ratios of the peaks in the two regions
by a uniform noise level $\sigma_0=0.02$, the value used in counting peaks from simulated maps, to obtain the
rescaled number density of peaks $n_{peak}^{II}$ and $n_{peak}^{III}$, respectively.
The average areas of the two regions $S^{II}$ and $S^{III}$ over a map are computed from the $128$ reconstructed maps with masks
by $S^{II}=(\sum_{i=1}^{128} S^{II}_i)/128$ where $S^{II}_i$ is the area of Part II in map $i$, and similarly for $S^{III}$.
The theoretical predictions for the total number of peaks in each S/N bin (with $\sigma_0=0.02$) over a map is
then calculated by $n_{peak}^{II}S^{II}+n_{peak}^{III}S^{III}$.
These theoretical predictions are then compared with the corresponding mean `observed' numbers of peaks obtained by averaging over
the $128$ counted numbers of peaks directly from the reconstructed convergence maps with masks after the exclusion of Part I regions.
We note that in this treatment, the effective usable area over a map $S^{II}+S^{III}\approx 8\deg^2$, in comparison with $\sim 7\deg^2$ and
$\sim 6\deg^2$ for the pure rejection analyses with the rejection areas of $1.5$ times and $2$ times of the mask size around each mask, respectively
(see Fig.~\ref{fig:rejection}.)

The results are shown in Fig.~\ref{fig:final} with the left panel for the peak counts and the right panel for the fitting results
from our two-noise-level model. The blue, and red histograms in the left panel correspond to the results from the reconstructed convergence maps with masks
excluding Part I regions, and the theoretical prediction from the two-noise-level model, respectively, where the signal-to-noise ratio
$\nu$ in the horizontal axis is defined with $\sigma_0=0.02$. It is seen that the theoretical predictions agree well with the simulation results.
The right panel presents the corresponding constraint for $(\Omega_m, \sigma_8)$. Comparing to the result shown in the left panel of Fig.~\ref{fig:maskfitting},
we see that our treatment here works well and improves the fitting dramatically with a much reduced bias.
%Thus the two-noise-model works very well and can potentially be applied in cosmological studies with weak lensing peak counts in the presence of masks.

\section{Summary and discussion}

In this paper, we analyze the mask effects on weak lensing convergence peak statistics and the consequent cosmological parameter
constraints from weak lensing peak counts.
We run large sets of ray-tracing simulations to generate base convergence and shear maps assuming the source redshift $z_s=1$.
By randomly populating source galaxies with intrinsic ellipticities, we perform convergence reconstruction from $\langle\boldsymbol \epsilon\rangle$,
the smoothed field of the `observed' ellipticities of source galaxies, for cases without and with masks, respectively. The mask size distribution
from \cite{Shan2012} is adopted. We then investigate in detail the mask effects on weak-lensing peak counts by comparing the results from the two cases.
Their influences on cosmological parameter constraints derived from peak abundances are further studied using the peak model of F10
including the noise effects.  The validity of this model in terms of the cosmological dependence of peak abundances are tested with simulations.
% are generate with the source redshift $z_s=1$. We generate source galaxy catalogs by randomly populating galaxies on the
%source plane with $n_g=30\hbox{ arcmin}^{-2}$. A random intrinsic ellipticity based on the truncated Gaussian distribution
%with the total $\sigma_{\epsilon}=0.4$ is assigned to each galaxy. The mask catalogs are produced according to the mask size
%distribution of CFHTLS. \textbf{For each of the $128$ base weak lensing simulations, we generate $1$ noisy maps without/with
%masks with different assigned intrinsic ellipticities and different masks in the masked case.} \textbf{We therefore have totally $128$
%convergence maps for each case.} We use the theoretical model of Fan et al. (2010) for weak lensing peak counts to analyze the
%cosmological parameter constraints in different cases. The model takes into account fully the noise effects arising from intrinsic
%ellipticities so that we can use directly peaks from convergence maps in cosmological studies without the need to find true peaks
%associated with clusters of galaxies. We test the model applicability with the `observed data' from simulations without masks and find
%that the derived cosmological parameters agree with the fiducial input of the simulations very well without significant bias.

Our main results are summarized as follows.

(1) The occurrence of masked regions reduces the number of usable source galaxies and therefore increases the noise in the regions
around masks. This in turn leads to systematic increases of the number of high peaks and consequently a significant bias in
cosmological parameters constrained from weak lensing peak counts. The larger the masked area fraction is, the larger the effects are.

(2) We find that the strongly affected region around a mask is about $1.5\sim 2$ times of the mask size. Excluding such regions in
peak counting can eliminate largely the mask effects and therefore reduce the bias in cosmological parameter constraints significantly
%derived from weak lensing peak abundance significantly.

(3) We develop a two-noise-level model that treats the mask affected regions separately. This model
can account for the mask effects on weak lensing peak counts very well except for very large masked regions with
radius larger than $3\arcmin$ where the noise cannot be suitably modeled as a Gaussian random field. These very
large masks need to be excluded in peak analyses. Then the constraints on cosmological parameters
based on the two-noise-level model improve dramatically comparing to the large bias from the model with a uniform noise.

In our analyses, we apply the Kaiser-Squires method for the nonlinear convergence reconstruction with a Gaussian filter.
For the maximum-likelihood reconstruction method \citep{Bar1996}, we expect that the mask effects on the reconstructed convergence field and
the peak counts are qualitatively similar to the results shown in this paper although quantitative studies are still needed.
For other methods, such as the multi-scale entropy restoration filtering, namely MRLens \citep{Starck2006}, the mask effects
can be different and detailed analyses should be carried out when a specific reconstruction method is used.
It is also noted that while qualitatively similar mask effects on peak abundances are expected,
different filter functions used in the convergence reconstruction and different peak binning methods used in the
analyses can lead to quantitatively different results.

The model of F10 and the improved two-noise-level model for the mask effects contain
only the noise effects without including the projection effects of large-scale structures and the complex mass distribution of dark matter halos.
While the noise from intrinsic ellipticities is the dominant source of errors in weak-lensing peak analyses
and our results show that the model(s) can indeed give rise to very good descriptions of the peak counts,
future large surveys aiming at high precision cosmological studies need more accurate modeling of the peak counts theoretically.
We will explore further improvements of the model carefully in our future studies.

%We note that the two-noise-level model works well to take into account the mask effects on weak lensing peak counts. Detailed inspection
%finds, however, that the model prediction for the dependence of peak counts on the peak height is slightly steeper
%than that from simulations. To improve the model, we may need to treat the mask affected regions more accurately by
%further dividing them into different sub-regions with different noise levels. We will investigate along this line of approach
%thoroughly in the future.

\section{Acknowledgement}
We thank the referees for the comments that help to improve the paper significantly.
We are grateful for the discussions with David Wittman that stimulate the studies on the mask effects, and with Huanyuan Shan and Ran Li. We also sincerely thank the support of Hu Zhan from National Astronomical Observatories, Chinese Academy of Sciences.
This research is supported in part by the NSFC of China under grants 11333001, 11173001 and 11033005, and the 973 program No. 2007CB815401.

%\newpage

\clearpage
\end{document}